\documentclass[preprint]{aastex}
\usepackage{amsmath,amsbsy}
\newcommand{\col}[1]{\mbox{Col.\ #1}}
\newcommand{\Col}[1]{\mbox{Column\ #1}}
\newcommand{\se}[1]{\mbox{Sect.\ \ref{sec:#1}}}
\newcommand{\ses}[2]{Sects.\ \ref{sec:#1} and \ref{sec:#2}}
\newcommand{\Se}[1]{\mbox{Section\ \ref{sec:#1}}}
\newcommand{\eq}[1]{\mbox{Eq. \ref{eq:#1}}}
\newcommand{\Eq}[1]{\mbox{Equation\ \ref{eq:#1}}}
\newcommand{\fg}[1]{\mbox{Fig.\ \ref{fig:#1}}}
\newcommand{\eqs}[2]{Eqs.\ \ref{eq:#1} and \ref{eq:#2}}

\newcommand{\Fg}[1]{\mbox{Figure\ \ref{fig:#1}}}
\newcommand{\Tb}[1]{\mbox{Table\ \ref{tab:#1}}}
\newcommand{\app}[1]{\mbox{Appendix\ \ref{app:#1}}}
\newcommand{\ie}{i.e.,}
\newcommand{\vs}{vs.}
\newcommand{\eg}{e.g.,}
\newcommand{\etc}{etc.}
\newcommand{\cf}{cf.}
\newcommand{\simu}[1]{\texttt{#1}}

\newcommand{\sumi}{(i)}
\newcommand{\sumii}{(ii)}
\newcommand{\sumiii}{(iii)}
\newcommand{\sumiv}{(iv)}

\newcommand{\figw}{0.49\textwidth}
\newcommand{\figww}{0.8\textwidth}
\newcommand{\arrayii}[2]{\left( \begin{array}{c} #1 \\ #2 \\ \end{array} \right)}
\begin{document}
\title{Accretion among preplanetary bodies: the many faces of runaway growth}
\author{C. W. Ormel}
\affil{Max-Planck-Institute for Astronomy, K\"onigstuhl 17, 69117 Heidelberg, Germany}
\affil{Astronomisches Rechen-Institut, Zentrum f\"ur Astronomie der Universit\"at Heidelberg, M\"onchhofstr.\ 12-14, 69120 Heidelberg, Germany}
\email{ormel@mpia.de}

\author{C. P. Dullemond}
\affil{Max-Planck-Institute for Astronomy, K\"onigstuhl 17, 69117 Heidelberg, Germany}
\email{dullemon@mpia.de}

\and

\author{M. Spaans}
\affil{Kapteyn Astronomical Institute, P.O. Box 800, 9700 AV, Groningen, The Netherlands}
\email{spaans@astro.rug.nl}

\begin{abstract}
When preplanetary bodies reach proportions of $\sim$1 km or larger in size, their accretion rate is enhanced due to gravitational focusing (GF).   We have developed a new numerical model to calculate the collisional evolution of the gravitationally-enhanced growth stage.  The numerical model is novel as it attempts to preserve the individual particle nature of the bodies (like $N$-body codes); yet it is statistical in nature since it must incorporate the very large number of planetesimals.  We validate our approach against existing $N$-body and statistical codes.  Using the numerical model, we explore the characteristics of the runaway growth and the oligarchic growth accretion phases starting from an initial population of single planetesimal radius $R_0$.  In models where the initial random velocity dispersion (as derived from their eccentricity) starts out below the escape speed of the planetesimal bodies, the system experiences runaway growth.  We associate the initial runaway growth phase with increasing GF-factors for the largest body.  We find that during the runaway growth phase the size distribution remains continuous but evolves into a power-law at the high mass end, consistent with previous studies.  Furthermore, we find that the largest body accretes from all mass bins; a simple two component approximation is inapplicable during this stage.  However, with growth the runaway body stirs up the random motions of the planetesimal population from which it is accreting.  Ultimately, this feedback stops the fast growth and the system passes into oligarchy, where competitor bodies from neighboring zones catch up in terms of mass.  We identify the peak of GF with the transition between the runaway growth and oligarchy accretion stages. Compared to previous estimates, we find that the system leaves the runaway growth phase at a somewhat larger radius, especially at the outer disk.  Furthermore, we assess the relevance of small, single-size fragments on the growth process.  In classical models, where the initial velocity dispersion of bodies is small, these do not play a critical role during the runaway growth; however, in models that are characterized by large initial relative velocities due to external stirring of their random motions, a situation can emerge where fragments dominate the accretion, which could lead to a very fast growth. 
\end{abstract}

\section{Introduction}
One of today's key questions in planetary science is to understand the processes through which $\mu$m-size dust grains are converted into the $\sim$10$^3$ km-size bodies that constitute the rocky planets as well as the cores of the giant planets of our solar system.  Nowadays this collection has been enormously expanded with the discovery of hundreds of extrasolar planets (see \citealt{UdrySantos2007} for a review).  It is a challenge to understand a transformation process that spans more than 35 orders of magnitude in mass, especially since observational data are rather limited.  Although grain growth in protoplanetary disks seems to be a robust process, as judged, for example, from (sub)millimeter studies \citep{NattaEtal2007,LommenEtal2009}, the observational signature of macroscopic bodies becomes weaker and weaker with their size.  In the solar system, important constraints include the remnants of objects that did not accrete into planets, like the bodies that constitute the Kuiper and Asteroid Belt as well as the meteoritic records that can be studied on Earth.   

In the core-accretion paradigm of planet formation \citep{MizunoEtal1978,HayashiEtal1985,LissauerStewart1993,PollackEtal1996,DominikEtal2007} two key ingredients play a crucial role in this transformation: molecular forces (surface forces) and gravity.  The former dominates the behavior of dust particles at small scales, the latter at larger scales.  At the high densities that characterize protoplanetary disks, micron size dust particles will quickly cluster into aggregates -- a process already observed in the cores of dense molecular clouds \citep{SteinackerEtal2010}.  Bodies of km-size have a gravity large enough to bind material and the same process allows these bodies to efficiently accrete each other.  In this study we focus on the gravitationally-dominated regime. However, since it determines the initial conditions, we first briefly review theoretical efforts to overcome the problematic intermediate size range.  

For meter-size boulders surface forces seem to be too weak to act as an efficient sticking agent (\citealt{BlumWurm2008}).  Furthermore, the interaction with the (turbulent) gas especially affects meter-size particles.  Whereas its damping properties are conducive to grow micron-size grains, the drag exerted on boulders causes them to drift inwards (and collide) at relative velocities of $\sim$10 m s$^{-1}$, except perhaps in some special settings like high-pressure environments \citep{KretkeLin2007,BrauerEtal2008i}.  At these high velocities, laboratory dust collision experiments indicate these dust aggregates fragment rather than stick, or that perhaps growth already stops at the $\sim$mm size \citep{GuettlerEtal2010,ZsomEtal2010}.  

Gravity has therefore been invoked to leap-frog the problematic meter-size regime. \citet{GoldreichWard1973} argue that once the dispersion of dust particles drops below the threshold for gravitationally stability, the collapse produces $\sim$km-size parent bodies.  However, this mechanism ignores the role of gas drag, and of (self-induced) turbulence which will undoubtedly develop as a result of the angular velocity mismatch between the dust-dominated midplane layer and the pressure-supported gas \citep{Weidenschilling1980}.  In what is perhaps a surprising twist, recent studies have instead appealed to turbulence as a means to \textit{concentrate} particles.  For large, meter-size boulders the interplay with the gas leads to pile-ups of these particles (a.k.a.\ the streaming instability, \citealt{YoudinJohansen2007}) and numerical simulations have shown that these particles concentrate in clumps, which subsequently collapse to form bodies of $\sim$10$^2$\ km to perhaps thousands of km in size \citep{JohansenEtal2007,JohansenEtal2009}.  As another flavor, \citet{CuzziEtal2008} argue that the intermittent properties of turbulence allow mm-size particles to concentrate in regions of overdensities of perhaps $10^3$--$10^4$.  Once captured by gravity, these clumps slowly sediment to form the first generation of sandpile planetesimals, in the $\sim$10--$10^2$ km size range \citep{CuzziEtal2010}.

Thus, it is still unclear what the outcome of the primary accretion stage will be, \ie\ the size distribution and the timescales.  Recent work has tried to constrain planetesimal formation scenarios by comparing the outcome of models that cover the gravitationally-dominated phase with the present day size distribution of the Asteroid Belt \citep{MorbidelliEtal2009}.  However, \citet{MorbidelliEtal2009} arrived at the conclusion that the Asteroid Belt population -- in particular the `bump' in the size distribution that is observed at at $\sim$100 km \citep{BottkeEtal2005} -- is quite incompatible with accretion models, implying that it must have been the direct outcome of the primary accretion process (but see \citealt{Weidenschilling2010} for a different interpretation). These works reveal the importance and motivation of modeling the gravity-dominated phase: its implications affect both the primary accretion process as well as the later stages of planet formation. 

After the formation of planetesimals growth becomes accelerated due to gravitational focusing (GF). GF increases the cross section for collisions by a factor $\sim$$(v_\mathrm{esc}/w)^2$ over the geometrical cross section, where $w$ is the relative velocity between the bodies and $v_\mathrm{esc}$ the escape velocity.\footnote{A list of symbols and key abbreviations is provided in \app{symbols}.}  For $w < v_\mathrm{esc}$ the accretion rate scales superlinearly with mass, $dM/dt \propto M^\kappa$, with $\kappa = 4/3$, which causes bodies to separate in terms of mass over time \citep{KokuboIda1996}.  This phenomenon is better known as runaway growth \citep{GreenbergEtal1978,WetherillStewart1989}.  The growth of the runaway body is relatively fast as long as the velocity dispersion $v$ of the reservoir of (smaller) bodies from which it is accreting stays low, such that GF remains efficient, \ie\ $v\sim w \ll v_\mathrm{esc}$.  However, GF does not only increase the collisional cross section, it also increases the rate of \textit{collisionless} encounters, which dynamically heat the system, increasing $w$, and decreasing the focusing.

Therefore, at some point runaway growth accretion will run out of steam.  This transition has been stated by \citet{IdaMakino1993} to take place at the point where
\begin{equation}
  2\Sigma_M M > \Sigma_m m,
  \label{eq:rg-oli-trans}
\end{equation}
with $m$ and $\Sigma_m$ the mass and surface density of planetesimals and $M$ and $\Sigma_M$ that of the runaway body(ies).  Physically, this definition implies that the transition takes place at the point where the velocity dispersion of the planetesimal swarm is determined by the runaway body(ies).  The consequence is that the growth rate ($dM/dt$) is reduced, although the growth mode stays locally runaway.\footnote{See \se{rg-def}, where we define runaway growth more precisely.}  However, competitor bodies that are spatially separated are not affected by this heating; runaway growth in these zones is therefore faster than in the heated zones until these zones are also heated.  The result is that (big) bodies that are dynamically separated tend to converge in terms of mass, a situation referred to as \textit{oligarchy} \citep{KokuboIda1998}.  This dichotomy in the population -- oligarchs ($M$) and `field planetesimals' ($m$) -- is a quite natural prediction of this process. Consequently, the oligarchic two group approximation is frequently used as starting point for subsequent (semi-analytical) studies \citep{ThommesEtal2003,GoldreichEtal2004,Chambers2006,Chambers2008,FortierEtal2007,BruniniBenvenuto2008}. 

To ensure sufficient growth within the lifetime of the gas disk ($\sim$10$^6$ yr) the random motions of planetesimals $v$ need to be damped during the oligarchy stage.  When gas drag acts as the cooling mechanism it can be shown that the radius of the protoplanet $R_\mathrm{pp}$ increases only linearly with time, $R_\mathrm{pp}\propto t$, which for the outer disk means that the growth timescale becomes dangerously close to that of the gas disk lifetime \citep{KokuboIda2002,ThommesEtal2003,Chambers2006}.  Additional damping may be achieved through fragmentation \citep{Chambers2008,KenyonBromley2009} or new, dynamically cold, reservoirs of planetesimals could be tapped by the migration of the protoplanets \citep{TanakaIda1999,MordasiniEtal2009}.  On the other hand, gap formation and/or gravitational scattering could result in a strongly inhomogeneous disk in which accretion is suppressed \citep{Rafikov2003ii,LevisonEtal2010}.  Most of these models have used the assumption that the size distribution can be approximated by two populations (protoplanets and planetesimals). However, the validity of this assumptions (and others) depends ultimately on our understanding of the outcome of the runaway growth stage.

Runaway growth (RG) calculations are quite challenging.  The most straightforward way to assess the outcome of RG is by $N$-body simulations.  
However, $N$-body simulations, also suffer from severe computational constraints: $N$ is restricted (typically to $\sim$$10^4$) and the dynamic range ($\sim$$M/m$) these simulations can achieve is necessarily rather limited \citep{KokuboIda1996,KokuboIda2000,BarnesEtal2009}.   Clearly, to follow RG over a larger range one has to turn to statistical simulations  \citep[\eg][]{GreenbergEtal1978,GreenbergEtal1984,WeidenschillingEtal1997,KenyonLuu1998,InabaEtal2001,Glaschke2006}.  For a proper calculation at least four parameters should be followed: the mass $m$, semi-major axis $a$, eccentricity $e$, and inclination $i$.  In most models, $e$ and $i$ are assumed to follow a distribution with parameters depending on mass and, if implemented, radial position only.  The distribution function is altered by collisional (accretion, fragmentation) and collisionless (viscous stirring, dynamical friction) processes and can be followed by numerical integration.  A drawback of this approach is, however, that the distribution functions rely on large numbers: the number of particles of each type $(m,a,e,i)$ should always be much greater than unity,  whereas, as we saw above, in RG/oligarchy the distribution becomes \textit{discrete}. In many codes the most massive bodies are therefore followed individually to incorporate the `local nature' of accretion \citep{WeidenschillingEtal1997,BromleyKenyon2006,Glaschke2006}. 

In this paper we describe a new method for the modeling of runaway and oligarchic growth. It is a Monte Carlo method in which each Monte Carlo `particle' either describes a single body or an entire swarm of small bodies. We obviously start with the latter: when all bodies are still $\sim$km size planetesimals each Monte Carlo particle represents a swarm of millions of planetesimals.  However, the code preferentially favors bodies of larger mass; by the time a runaway body begins to dominate over the rest, it has already become a single Monte Carlo particle, and the individual particle nature of this runaway body is then automatically taken care of.  In this way the transition from a fluid of planetesimals to a system with a few individual runaway bodies is properly treated. In fact, this would even allow a smooth transition to the next step of realism: an $N$-body simulation, where the runaway bodies (oligarchs) would be treated as $N$-body particles. But we leave this next step to a later publication.

With this new tool we embark on a parameter study of RG, in which the initial conditions, \eg\ the mass and velocity distribution are varied.  We include key physical processes like dynamical friction, gas drag, fragmentation, and turbulent stirring of bodies.  Furthermore, we resolve the spatial dimension of the disk.  Typically, simulations are followed until the biggest bodies reach $\sim$$10^3$ km.  We will define the timescale for runaway growth and identify the point where RG has been superseded by oligarchy.  More generally, we assess the nature of gravitationally-dominated accretion under variation of the physical conditions. In particular, we address the role of fragmentation by adopting a very simple model where bodies colliding at velocities above their mutual escape speed convert a fraction of their mass into mm-size fragments.

\Se{geo-model} presents the features of the multi-zone collision model. The collision model is validated in \se{validation}.  Readers more interested in the results may jump directly to \se{rg-oli}, which outlines the key characteristics of the runaway growth and oligarchy accretion phases.  \Se{rgsims} presents our parameter study, where the accretion behavior is studied under variation of the physical conditions.  \Se{discussion} discusses some implications, while \se{summary} summarizes the key results.

\section{The collisional evolution model}
\label{sec:geo-model}
In this section we address the key elements of the collision model. \Se{key-def} provides some preliminaries. In \se{MC} we describe our Monte-Carlo model, with which we calculate the time-evolution of the system.  The zonal setup of the program is discussed in \se{multi-zone}.  \Se{interactions} introduces the interaction radii $R_\mathrm{int}$ for the three velocity regimes under consideration. \Se{colmod} outlines how collisions are treated. Finally, \se{merits} discusses merits and drawbacks of our approach. 

\subsection{Preliminaries}
\label{sec:key-def}
For planetesimals -- bodies which we consider to be of $\sim$km-size or larger -- we use the epicycle approximation to describe their orbital motion, in which the velocity is the vectorial sum of the Keplerian velocity $v_k(a)$ corresponding to its semi-major axis $a$ and a random component of magnitude $v$.  When discussing interactions between two bodies of different mass we use $v_M$ for the random velocity of the largest body and $v_m$ for that of the smallest.  The orbital eccentricity is related to $v$ as $e \simeq v/v_k$; similarly the random velocity in the vertical direction, $v_z$, is related to inclination $i$ as $v_z/v_k \simeq \sin i \approx i$ for $i\ll 1$.

The Hill sphere $R_h$ of a body of mass $M$ and its corresponding Hill velocity $v_h$ are defined as
\begin{equation}
  R_\mathrm{h} = a \left( \frac{M}{3M_c} \right)^{1/3}; \qquad v_h = R_h \Omega 
  \label{eq:Rh}
\end{equation}
with $\Omega$ the local orbital frequency and $M_c$ the mass of the central star.  The Hill radius approximately represents the distance over which 3-body interactions (the third body being the sun) become important.   Using the escape velocity of the body, $v_\mathrm{esc} = \sqrt{2GM/R}$, where $G$ is Newton's constant, we find the useful auxiliary relation
\begin{equation}
  R v_\mathrm{esc}^2 = 6 R_h v_h^2,
  \label{eq:RRh-rel}
\end{equation}
which we will frequently employ.  When discussing interactions between two bodies one should replace the above definitions by combined quantities, for which $M = M_1+M_2$ and $R=R_s=R_1+R_2$ but this leaves \eq{RRh-rel} unaffected.

Another useful dimensionless number is $\alpha$, the ratio between the physical radius of a body and its Hill sphere \citep[\cf][]{GoldreichEtal2004},
\begin{equation}
  \alpha \equiv \frac{R}{R_h} = \frac{R}{a} \left( \frac{3M_\odot}{M} \right)^{1/3} = 7.5\times10^{-3} \left( \frac{M_c}{M_\odot} \right)^{-1/3} \left( \frac{a}{\mathrm{AU}} \right)^{-1} \left( \frac{\rho_s}{\mathrm{1\ g\ cm^{-3}}} \right)^{-1/3},
  \label{eq:alpha}
\end{equation}
where $\rho_s$ is the internal density of the body.  Using \eq{RRh-rel} we find $v_\mathrm{esc}=v_h\sqrt{6/\alpha}$.  

\subsection{The Monte Carlo collision model: interactions between \textit{representative} bodies (RBs)}
\label{sec:MC}
\subsubsection{Computational and physical particles}
Most statistical models involving planetesimal accretion rely on the concept of the distribution function to compute the collisional and dynamical evolution of a system.  In this study, however, we calculate the evolution using a Monte Carlo method: a particle-based approach but still statistical in nature (rather than $N$-body).  At the core of the method is the concept of `representative bodies' (RBs).  These are the particles the computer program uses as a proxy for the full particle distribution of the physical system.  One can say that each RB represents a group (or swarm) of physical particles.  Because of their overwhelmingly large numbers, the behavior of each individual body cannot be followed; but with a limited \textit{representative} sample -- the RBs -- a good census of the distribution can be obtained.

The number of physical particles a RB represents is denoted $N_g$, the group size.  There is complete freedom in choosing $N_g$; indeed, it will be a different number for each RB.  In the computer program it is assumed that the physical particles corresponding to the RB share identical properties, \ie\ the same masses, velocity dispersions, semi-major axis, \etc\  However, we do assume that the $N_g$ particles are homogeneously distributed over the part of the simulation space the RB traverses; \ie\ if the scaleheight of the RB is $h_z$, the horizontal dispersion $h_x$, and the semi-major axis $a$, we assume that the $N_g$ physical bodies that are represented by the RB share the same $a, h_x, h_z$ but that their phase angles characterizing the orbit are randomly distributed.  It is important to point out that effects like resonances or shepherding \citep[\eg][]{WeidenschillingDavis1985,Patterson1987,MandellEtal2007} are not treated in our approach. 

Each representative body (RB) is characterized by four independent properties: \sumi\ mass $m$; \sumii\ planar velocity dispersion (resulting from their eccentricity) $v$; \sumiii\ vertical velocity dispersion (as resulting from their inclination) $v_z$; \sumiv\ semi-major axis $a$.  Using these properties other particle properties are obtained, like the radius, $R=(3m/4\pi \rho_s)^{1/3}$, or the escape velocity, $v_\mathrm{esc}$.  Each RB is further identified by a unique $N_g$.

\subsubsection{Choice and adaptation of $N_g$}
How is the relation between the physical particles and the RBs determined, \ie\ the value of $N_g$?  `Traditional' MC-methods have $N_g=1$: each computation body has a 1-1 correspondence to a physical body.  This is, for example, sometimes used in aerosol coagulation studies \citep[\eg][]{Gillespie1975}.  However, for astrophysical purposes this is clearly inapplicable since then we only study the behavior of a limited number of physical bodies.  If $N_\mathrm{rb,0}$ is the initial number of RBs and $m_0$ the initial mass of the bodies, $N_g=1$ implies that we only simulate a mass $M_\mathrm{tot} = N_g N_\mathrm{rb,0} m_0 = N_\mathrm{rb,0} m_0$, which, due to the modest amount of RBs computers can handle, results only in a very limited potential for growth.  Clearly, $N_g \gg 1$ is required, at least initially.

In previous studies we have experimented with algorithms for choosing the proper group size $N_g$ \citep{OrmelSpaans2008}.  An algorithm in which the RBs represents a fraction of the total mass of the system ($M_\mathrm{tot}$), such that $N_g \propto M_\mathrm{tot}/m$ with $m$ the mass of the RB, is in most situations a good strategy \citep{ZsomDullemond2008,OrmelSpaans2008}.  However, it turns out that this mass sampling fails in systems that experience runaway growth.  In such systems, a runaway body will form that separates from the continuous particle distribution.  However, at the point of separation the mass of the single runaway body is typically negligible compared to $M_\mathrm{tot}$ \citep{Wetherill1990,MalyshkinGoodman2001}.  For this reason, it is imperative that collision models resolve the high-mass tail of the distribution very well.  This `unequal' mass sampling is the idea behind the `distribution method' \citep{OrmelSpaans2008}, in which the RBs are distributed equally in terms of log mass, with the result that the high mass bodies are comparatively much better represented.  In practice $N_g=1$ is usually reached for the large bodies, whereas for the low-mass ($\sim$$m_0$) particles $N_g\gg1$.  Perhaps counter intuitively, the total group mass of a low-$m$ RB ($N_g m_0$) is typically much larger than the mass of the high-$m$ RB that has $N_g=1$.

In short, $N_g$ is a function of the RB's mass and since it is based on the \textit{current} distribution, also a function of time.  Furthermore, we recognize that in spatially isolated regions the distribution could evolve differently and let $N_g$ be a function of semi-major axis as well; hence, $N_g = N_g(m,a,t)$.  The way how the distribution method operates is discussed in \app{app-rg}.  For a RB, we now distinguish between $N_g$, the number of bodies it represents at a certain point in the code, and $N_g^\ast$, the number of representative bodies it \textit{should} have according to the adopted algorithm for choosing $N_g$ (here: the distribution method, \app{app-rg}).  We desire that $N_g = N_g^\ast$; however, $N_g^\ast(m,a,t)$ for a given RB varies over the course of the simulation run.  How is the desired group size achieved? 

Let us illustrate.  Suppose that the physical bodies associated with the RB undergo many accretion events such that they move up the mass hierarchy: $m$ increases with respect to a characteristic mass of the system.  The distribution method will then signify that they should be better resolved; $N_g^\ast$ decreases and becomes less than $N_g$, say $N_g^\ast = N_g/2$.  The RB then splits into two identical RBs with $N_g = N_g^\ast$, which from now on evolve differently.  Likewise, if the RB is `inert' (no accretion events) and declines in the mass-hierarchy, $N_g^\ast$ will increase, perhaps becoming larger than $N_g$.  We then say that the bodies are `under resolved'.  Such a situation is forbidden: allowing it would mean that the total number of representative bodies will increase indefinitely and strain computational resources.  Another RB, sufficiently close in phase space, is sought to which the first RB is combined, averaging over their properties.  However, this averaging is only done when the particle properties are indeed very close, $v$, $v_z$, and $m$ should each be within 5\% or less; otherwise the RB in question is kept in the program as a regular RB until such a situation does materialize.

Due to these procedures, it is clear that the current number of RBs, $N_\mathrm{rb}$, fluctuates with time.  However, on average it is set by a target value -- say, $N_\mathrm{rb}^\ast$ -- and this determines the \textit{resolution} of the simulation.  How $N_g^\ast$ is determined form $N_\mathrm{rb}^\ast$ is explained in \app{app-rg}.

\subsubsection{Interactions between the RBs}
How group encounters are implemented is described in detail in \citet{OrmelSpaans2008} but let us illustrate the situation for a collision between two different RBs each representing $N_{g1}$ and $N_{g2}$ physical particles, respectively.  The total physical collision rate is $\lambda_{12} = N_{g1}N_{g2} \Delta v_{a,12} \sigma_\mathrm{inc,12} /V_\mathrm{int}$, where $\Delta v_{a,12}$ is the relative approach velocity, $\sigma_\mathrm{inc,12}$ the interaction cross section, and $V_\mathrm{int}$ the volume involved. Interactions among the $N_g$ physical particles within a representative body are also accounted for and occur at a rate $\lambda_{11} = N_{g1}(N_{g1}-1) \Delta v_{a,11} \sigma_\mathrm{inc,11} /2V_\mathrm{int}$. The quantities $\Delta v_a, \sigma_\mathrm{int}$, and $V_\mathrm{int}$ are all determined by the properties of the two RBs.  Assuming, without loss of generality, that $N_{g2} \ge N_{g1}$, the group collision is characterized by $N_\mathrm{int} = N_{g2}$ interactions in total with RB \#1 accreting $N_\mathrm{idv} = N_{g2}/N_{g1}$ physical particles of the second group.  The collision then augments the mass of the first particle by an amount $m_2 N_\mathrm{idv}$ and this becomes the new mass of RB \#1 ($m_1\rightarrow m_1+N_\mathrm{idv}m_2$). The fact that typically $N_\mathrm{int} \gg 1$ causes the code to speed up significantly.

For $m_2 \ll m_1$ and $N_\mathrm{idv}\gg1$ there is an important caveat, however. It is possible that $N_\mathrm{idv}m_2$ is of the same order as $m_1$, in which case we would instantaneously (\ie\ during 1 group collision) increase the mass of RB \#1 by an amount of the order of its own mass, whereas in reality this occurs gradually.  This is undesired since a sudden unphysical jump in the properties of RB \#1 and in its collision rates with the other RBs is applied.  Therefore, in the program we let the first (more massive) RB accrete a mass of at most $f_\epsilon m_1$ and limit $N_\mathrm{idv}$ accordingly, $N_\mathrm{idv} \sim f_\epsilon m_1/m_2$.  This means that only a fraction of the physical particles of RB \#2 are involved in the group collision; the group is split.  Note that the increase in the number of RBs by splitting up a single RB may cause $N_g$ to fall below $N_g^\ast$ (see above).  The choice of $f_\epsilon$ reflects the computational cost.  In this study, we adopt $f_\epsilon=5\times10^{-2}$ We have checked that our results are insensitive upon variation of $f_\epsilon$ by a factor of two.  

\subsubsection{Summary of the collision code}
\begin{figure}[t]
  \centering
  \includegraphics[width=\textwidth,clip]{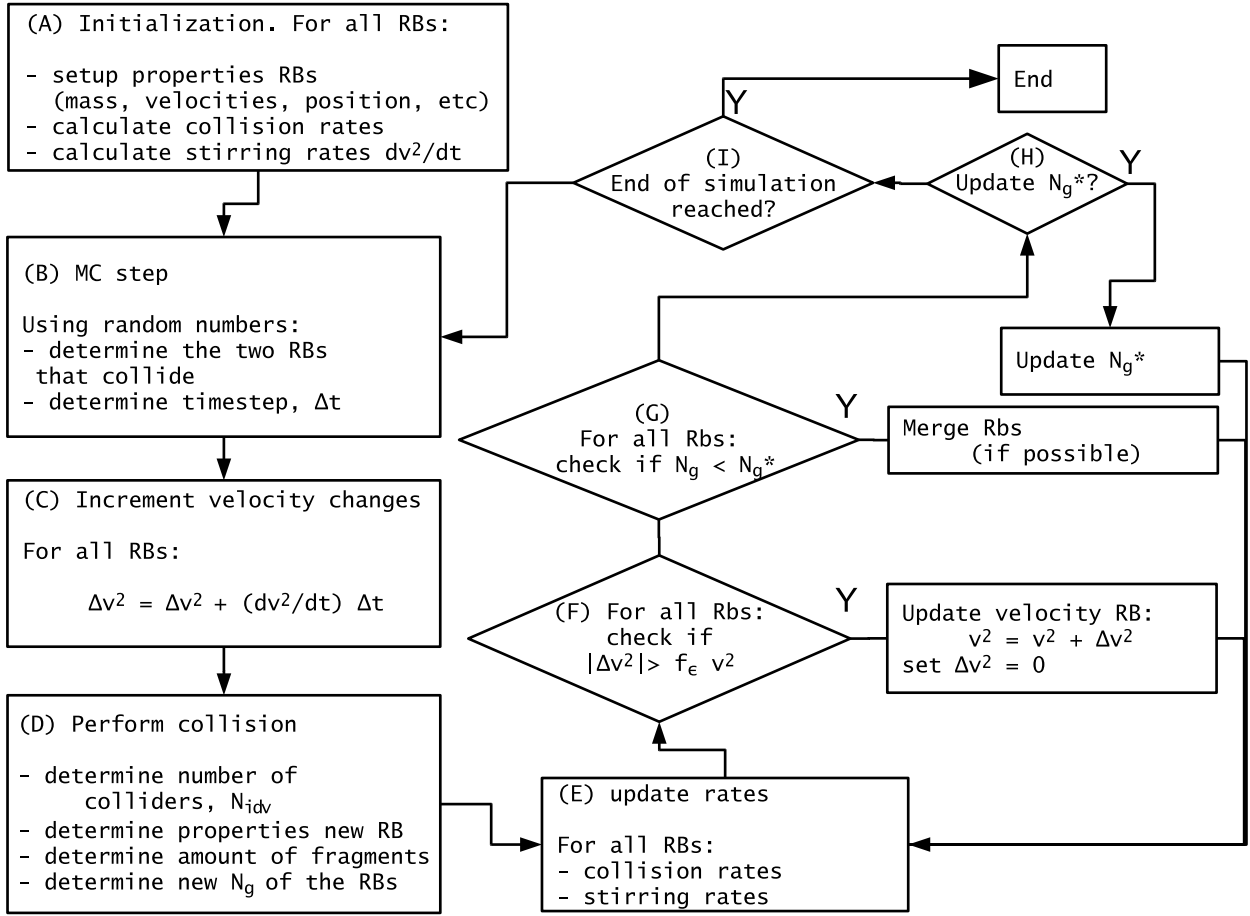}
  \caption{\label{fig:flowch}Flowchart describing the key steps of our collisional evolution model.}
\end{figure}
The flowchart of \fg{flowch} summarizes the several steps of the program.  The flowchart is intended to be schematic; it does not do justice to the full complexity of the underlying algorithm.  First, the properties of each RB are defined, \ie\ their masses, random velocities (inclination and eccentricity), and positions are assigned.  We distinguish between collision rates and stirring rates, the latter determine the evolution of the random velocities.  Steps B--D constitute the core of the program.  Here, the collision partners are determined, together with the time step and the corresponding velocity changes.  As explained below, in \se{merits}, we keep track of the random velocity change ($\Delta v^2$) of every RB by integrating its stirring rate ($dv^2/dt$) over time.  After the collision has been performed in D, the particle properties have changed, which requires us to update the stirring and collision rates of all RBs in step E.  The following steps, F--I, check several criteria.  In F we check whether the RBs' random velocities have to be updated.  If true, $\Delta v^2$ for the RB in question is put to zero.  In G we check whether RBs have become under resolved, after which we merge RBs using the procedure described above.  In H, finally, we check whether the function that determines the group size, $N_g^\ast$, must be renewed.  If any of these are positive, particle and stirring rates must again be calculated.  After all theses criteria have been met, the state returns to B with a new cycle of the program.

\subsection{Spatial differentiation: the multi-zone setup}
\label{sec:multi-zone}
\begin{figure}[t]
  \centering
  \includegraphics[width=\figww,clip]{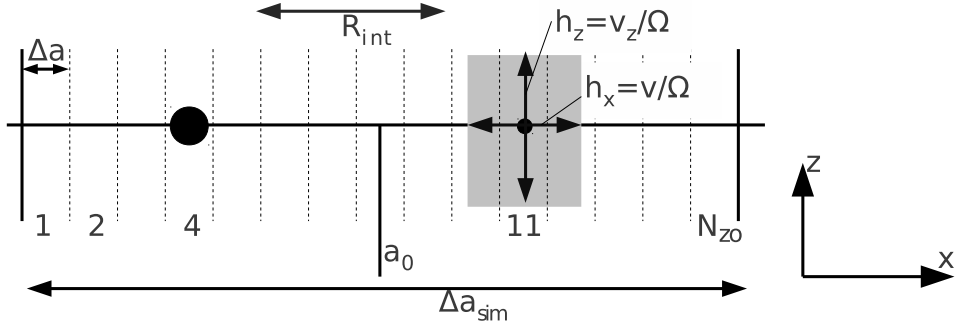}
  \caption{\label{fig:geopic2}Sketch of the multi-zone setup.  Symbols signify: $\Delta a$, resolution width; $a_0$ semi-major axis; $a_\mathrm{sim}$, simulation width; $N_\mathrm{zo}$, number of zones, $h_x,h_z$ horizontal/vertical scaleheights, $R_\mathrm{int}$, interaction radius.  The scalewidth of the bodies is determined by the resolution of the grid $\Delta a$, $h_x = \textrm{max}(v/\Omega, \Delta a)$.}
\end{figure}
Our code is multi-zone, with which we mean that RBs are assigned a particular semi-major axis on which their interaction radius depends.  This situation is illustrated in \fg{geopic2}.  Here, we refer to the radial direction of the disk (the semi-major axis, $a$) as the $x$-direction and use $z$ for the vertical direction.  There are $N_\mathrm{zo}$ zones each spanning a width $\Delta a$.  The code then simulates a patch $a_\mathrm{sim} = N_\mathrm{zo} \Delta a$ in semi-major axis, centered at $a_0$.  The RBs are placed at the center of the zones but (as mentioned before) assumed to be randomly distributed over the width of the zone, $\Delta a$, which sets the minimum value of the scalewidth, $h_x$.  However, if the horizontal excursions due to their eccentricities exceed $\Delta a$, like with the RBs in zone 11 of \fg{geopic2}, their width is given by the eccentricities, or random velocities in the planar direction $v$, \ie\ $h_x = v/\Omega$.

Whether or not RBs can interact (\ie\ have non-zero rates) depends on whether their mutual excursions overlap (determined by $h_{x1}$ and $h_{x2}$ and their radial distance) \textit{and} on the interaction range, $R_\mathrm{int}$ (discussed in \se{interactions}).  For example, if $R_\mathrm{int} = 0$ the RBs in zones 4 and 11 of \fg{geopic2} do not overlap and all rates are zero.  On the other hand, if the RB in zone 4 is placed in zone 11 of \fg{geopic2} then they would perfectly overlap, and if it would be in zone 10 then there would be partial overlap.  Also, if the interaction radius is large, say, $R_\mathrm{int}=10 \Delta a$, then there would be perfect overlap again, despite the fact that the RBs do not cross.  In \app{calc-ints} the precise algorithm is presented.

The way we treat the spatial dimension here echoes many features, albeit implemented more simply, of the multi-zone model developed by \citet{SpauteEtal1991} and \citet{WeidenschillingEtal1997} (W97 in this section).  But there are differences and we briefly mention these.  In W97 when particles in bins are `promoted' to individual bodies, these are assigned a true orbit (including phase angles), which is not done here. Another major difference is that W97 assume `reflective' boundary condition, whereas in our treatment the boundaries are periodic: a body in the last zone $N_\mathrm{zo}$ lies computationally adjacent to the first zone.  This means that in our code there is (for the moment) no radial gradient in the physical conditions, \eg\ gas density, sound speed, \etc\  Our code is in that sense local.  Like W97 when bodies merge they are placed in a new zone corresponding to the position of their common center of mass (since we treat a discrete grid, random numbers determine the zone the merged body is assigned to).  However, for the remainder there is no spatial diffusion between the zones; \eg\ there is no radial orbital decay of (small) particles due to gas drag or evolution in semi-major axis due to gravitational scattering.  

Our implementation of the radial direction is (currently) not suited to model differences in evolution arising from global gradients in the physical conditions of the disk.  However, even on scales where the physical conditions can be approximated to be constant, the disk will become inhomogeneous due to the emergence of runaway bodies, and it is this effect that we intend to explore in this study.   Dynamical friction keeps these bodies rather cold and the low $h_x$ prevents them from overlapping.  They become mutually isolated.  A statistical `particle-in-a-box' description does not do justice to this situation.  Usually, single zone models resolve this problem by forcing the collision rates among the largest bodies to be zero \citep{WetherillStewart1993,InabaEtal2001} but this, perhaps somewhat ad-hoc prescription, is not adopted here.  In addition, a runaway body, when sufficiently  massive, starts to dynamically heat the planetesimal bodies, but only these in its neighborhood, which again makes the disk inhomogeneous (in phase space).\footnote{Here, `neighborhood' is determined by the extent of the viscous stirring radius, $R_\mathrm{vs}$, and can exceed the width of the zone, $\Delta a$.}  Thus, it is important to have a spatially resolved disk to assess the significance of these effects. 

How many zones are needed? Clearly, the more zones the better the spatial resolution and precision of the method. However, the number of zones will be limited by computational constraints because \sumi\ each zone requires a minimum number of computational particles to resolve the mass spectrum and \sumii\ we compute the interaction between all RB pairs (not only these of the same zone).  Using the properties of the system we define two key length scales.  The first determines the resolution of the simulation ($\Delta a$) and is set to $\sim$$2v_\mathrm{esc,0}/\Omega$, the spatial excursion the initial population of planetesimals (the bodies that contain most of the mass) would have if their random velocity equals the escape velocity, $v_\mathrm{esc,0}$.   Here we anticipate that even if the initial random motions $v_0$ are $\ll$$v_\mathrm{esc,0}$ planetesimal-planetesimal stirring, which occurs on a timescale much shorter than accretion, will quickly heat up the planetesimals to velocities on the order of their escape velocities \citep{Rafikov2003ii}.  However, if during runaway growth the dominant accretion mode (in terms of mass) is between the biggest bodies (as postulated by \citealt{MakinoEtal1998}) which are characterized by a very low $h_x$ we may still have oversampled the radial dimension.  For these reasons we will test the dependence on resolution (\se{conv}). As the nominal resolution we use 
\begin{equation}
  \Delta a = \frac{2v_\mathrm{esc,0}}{\Omega} =2\sqrt{\frac{8\pi G \rho_s}{3}} \frac{R}{\Omega} = 5\times10^{-4}\ \mathrm{AU}\ \left( \frac{\rho_s}{\mathrm{1\ g\ cm^{-3}}}\right)^{1/2} \left( \frac{R_0}{10\ \mathrm{km}} \right) \left( \frac{a}{\mathrm{AU}} \right)^{3/2}
  \label{eq:ares}
\end{equation}
Note that this implies a much finer grid than used by \citet{WeidenschillingEtal1997}, where $\Delta a =0.01$ AU is adopted.

Similarly, we can set a condition for the total radial width of the simulation.  Here, we anticipate that an oligarch or runaway body of final mass $M_f$ dominates a region several times its Hill sphere, \eg\ $5R_\mathrm{h,f}$ \citep{KokuboIda1998}.  Such a length scale should fit comfortably within the total width of the simulation.  Dividing the two lengths scales, $5R_\mathrm{h,f}$ and $\Delta a$, gives the minimum required number of zones, $N_\mathrm{zo}^\mathrm{min}$,
\begin{equation}
  N_\mathrm{zo}^\mathrm{min} \simeq \frac{5R_\mathrm{h,f}}{\Delta a} \approx 9 \rho_s^{-1/6} \left( \frac{R_\mathrm{f}/R_0}{100} \right) \left( \frac{a}{\mathrm{1 AU}} \right)^{-1/2}
  \label{eq:Nz}
\end{equation}
where we used that $R_\mathrm{h,f}/R_{h,0} = R_f/R_0$, $M=4\pi \rho_s R^3/3$ and \eq{ares}.  In order to discriminate between an oligarchic (several big bodies) or a runaway (one big body) outcome, $N_\mathrm{zo}$ should be chosen several factors larger than $N_\mathrm{zo}^\mathrm{min}$.  From \eq{Nz} it follows that for the same amount of growth, the inner disk requires more zones than the outer disk.

\subsection{The interactions: collisions, dynamical friction, and viscous stirring}
\label{sec:interactions}
In our approach the collisional and dynamical evolution of the system follows from simple geometrical principles.  For collisions this is a well-tested approach; the physical radii and relative velocity of two bodies directly determine its collision probability, or collision rate.  We now extend this line of thinking and define an \textit{interaction} cross-section (or radius, $R_\mathrm{int}$) also for \textit{collisionless} encounters.  This may seem presumptuous since the gravitational interaction, being a long-range force, formally extends over an infinite distance; \ie\ at any given time a planetesimal feels the force of many (strictly speaking: all) planetesimals. In this section we will only treat the close encounters that have the strongest influence on the orbit of a planetesimal, which result in a (finite) cross section for interaction.  But for the final calculation of the stirring rates we will add a Coulomb factor to also include the more distant interactions (\app{stirring-plots}).  We will introduce two cross sections (or interaction radii $R_\mathrm{int}$) for the collisionless encounters: viscous stirring, $R_\mathrm{vs}$, and dynamical friction, $R_\mathrm{df}$.  

The interaction radii serve a twofold goal: \sumi\ they determine the cross section of the interactions, which enter in the collision/encounter probability; and \sumii\ they determine whether bodies at different semi-major axes mutually influence each other.  Every RB-pair is characterized by a unique $R_\mathrm{int}$ and also a unique interaction outcome, \eg\ the change in velocity is the same for all physical particles represented by the RB. 

The geometrical approach reflects \citet{GoldreichEtal2004} in their analytical study of oligarchic growth.  It provides a very insightful treatment of how a population of planetesimals of mass $m$ and a population of oligarchs of mass $M$ with $m\ll M$ mutually influence each other.  However, whereas \citet{GoldreichEtal2004} treats a two components system, our model contains $N_\mathrm{rb}$ groups, amounting to $\sim$$N_\mathrm{rb}^2$ interactions, $N_\mathrm{rb}$ per RB.  Our treatment is therefore numerical, but the underlying principle is the same: the individual interactions between the groups follow from geometrical principles. 

We consider three different types of interactions:
\begin{enumerate}
  \setlength{\itemsep}{0mm}
  \item Collisions: accretion, bouncing, or fragmentation.  The \textit{collisional radius} is denoted, $R_\mathrm{col}$.
  \item Dynamical friction. This occurs when particles experience a gravitational interaction but do not collide.  This process leads to momentum exchange between the particles at an impact parameter $b=R_\mathrm{df}$.  In this work, $R_\mathrm{df}$ is given by the condition that the deflection angle $\theta$ is large, $\theta \sim 1$. Then, the encounter can be approximated as an 1D elastic collision.
  \item Viscous stirring. Apart from the exchange of momentum through dynamical friction, the nature of the encounter is such that energy can be extracted from or added to the Keplerian potential (see \app{vs}).  This process is known as viscous stirring and operates at an interaction radius $R_\mathrm{vs}$.  The definition for the viscous stirring radius is set by the condition that the encounter changes the random velocity of the lightest particle $v_m$ by a similar amount: $\Delta v_m \sim v_m$.
\end{enumerate}

The distinction between the dynamical friction and viscous stirring interactions should not be interpreted as meaning that these belong to two distinct encounters.  In contrast, a (single) encounter will both contribute to the friction as well as the stirring.  We simply dissect collisionless encounters into a part that preserves the random energy (dynamical friction) and a part that does not (viscous stirring) \citep{Ida1990}.

For the calculation of the interaction radii we first need to specify the relative random velocity $w$ (and $w_z$ for the vertical direction), which is a function of the random velocities $v_m$ and $v_M$.  Usually, the velocity of the lightest body $v_m$ is larger than the velocity of the heaviest body $v_M$ due to dynamical friction but the situation may be different, \eg\ in the presence of gas drag.  The relative velocity at the point where the interaction takes place will in reality depend on the phase angles of the interacting bodies, and $w$ will generally follow a distribution in velocity.  These subtleties are ignored here and, as a crude approximation, $w$ is simply taken equal to the maximum random velocity, \ie\
\begin{equation}
  w = \mathrm{max}(v_m; v_M),
  \label{eq:w}
\end{equation}
The relative random velocity $w$ defines three velocity regimes:
\begin{itemize}
  \item the superescape regime, $w>v_\mathrm{esc}$ and $v_a=w$;
  \item the dispersion-dominated (d.d.) regime, $2.5v_h<w<v_\mathrm{esc}$, and $v_a=w$;
  \item the shear-dominated (s.d.) regime, $w<2.5v_h$, and $v_a=3b\Omega/2$.
\end{itemize}
Here, $v_a$ is the velocity at which the bodies approach each other.  In the superescape and d.d.-regimes we have that the approach velocity, $v_a$, equals $w$.  However, if $w$ becomes smaller than $v_h$, the s.d.-regime, the approach of the particles is determined by the Keplerian shear, $v_a\approx 3b\Omega/2$.  Therefore, the gravitational regime ($w<v_\mathrm{esc}$) splits into a d.d.-regime ($v_h\lesssim w \lesssim v_\mathrm{esc}$) and a s.d.-regime ($w \lesssim v_h$), see \fg{rintrad}.  Numerical studies have shown that when $w/v_h \ll 1$ particles at impact parameters $b=2.5R_h$ will enter the Hill sphere \citep[\eg][]{Nishida1983,PetitHenon1986,IdaNakazawa1989,GreenbergEtal1991}.  We adopt $w=2.5v_h$ as the boundary separating the s.d.- and d.d.-regimes.

Similar to \eq{w}, we will define $w_z = \mathrm{max}(v_{m,z}; v_{M,z})$ as the relative velocity in the vertical direction.  From this, we define the \textit{effective scaleheight} $h_\mathrm{eff} = w_z/\Omega$.  Simplifying arguments like these are very common for the geometrical approach: they are perhaps not formally correct but for the moment they satisfy our purpose of a model that is accurate within factors of unity.  Using the quantities for the interaction radii $R_\mathrm{int}$, the scaleheight $h_\mathrm{eff}$, and the velocity change upon interaction $\Delta v^2_\mathrm{int}$ we are able to construct interaction rates that only depend on these geometrical quantities.  For example, for the stirring rates we obtain (see \eq{stir})
\begin{equation}
  \frac{dv^2}{dt} = \frac{\pi R_\mathrm{int} R_z v_a}{2 h_\mathrm{eff}} N_\mathrm{sj} \Delta v^2_\mathrm{int},
  \label{eq:dv2dt}
\end{equation}
where $R_z$ is the interaction range in the vertical direction, $v_a$ the approach velocity (see below), and $N_\mathrm{sj}$ the column density of perturbers.  \Eq{dv2dt} can be compared with expressions that follow from more sophisticated studies.  In \app{calib} we quantify by how much our model is off, and adjust the collision rates that follow from our expressions accordingly, \ie\ by inclusion of order-of-unity calibration constants into expressions like \eq{dv2dt}.  

\begin{figure}[tbp]
  \centering
  \includegraphics[width=0.6\textwidth, clip]{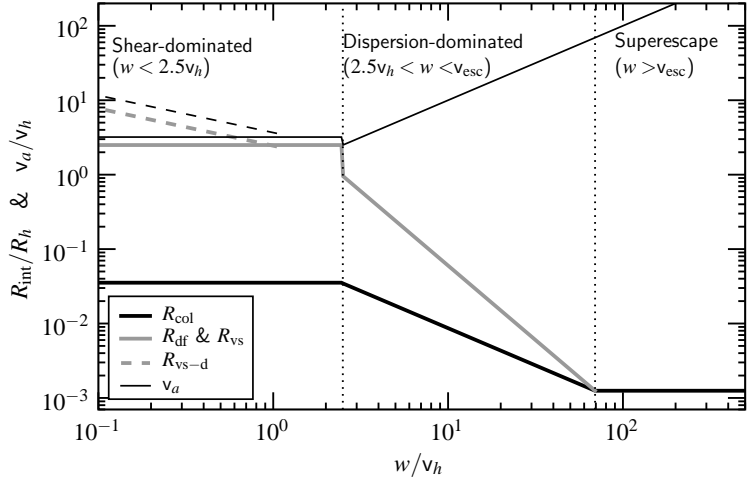}
  \caption{\label{fig:rintrad}The interaction radii $R_\mathrm{vs}$, $R_\mathrm{vs-d}$, $R_\mathrm{df}$, $R_\mathrm{col}$, and the approach velocity $v_a$ as function of the relative velocity $w$ between the interacting bodies.  Radii are normalized to the Hill radius $R_h$ and velocities to the Hill velocity $v_h = R_h\Omega$ of the largest body. The dashed lines represent $R_\mathrm{vs}$ and $v_a$ for distant interaction in the s.d.-regime.  We have adopted $\alpha=R_s/R_h=1.25\times 10^{-3}$. }
\end{figure}

We now provide expressions for $R_\mathrm{col}, R_\mathrm{df}$, and $R_\mathrm{vs}$ in these regimes.  These are summarized in \Tb{Rs-table} and \fg{rintrad}.

\begin{deluxetable}{lllll}
  \tablecaption{\label{tab:Rs-table}Summary of the \textit{interaction radii $R_\mathrm{int}$} for collisional encounters (collisions) and collisionless encounters (dynamical friction and viscous stirring).}
  \tablewidth{0pt}
  \tablehead{ 
    Velocity regime  & \multicolumn{4}{l}{Interaction radii, $R_\mathrm{int}$} \\ \cline{2-5}
    & Collisions   & Dynamical friction      & \multicolumn{2}{l}{Viscous stirring} \\
    &               &                         &   close & distant \\ \cline{4-5}
                                                & $R_\mathrm{col}$                            & $R_\mathrm{df}$                                 & $R_\mathrm{vs}$ & $R_\mathrm{vs-d} $
    }
    \startdata
                                                
    $w>v_\mathrm{esc}$                          & $R_s$                                         & $0$                                             & $0$   \\
    $v_\mathrm{h} \lesssim w \lesssim v_\mathrm{esc}$ & $(6R_sR_h)^{1/2} (v_h/w )$                    & $6R_h (v_h /w)^2$                               & $6R_h (v_h^2 /v_m w)$ \tablenotemark{a} \\
    $w \lesssim v_\mathrm{h}$                   & $\alpha^{1/2} R_h$ \tablenotemark{b}        & $2.5R_\mathrm{h}$                               & $2.5R_h$      & $R_\mathrm{h} (6v_\mathrm{h}/v_m)^{1/2}$ \\
    \enddata
    \tablenotetext{a}{Valid only when $R_\mathrm{vs}/w < \Omega^{-1}$. Otherwise, the expression $R_\mathrm{vs-d}$ for the shear-dominated regime applies}
    \tablenotetext{b}{This indicates the effective collision radius.  The more general approach, which more accurately takes care of the spatial dimension, is presented in \app{Hill}. }
  \tablecomments{$w$ is the relative velocity in the d.d.-regime, $v_m$ the random velocity of the smallest particle, and $v_h$ the Hill velocity of the largest particle.  \Eq{RRh-rel} has been applied for the expressions in the d.d.-regime.\\ 
  }
\end{deluxetable}

\subsubsection{The superescape regime, $w>v_\mathrm{esc}$}
Gravitational focusing (GF) is unimportant and all interactions are collisional $R_\mathrm{col}=R_s = R_1 + R_2$,  $R_\mathrm{df} = R_\mathrm{vs} = 0$.

\subsubsection{The dispersion-dominated regime, $v_h \lesssim w \lesssim v_\mathrm{esc} $}
GF increases the collisional radius over the geometrical radius.  The approach velocity is $w$ but at impact the velocity is (at minimum) $v_\mathrm{esc}$. Angular momentum conservation yields that the corresponding impact parameter at infinity is
\begin{equation}
  R_\mathrm{col} = R_s \frac{v_\mathrm{esc}}{w}.
  \label{eq:Rcol2b}
\end{equation}

Similarly, the criterion for dynamical friction is that the deflection angle changes over a large angle, or that $f_\mathrm{g} \Delta t \sim w$, where $f_g \approx G(M_1+M_2)/b^2$ is the gravitational force.  Using $\Delta t \sim b/w$ the corresponding impact parameter is therefore 
\begin{equation}
  b = R_s \left( \frac{v_\mathrm{esc}}{w} \right)^2 \equiv R_\mathrm{df}.
  \label{eq:Rdf}
\end{equation}
More formally, this impact parameter corresponds to a deflection angle of $w$ by $\pi/4$ \citep{BinneyTremaine2008}.

The reader could (correctly) argue that there is a certain level of arbitrariness in choosing $R_\mathrm{df}$. For example, a deflection angle of $\theta = \pi/2$ would amount to a dynamical friction radius that is only half that of \eq{Rdf},  while for our formal definition of dynamical friction stated above -- that it can be considered as a 1D elastic collision -- we would need $\theta = \pi$ and $b$ should be much smaller.  However, \eq{Rdf} does give an indication of the scale at which strong interactions ($\theta \sim 1$) become important.  To complement the approach, as mentioned above, it is required to compute the resulting stirring rates ($dv^2/dt$, \eq{dv2dt}), compare these with existing literature treatments (\app{calib}), and, if necessary, to adjust the rate by the order-of-unity calibration constants.  Somewhat surprisingly, the combination of \eq{Rdf} and the `elastic 1D collision' model (\app{stir-delv2}) turns out to match very well the analytical result (\app{stirring-plots}).

The criterion for viscous stirring is that the change in the random velocity of the \textit{lightest} particle is significant, $\Delta v_m \sim v_m$. Thus, we solve $f_\mathrm{g} b/w = v_m$ to obtain $b = R_\mathrm{vs} = R_s v_\mathrm{esc}^2/v_m w$, the radius for viscous stirring in the dispersion dominated regime.  In the (usual) case that $v_M<v_m=w$ this equals the dynamical friction radius $R_\mathrm{df}$, but if $w=v_M>v_m$, $R_\mathrm{vs} > R_\mathrm{df}$.

\subsubsection{The shear-dominated regime, $w \lesssim v_h$; distant interactions}
In the s.d.-regime the approach velocity $v_a$ is given by the Keplerian shear instead of $w$. For these interactions the solar gravity cannot be neglected and the interaction includes three bodies.  Particles approaching at distances $b \lesssim R_h$ will not enter the Hill sphere \citep{IdaNakazawa1989,GreenbergEtal1991}; instead, their trajectories strongly bend and the particles move away on horseshoe orbits. However, particles at slightly larger impact radii do enter the Hill sphere such that the characteristic impact radius is $\sim$$R_h$.  Following numerical and theoretical studies \citep[\eg][]{GreenbergEtal1991} we put the radius for entering the Hill sphere at $2.5R_h$.  Similarly, the average approach velocity for particles entering the Hill sphere is calculated to be $3.2v_h$ (see \app{Hill}). 

Since particles at impact parameters $b=2.5R_h$ can be accreted, we put $R_\mathrm{col} = 2.5R_h$.  However, not every Hill-penetrating encounter results in a collision.   In \app{Hill} we calculate the hit probability, $f_\mathrm{hit}$, with which expressions as the accretion rate must be supplemented.  Alternatively, as a 0th-order approximation, we can define an `effective impact parameter' by assuming the 2-body regime (\eq{Rcol2b}) and a relative velocity of $\sim$$2.5v_h$.  Then, $b_\mathrm{col} = R_\mathrm{col}^\mathrm{2-body}(w=2.5v_h) \approx \sqrt{RR_h} = \alpha^{1/2}R_h$.  This is what has been plotted in \fg{rintrad} but we emphasize that the program uses the $R_\mathrm{col}$--$f_\mathrm{hit}$ `route', since this more accurately takes account of the spatial structure (\app{Hill}).

Any particle that enters the Hill sphere experiences a strong interaction, such that $R_\mathrm{df} = 2.5R_h$.  For viscous stirring, on the other hand, significant stirring ($\Delta v_m \sim f_g \Delta t \sim v_m$) already takes place at larger impact radii, $R_\mathrm{vs} > R_h$.  Since the interaction timescale in this regime is set by the Keplerian shear, $\Delta t \sim \Omega^{-1}$, the resulting impact parameter for these encounters becomes $b \sim \sqrt{v_\mathrm{esc}^2 R /v_m\Omega} = R_\mathrm{h} \sqrt{6v_\mathrm{h}/v_m} \equiv R_\mathrm{vs-d}>R_\mathrm{h}$. These are long-range forces that gravitationally perturb particles on non-crossing orbits \citep{Weidenschilling1989}.

Therefore, we distinguish between two viscous stirring radii in the s.d.-regime.  When particles are capable to enter the Hill sphere, stirring is very efficient, because the velocity of these particles gets boosted to $v_h$, which can be $\gg$$v_m$.  For these particles $R_\mathrm{vs}=2.5R_h$, $\Delta v_m^2 \sim v_h^2$, and $v_a=3.2v_h$.  Otherwise, the viscous stirring radius is set to $R_\mathrm{vs-d}$ with an accompanying velocity change of (only) $\Delta v \sim v_m$, and $v_a = 3b\Omega/2 = 3R_\mathrm{vs-d}\Omega/2$ (see \fg{rintrad}).  In the s.d.-regime  we have that $R_\mathrm{int} \gg h_\mathrm{eff}$ and therefore $R_z = h_\mathrm{eff}$.  Inserting these expressions into \eq{dv2dt} we see that $dv_m^2/dt \propto R_h v_h^3$ for Hill-penetrating encounters, while $dv_m^2/dt \propto R_h v_h^2 v_m$ for distant interactions.   Thus, despite the fact that both $v_a$ and $R_\mathrm{int}$ are larger for the distant encounters, their heating rates ($dv^2/dt$) are less than the close, Hill penetrating encounters due to the `boost' $\Delta v_m\sim v_h$ the particles receive in the latter case.  However, particles separated at impact parameters $b>2.5R_h$ can only be stirred by the distant interactions.

\subsection{The collision model, gas drag, and fragmentation}
\label{sec:colmod}
In this study we adopt a collision model that contains key physical processes like accretion, fragmentation, and bouncing but is overall characterized by its simplicity.

\subsubsection{Disk properties and gas drag}
\label{sec:diskprops}
\begin{deluxetable}{lllll}
  \tablecaption{Adopted disk properties \label{tab:disk}}
    \tablehead{
    \colhead{Parameter} &  \colhead{Symbol} & \colhead{1 AU}    & \colhead{6 AU}     & \colhead{35 AU} \\
    }
    \startdata
    Bodies' internal density      & $\rho_s$              [g\ cm$^{-3}]$  &3.0      &1.0        &1.0 \\
    Ratio $R/R_h$                 & $\alpha$                              &$5.2(-3)$&$1.3(-3) $ &$2.1(-4)$\\
    Solid surface density         & $\Sigma$              [g\ cm$^{-2}$]  &$16.7    $&$2.0     $ &$0.2    $\\
    Dust-to-gas ratio             &                                       &86       & 56        & 56 \\
    Sound speed                   & $c_\mathrm{g}$        [cm\ s$^{-1}$]  &$1.0(5) $&$6.2(4)  $ &$4.1(4) $ \\
    Gas density                   & $\rho_\mathrm{g}$     [g\ cm$^{-3}$]  &$1.4(-9)$&$9.5(-12)$ &$1.1(-13)$\\
    Nebula pressure par.          & $\eta$                                &$1.8(-3)$&$4.4(-3) $ &$1.1(-2)$ \\
    Drag coefficient              & $C_D$                                 &$0.44   $&$0.44    $ &$\ge 0.44$ \\
    \enddata
  \tablecomments{
    Properties characterizing the physical conditions at 1, 6, and 35 AU. 
  }
\end{deluxetable}
\Tb{disk} list the adopted disk properties.  We will run simulations at three distinct disk radii.  The mass of the central star is put at $1\ M_\odot$.  The parameters for the sound speed $c_\mathrm{g}$ and the nebula pressure parameter $\eta \sim (c_g/v_k)^2$ are adopted from \citet{NakagawaEtal1986}, following the minimum mass solar nebula profile \citep{Weidenschilling1977i,HayashiEtal1985}.  However, we vary the solid surface density $\Sigma$ and dust-to-gas ratio to enable a comparison with the studies of \citet{InabaEtal2001} (for 1 AU) and \citet{KenyonLuu1998} (at 35 AU).  Therefore, the underlying density structure does not strictly follow a power-law.  Since we do not treat a global disk configuration, these deviations are not critical.

In simulations with gas drag we apply a deceleration to the particle's velocity evolution on top of the accelerations that follow from gravitational encounters. We use the modified expressions of \citet{AdachiEtal1976} as written down by \citep{InabaEtal2001}:
\begin{subequations}
  \label{eq:stirring-gd}
  \begin{equation}
    \left( \frac{dv^2}{dt} \right)_\mathrm{drag}  = -\frac{2v^2}{t_\mathrm{drag}} \sqrt{ \frac{9E^2}{4\pi} e^2 +\frac{1}{\pi} i^2  +\frac{9}{4} \eta^2 },
  \end{equation}
  \begin{equation}
    \left( \frac{dv^2_z}{dt} \right)_\mathrm{drag}  = -\frac{v_z^2}{t_\mathrm{drag}} \sqrt{   \frac{E^2}{\pi} e^2 + \frac{4}{\pi} i^2 +\eta^2 },
  \end{equation}
\end{subequations}
where $t_\mathrm{drag} = 8\rho_s R /3C_D\rho_g\Omega a$, is the particle friction time, $E=1.211$, and $C_D$ the drag coefficient.  The drag coefficient equals $C_D = 0.44$ for large bodies but small bodies in the 35 AU simulations follow a different Stokes drag regime for which $C_D$ is larger \citep{Weidenschilling1977}.

\subsubsection{Collisions and fragmentation behavior}
We assume that planetesimal bodies are rubble piles consisting of much smaller particles (fragments) of mm size.  Due to the porous spaces, the internal density of the rubble piles is fixed at a low value of $\rho_s = 1\ \mathrm{g\ cm}^{-3}$, except of the models at 1 AU where we have put $\rho_s = 3\ \mathrm{g\ cm}^{-3}$ to facilitate the comparison with \citet{InabaEtal2001}.  Within a single (head-on) collision between two rubble piles the fragments will undergo many more collisions and dissipate a lot of the collision energy, which renders the overall collision very inelastic.  Thus, although the individual coefficient of restitution $\epsilon_1$ between two fragments is usually on the order of $\sim$0.5 \citep[\eg][]{HeisselmannEtal2010}, we assume that collisions between two rubble piles can be modeled with a net effective coefficient of restitution ($\epsilon$) that is much lower, $\epsilon = 0.01 \ll 1$.  According to \citet{GreenbergEtal1978} this value for the (effective) coefficient of restitution corresponds to loosely bounded regolith or weak material.  In reality, $\epsilon$ will further depend on the impact parameter and more sophisticated collision models are needed \citep{LeinhardtEtal2000,LeinhardtRichardson2002}.  

In our model, fragmentation only occurs when the relative velocity exceeds the escape velocity, $v_\mathrm{esc}$.  As a very simple prescription we assume the fragments are of the same size \textit{and} very small, $r_\mathrm{f} \ll R_0$.  Furthermore, we assume that the collision dissipates the majority of the collision energy and that only an energy $\epsilon E_\mathrm{col}$ remains to eject the fragments, with $E_\mathrm{col}= m_1m_2 v_a^2/2(m_1+m_2)$ the impact energy\footnote{Since we treat the $v_a>v_\mathrm{esc}$ case, we neglect the focusing term for the impact energy.} and $\epsilon = 0.01$.  However, there is no energy required to break the material since the fragments are already loosely bound.  Therefore, a mass fraction $f_\mathrm{frag} = \epsilon E_\mathrm{col}/(m_\mathrm{tot} v_\mathrm{esc}^2/2)$, with $m_\mathrm{tot}=m_1+m_2$ the combined mass of the collision partners, ends up as fragments with the mass of the main body being reduced correspondingly, $M=m_\mathrm{tot}(1-f_\mathrm{frag})$. The choice of these parameters yields an impact strength for destruction of $Q_\mathrm{D} \simeq v_\mathrm{esc}^2/2\epsilon \approx 10^5\ (R/\mathrm{km})^2\ \mathrm{erg\ g^{-1}}$, which falls within the range of several proposed strength curves in this gravitationally-dominated regime \citep{BenzAsphaug1999}.

We will treat two values for the fragment size, $a_\mathrm{fr} \approx 1$ mm (chondrule size particles) and $a_\mathrm{frag}=10$ cm (boulder-type particles).  Fragments are not allowed to accrete among themselves but can be accreted by larger bodies.  Next, we recognize that gas drag will influence the approach velocity of the fragments.  We distinguish between two situations: \sumi\ strong drag and \sumii\ weak drag.  Strong drag occurs in the 1 and 6 AU simulations: the fragments are tied to the gas and move at a fixed relative velocity of $v=\eta v_k$, corresponding to the subkeplerian gas velocity, and $v_z = 0$.  This means in practice that the re-accretion of fragments is suppressed since their approach velocity $v_a$ is rather large with little or no GF.  However, in the 35 AU models we relax the strong coupling assumption and model the dynamical behavior of the fragments in exactly the same way as the big bodies.  Then, the gas drag only has a (slight) damping effect with the cooling being dominated by mutual (inelastic) collisions among the fragments.  Since these collisions are abundant, these are usually very effective to dissipate any random motion, despite the stirring of the big bodies to which the fragments are also subject to.  

\subsubsection{Turbulent stirring}
\label{sec:Tstir}
Planetesimal eccentricities are excited due to gas density fluctuations in the disk caused by turbulence \citep{LaughlinEtal2004,Nelson2005}.  This results in an eccentricity evolution of \citep{OgiharaEtal2007,IdaEtal2008}
\begin{equation}
  e \sim 0.1 \gamma \left( \frac{\Sigma_g}{\Sigma_{g1}} \right) \left( \frac{a}{AU} \right)^2 \left( \frac{t}{T_K} \right)^{1/2}
  \label{eq:IdaEtal2008}
\end{equation}
where $\Sigma_\mathrm{g1} = 2400\ \mathrm{g\ cm^{-2}}$.  Inserting this value for $\Sigma_{g1}$, $T_K = 2\pi/\Omega$ for the orbital period, and squaring gives
\begin{equation}
  \frac{de^2}{dt} = 1.6\times10^{-3} \gamma^{2} \left( \frac{\Sigma_g a^2}{90\ M_\oplus} \right)^2 \Omega,
\end{equation}
which in terms of velocity units $v=ea\Omega$ reads
\begin{equation}
  \frac{dv^2}{dt} = 3.5\times10^{-5}\ \mathrm{cm^2\ s^{-3}} \left( \frac{\gamma}{10^{-4}} \right)^2 \left( \frac{\Sigma_g a^2}{100\ M_\oplus} \right)^2 \left( \frac{a}{\mathrm{AU}} \right)^{-5/2}.
\end{equation}
In the above expressions the dimensionless $\gamma$ embodies the strength of the turbulent density fluctuations.  Using $\gamma$ in the range of $10^{-2}$--$10^{-3}$ \citet{IdaEtal2008} show that most planetesimal collisions result in destruction \citep[\eg][]{MorbidelliEtal2009}.  These large values for $\gamma$ were based on global simulations involving the magneto-rotational instability (MRI). However, subsequent \textit{local} shearing box simulations \citep{YangEtal2009} indicated a lower value, $\gamma \sim 10^{-4}$, despite the fact that these MRI simulations gave a rather high turbulent-alpha parameter of $\alpha_T \approx 10^{-2}$.  Therefore, we will adopt both $\gamma=10^{-3}$ and $\gamma=10^{-4}$ when including turbulent stirring in our simulations.  However, provided an sufficient shielding by (sub)$\mu$m size grains, the turbulence may be significantly suppressed in the interior regions of the disks (dead zones) \citep[\eg][]{TurnerSano2008}, and turbulent stirring may perhaps not be so effective.

\subsection{Merits and drawbacks of the collision model}
\label{sec:merits}
We end this section with an assessment of the collision model, discussing its strengths and weaknesses, and sketch avenues for future extensions.  As the strengths of the collision model we list:
\begin{enumerate}
  \item the particle nature of the geometric model;
  \item the large dynamic range concerning runaway growth;
  \item the presence of stochastic effects due to the Monte Carlo noise;
  \item the intuitive nature of the geometric model.
\end{enumerate}
Here, the first three points are related to the numerical model (\se{MC}), whereas the last concerns the interactions (\se{interactions}).  We find the way we treat interactions, which is inspired by \citet{GoldreichEtal2004} 2-group's approximation, more intuitive than the rather complex (but perhaps more precise) formalisms of existing statistical programs \citep[\eg][]{InabaEtal2001}.  The advantage of the geometrical approach is that it identifies the critical mechanisms that drive the evolution in a transparent way -- just because the key ingredients are all physically-intuitive properties like length scales and velocity changes.  However, this is primarily a matter of taste; our numerical model would work just as well with the more formal expressions for the interaction rates.

Arguably the biggest advantage of our Monte Carlo approach is that it deals with particles -- the representative bodies (RBs) -- which properties are independent of each other.  For example, we can have two RBs with the same mass but with different velocities due to a sudden stochastic encounter.  This is quite different from the usual mass-binning methods, where the mass of the bin determines all other properties.  For the MC-method it is also easier to include more particle properties, for example, the internal structure of the bodies (molten or primordial) without any direct increase in the computational costs.  Its particle nature and the many (independent) properties with which RBs can be quantified render the collision model especially attractive. 

There are many codes that mix statistical and discrete elements \citep{WeidenschillingEtal1997,BromleyKenyon2006,Glaschke2006}.  Usually this involves a transition mass at which bodies are `promoted' to an $N$-body particle or a discrete particle.  Our model, in terms of the representative bodies, offers perhaps a more natural way to implement the transition; \ie\ individual bodies are those which have $N_g=1$ and this does depend on mass.  Indeed, $N_g=1$ bodies are formed pretty early in the course of the simulation run, a prerequisite to properly follow the runaway growth process. 

As the key drawbacks/omissions we list:
\begin{enumerate}
  \item the inefficiency of the Monte Carlo algorithm;
  \item the difficulty to model (semi) steady-state systems;
  \item the limited spatial diffusion of particles.
\end{enumerate}
The most severe disadvantage of MC-methods is that they are computationally very inefficient: of the $\sim$$N_\mathrm{rb}^2$ collision rates, only $\sim$$N_\mathrm{rb}$ are used.  A run of our model, although much faster than $N$-body, takes much longer than statistical methods based on the Smoluchowski (mass-binning) approach and is additionally rather noisy (see \citealt{OkuzumiEtal2009} for further discussion).  As a consequence, we were unable to explicitly calculate the velocity distribution within a swarm of bodies  (\ie\ to relax the assumption of a fixed distribution; see \eq{rayleigh}) since this would have required too many RBs.

For the same reason, it is too demanding to treat collisionless encounters also on an event-based approach.  Due to their increased gravitational focusing (\fg{rintrad}) collisionless encounters are more frequent than collisional interactions by several orders of magnitude, especially when the system relaxes to a quasi-steady state in velocity space. The problem is that the MC code does not recognize that the collective effect of these encounters cancels out, but instead resolves the strongly fluctuating velocities of the bodies, which render the code very inefficient.  Therefore, the random velocities -- eccentricities and inclinations -- of the swarms are updated in a continuous fashion, as described in \fg{flowch}.   For every RB the cumulative effect of all $N_\mathrm{rb}$ interactions is calculated, resulting in a stirring rate, $dv^2/dt$, and a (cumulative) velocity change $\Delta v^2 (t)$.  And only if the relative incremental change has exceeded its current value by a few percent (\ie\ $|\Delta v^2|/v^2 = f_\epsilon \simeq 0.05$) is the particle's velocity updated ($v^2\rightarrow v^2 + \Delta v^2$), together with all $\sim$$N_\mathrm{rb}$ interactions it is involved in.\footnote{We confirmed that the results were insensitive to a change of a factor 2 of the control parameter $f_\epsilon$.}

Thus, the velocity evolution of a swarm of bodies $v(t)$ evolves smoothly with time.  Although this is fine for the big bodies, strong scatterings are inherently discrete for small bodies.  This especially concerns the s.d.-regime, in which the velocity of small bodies jumps to $\sim$$v_h$ (which is possibly $\gg$$v$) after a single scattering; \ie\ its velocity-evolution is spiky rather than continuous.  With our MC-model we have the tools to address these drawbacks, however, and in future work we may switch to an event-based evolution of the velocity (provided the associated increase in computational effort can be overcome).

Although in the collision model we have resolved the radial direction of the bodies, any spatial diffusion of planetesimals or runaway bodies due to gravitational encounters \citep[\eg][]{TanakaIda1997,OhtsukiTanaka2003} were ignored.  Likewise, scattering by turbulent density fluctuations will further contribute to the radial mixing.  These effects may render the distribution more homogeneous over large distances than our zonal setup currently supposes.  On the other hand, if scattering by the oligarchs/runaway body dominates, the surface density distribution of the disk may also become inhomogeneous \citep{Rafikov2003ii,LevisonEtal2010}.  Gap formation is not included in the current setup of the model; but it could become important for the oligarchy stage.  To study these and later stages the width of the simulation $\Delta a_\mathrm{sim}$ has to be increased, which invalidates the local assumption we apply ($\Delta a_\mathrm{sim}/a \ll 1$).  On a global (AU) scale, the differences in the evolution timescale may stifle the onset of runaway growth in the outer disk,  due to the effect of long-term perturbations \citep{Weidenschilling2008}.  These effects cannot be self-consistently explored with our current setup.  However, we will assess the consequence of a pre-stirred population of planetesimals by considering the superescape regime in which the (initial) random velocity of bodies exceed their escape velocity $v>v_\mathrm{esc}$.  

\section{Model validation}
\label{sec:validation}
\subsection{Comparison with $N$-body: viscous stirring and dynamical friction}
\begin{figure*}[tbp]
  \centering
  \includegraphics[width=0.9\textwidth]{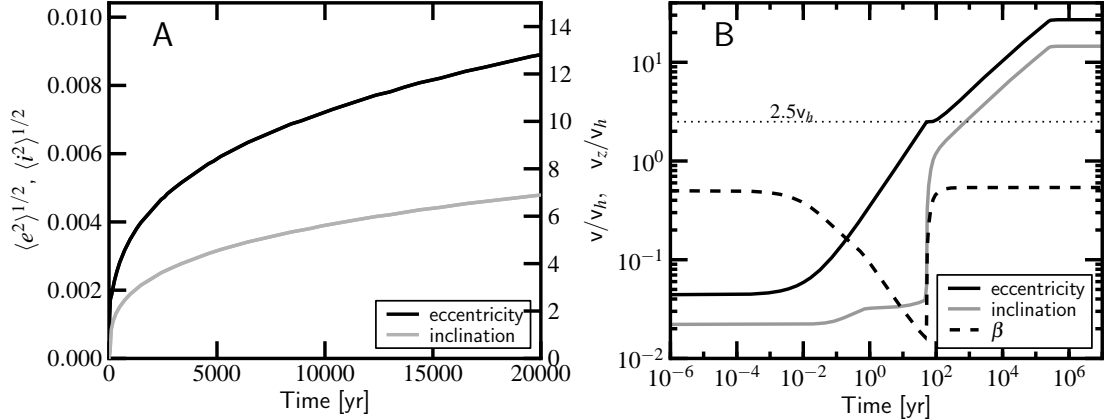}
  \caption{\label{fig:vs-test}Test of viscous stirring. The evolution of the random velocity component ($y$-axis) of a population of 800 equal-size ($m=10^{24}$ g) planetesimals as function of time ($x$-axis) for both a linear (\textit{left} panel) and logarithmic (\textit{right} panel) scaling. The planetesimals are distributed over an annulus of 0.056 AU centered at 1 AU. Eccentricity evolution is given by the \textit{black} curve, inclinations by the \textit{grey} curve. The \textit{dashed} curve in panel B gives the ratio of inclinations to eccentricities, or $\beta = v_z/v$.  The dotted horizontal line signifies the transition between the shear-dominated and dispersion-dominated regimes.  Compare with Fig.\ 4\ of \citet{StewartIda2000}.}
\end{figure*}
In this test, we copy the setup of the $N$-body simulations of \citet{StewartIda2000} (SI00 in this section).  In the SI00 $N$-body simulations the dynamical behavior among planetesimals is studied without accretion.  Therefore, we switch off accretion in our code ($R_\mathrm{col}=0$) and only treat dynamical friction and viscous stirring.  

\subsubsection{Equal mass system}
In the first test, 800 $m=10^{24}\ \mathrm{g}$ planetesimals are positioned in a narrow annulus of 0.056 AU at a distance of 1 AU from the sun.  Since bodies are of equal mass and no accretion takes place, the evolution of the system is determined by viscous stirring.  The bodies initially have a very low random velocity, $1\ \mathrm{cm\ s^{-1}}$, but this increases with time due to viscous stirring.  The resulting velocities are plotted in \fg{vs-test}, on a linear scale (panel A) and a logarithmic scale (panel B) and are normalized to the (combined) Hill velocity ($v_h \approx 21\ \mathrm{m\ s^{-1}}$).  Comparing \fg{vs-test}a with Fig.\ 4\ of SI00 we find that the curves match to the 10\% level (with most of the discrepancy in the inclination).  This confirms the validity of our viscous stirring expressions, at least for the dispersion-dominated regime which we are probing with this test. 

\Fg{vs-test}b shows the evolution resulting from our model over a more extended domain in time, which is instructive since it displays the underlying expressions of our geometrical model. In the s.d.-regime ($v\ll v_h$), it shows that eccentricities grow much faster than inclinations, causing the $\beta$ parameter ($\beta=v_z/v$) to decrease.  The reason for this behavior is that the interaction radius of viscous stirring $R_\mathrm{vs}$ is larger than the scaleheight of the planetesimals, $R_\mathrm{vs}\sim R_h \gg h_\mathrm{eff}$ and that therefore the interactions take place in a planar (2D) geometry, which suppresses the vertical stirring.  After $t\sim10\ \mathrm{yr}$, however, the interactions enter the d.d.-regime, $v>2.5v_h$.  At this stage the interaction geometry becomes three dimensional, which means that $\beta$ evolves towards its equilibrium value, $\beta \approx 0.5$. This effect initially stagnates the eccentricity stirring at the expense of the inclinations but after $t\sim10^2$ yr the equilibrium has been achieved and $v$ and $v_z$ evolve on similar timescales.  Finally, after $t\approx10^5$ yr the interactions have reached the superescape regime ($v=v_\mathrm{esc}$) and viscous stirring does not operate anymore. Bodies cannot be stirred above their mutual escape velocity.\footnote{In this test we have fixed the internal density of the bodies at $\rho_s=3\ \mathrm{g\ cm^{-3}}$.  However, if these are truly point sources (infinite $\rho_s$) the superescape regime does not exist and the flattening would take place.} 

\Eq{dv2dt} can be used to understand the qualitative behavior of the curves.  In the s.d.-regime, where the disk is thin, we have that $R_z=h_\mathrm{eff}$, $R_x\sim R_h$, $v_a \sim v_h$, and $\Delta v^2 = v_h^2$.  Since these quantities are all constant we find that the random velocities grow with the square root of time, $v\propto t^{1/2}$.  In the d.d.-regime, on the other hand we have that $R_xR_z \sim R^2 (v_\mathrm{esc}/v)^4$, $v_a/h_\mathrm{eff}$ constant, and $\Delta v^2 = v^2$. This results in $dv^2/dt \propto v^{-2}$ and therefore $v\propto t^{1/4}$.  These slopes are indeed observed in \fg{vs-test}b and agree with more detailed previous studies \citep[\eg][]{Ida1990,IdaMakino1992}.

\subsubsection{Two component system}
\begin{figure}[t]
  \centering
  \includegraphics[width=0.5\textwidth]{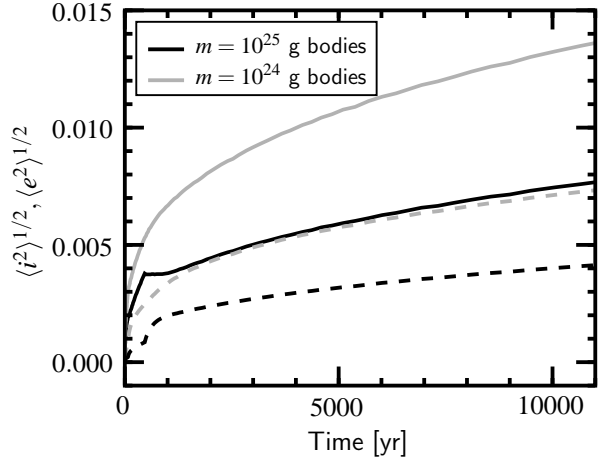}
  \caption{\label{fig:vs-testb}Like \fg{vs-test} but for a two component system of $m=10^{24}$ g (grey curves) and $m=10^{25}$ g bodies (black curves) with an equal amount of mass in both components.  The total surface density is again $\Sigma = 10\ \mathrm{g\ cm^{-2}}$.  Inclinations are given by dashed curves.  Compare with \citet{StewartIda2000}, their Fig.\ 9b.}
\end{figure}
Next, we calculate the evolution of a two component system, adopting the same parameters as before, but replacing half of the mass of the $10^{24}$ g bodies with bodies of $10^{25}$ g.  Thus, we have a two-component system in which we expect dynamical friction to operate.  Our results are displayed in \fg{vs-testb} and should be compared with Fig.\ 9b of SI00.  Again, we find a good match at the 10\%\ level.  Perhaps dynamical friction in our model seems to act a bit stronger than the $N$-body results suggest, but we do not consider the offset as critical.

Note the small plateau of the eccentricity curve of the $m=10^{25}$ g bodies near $t\sim10^3$ yr.  As discussed above this is due to the transition to the 3D regime.  In addition, our expressions have that $R_\mathrm{df}\ll R_\mathrm{vs}$ in the s.d.-regime, whereas $R_\mathrm{df}\sim R_\mathrm{vs}$ in the d.d.-regime.  Therefore, one can say that dynamical friction really starts to operate effectively from this point and this explains why the curves lie initially (around $t\sim0$) much closer together.  We remark, finally, that our treatment of the s.d./d.d.-transition regime is rather crude, which explains the erratic behavior of the curves at this point (see also our remarks toward the end of \app{stir-cal}).  In order to achieve a better match, SI00 and \citet{OhtsukiEtal2002} continue to empirically modify their analytical expressions, but we consider this beyond the scope of this work.

\subsection{Comparison with \citet{InabaEtal2001}}
\label{sec:Inaba}
Next, we compare the outcome of our numerical model including accretion with the statistical study of \citet{InabaEtal2001} (I01 in this section).  I01 presents a fully-consistent and highly accurate model, including both the dynamical as well as the collisional evolution.  I01 in turn compares their work against \citet{WetherillStewart1993} such that we in fact compare three statistical models against each other, see Fig.\ 9 of I01, where the cumulative number of bodies and the velocity structure are plotted.

\begin{figure}[tp]
  \centering
  \includegraphics[width=0.7\textwidth]{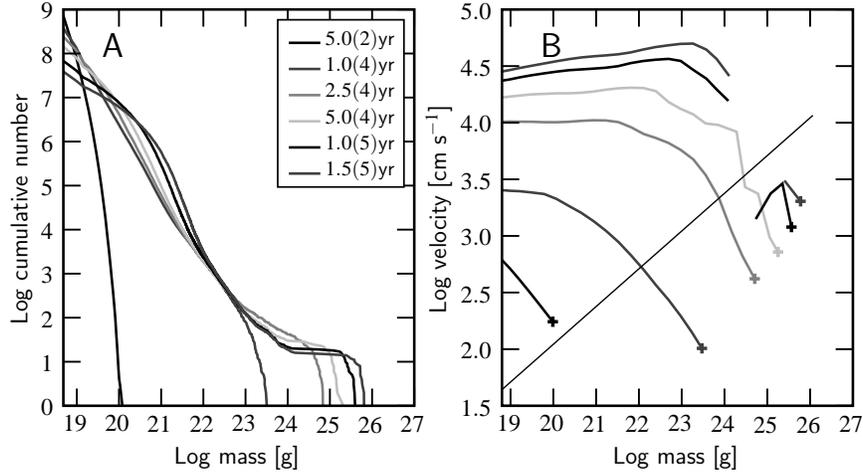}
  \caption{\label{fig:Inaba-1}The cumulative number distribution and velocity distribution at several times during the evolution at 1 AU.  Parameters are the same as in \citet{InabaEtal2001}, compare with their Fig.\ 9. Four simulations were combined, each with 1/4 the width of that of \citet{InabaEtal2001} at the nominal resolution width (\eq{ares}). }
\end{figure}
The simulation parameters are the following (after Table 1 of \citealt{WetherillStewart1993}).  The initial ($t=0$) distribution of bodies is monodisperse of mass $m_0=4.8\times10^{18}$ g and horizontal velocity $v=4.7\times10^2\ \mathrm{cm\ s}^{-1}$. The internal density of the bodies is fixed at $\rho_s = 3\ \mathrm{g\ cm^{-3}}$ and the gas density is $\rho_g = 1.2\times10^{-9}\ \mathrm{g\ cm^{-3}}$, conditions that correspond to a disk radius of 1 AU. The distribution is evolved until $t=1.5\times10^5$ yr by which time the bodies have grown to masses of $m\approx10^{26}$ g, corresponding to sizes of $R\approx2\times10^3$ km.

I01 and \citet{WetherillStewart1993} model a patch of the disk of 0.17 AU.  From \eq{ares} we obtain $\Delta a = 6.3\times10^{-4}$ AU as the nominal resolution width, which means that 270 zones must be included.  This proved to be a bit too demanding for the program since each zone must contain a minimum number of representative bodies.  As a solution, we have reduced the number of zones by a factor four ($N_\mathrm{zo}=67$) and computed four of these runs. Each simulation then models a total width $1/4$ of that of I01,  with the four (independent) runs being combined for the total.  The reduction in simulation width can be justified since the 67 zones is still larger than the minimum (\eq{Nz}); \ie\ the simulation width is still sufficiently wide to harbor $R=2\times10^3$ km bodies towards the end of the simulation.

The number of particles per zone equals $N_\mathrm{res}=600$ and the total number of representative bodies per run is $N_\mathrm{rb} = N_\mathrm{res} N_\mathrm{zo} \approx 40\, 000$.  Other combinations of $N_\mathrm{zo}, N_\mathrm{res}$ and $\Delta a$ will be investigated in the next section.  

\Fg{Inaba-1} shows the results.  Like Fig.\ 9 of I01 we show the cumulative number distribution, \ie\ the number of bodies of mass less than $m$, and the velocity distribution.  For the latter we have binned the RBs by mass and shown the mass-weighted planar velocity component.  We plot the distributions at the same times as in I01.  Note that the curves in the velocity plot are in our case occasionally a bit noisy, due to the low-$N$ statistics.  The gaps in the velocity plot are caused by the absence of RBs at these masses. 

Comparing the figures, we find an excellent match.  The general trends and shapes of the curves are in agreement.  There are minor differences but we do not think this invalidates either model.  Perhaps the biggest difference is the speed of the initial evolution, which by comparing the $t=10^4$ yr curves, can be seen to be faster in our case.  During the initial stages growth is rapid (\ie\ runaway growth) and the high-mass end of the distribution is a rather sensitive function of time, see \se{rg-indicators}.  Therefore, we do not consider the offset as critical to the validity of either model.

We conclude that with the choice of these values for $N_\mathrm{zo}, N_\mathrm{res},$ and $\Delta a$ we obtain a satisfactory match.  Next, we will test how sensitive the results are upon variation of these control parameters.

\subsection{Convergence tests}
\label{sec:conv}
\Tb{simruns} lists 14 simulation runs where the control parameters $N_\mathrm{zo}$, $N_\mathrm{res}$, and $\Delta a$ are varied.  The physical parameters like the gas density are kept the same as in the previous section.  When varying the parameters, we have kept the total number of RBs (=$N_\mathrm{zo} \times N_\mathrm{res}$) approximately the same; that is, an increase of $N_\mathrm{zo}$ by a factor of four is accompanied by a decrease of $N_\mathrm{res}$ by the same factor but for a given $N_\mathrm{zo}$ we run several models at different $N_\mathrm{res}$. In \Tb{simruns} the runs are listed by increasing number of zones, $N_\mathrm{zo}$.

\begin{table}[tp]
  \centering
  \caption{\label{tab:simruns}List of simulation runs to test the influence of the control parameters.}
  \begin{tabular}{llllllllll}
    \hline
    \hline
    id & $N_\mathrm{zo}$ & $N_\mathrm{res}$ & $N_\mathrm{run}$  & $\Delta a$ & $\Delta a_\mathrm{sim}$ & $T_\mathrm{rg}$ & $T_\mathrm{2k}$ & $\phi_\mathrm{vs,50}^\mathrm{min}$ \\
    &  & & & AU  & AU  & yr  & yr \\
    (1) & (2) & (3) & (4) & (5) & (6) & (7) & (8) & (9) \\
    \hline
1 & {$4$} & {$2500$} & {$1$} & {$4.2\times10^{-2}$} & {$0.17$} & {$1.2\times10^{3}$} & {$5.5\times10^{4}$} & {$2.9\times10^{-2}$} \\
2 & {$4$} & {$5000$} & {$1$} & {$4.3\times10^{-2}$} & {$0.17$} & {$1.1\times10^{3}$} & {$4.2\times10^{4}$} & {$4.7\times10^{-2}$} \\
3 & {$17$} & {$590$} & {$1$} & {$1.0\times10^{-2}$} & {$0.17$} & {$1.2\times10^{3}$} & {$9.6\times10^{4}$} & {$1.2\times10^{-1}$} \\
4 & {$17$} & {$1180$} & {$1$} & {$1.0\times10^{-2}$} & {$0.17$} & {$1.1\times10^{3}$} & {$9.0\times10^{4}$} & {$10.0\times10^{-2}$} \\
5 & {$17$} & {$2400$} & {$1$} & {$1.0\times10^{-2}$} & {$0.17$} & {$9.8\times10^{2}$} & {$1.2\times10^{5}$} & {$1.1\times10^{-1}$} \\
6 & {$67$} & {$150$} & {$1$} & {$2.5\times10^{-3}$} & {$0.17$} & {$1.1\times10^{3}$} & {$1.9\times10^{5}$} & {$3.0\times10^{-1}$} \\
7 & {$67$} & {$300$} & {$1$} & {$2.5\times10^{-3}$} & {$0.17$} & {$1.2\times10^{3}$} & {$1.4\times10^{5}$} & {$3.3\times10^{-1}$} \\
8 & {$67$} & {$600$} & {$1$} & {$2.5\times10^{-3}$} & {$0.17$} & {$1.2\times10^{3}$} & {$1.6\times10^{5}$} & {$3.3\times10^{-1}$} \\
9 & {$67$} & {$150$} & {$4$} & {$6.3\times10^{-4}$} & {$0.042$} & {$(1.2\pm0.1)\times10^{3}$} & {$(2.4\pm0.3)\times10^{5}$} & {$1.0\pm0.1$} \\
10 & {$67$} & {$300$} & {$4$} & {$6.3\times10^{-4}$} & {$0.042$} & {$(1.3\pm0.0)\times10^{3}$} & {$(2.2\pm0.4)\times10^{5}$} & {$(9.2\pm0.7)\times10^{-1}$} \\
11 & {$67$} & {$600$} & {$4$} & {$6.3\times10^{-4}$} & {$0.042$} & {$(1.3\pm0.1)\times10^{3}$} & {$(2.4\pm0.4)\times10^{5}$} & {$(9.4\pm0.4)\times10^{-1}$} \\
12 & {$251$} & {$40$} & {$4$} & {$1.7\times10^{-4}$} & {$0.042$} & {$(3.0\pm1.4)\times10^{3}$} & {$(2.6\pm0.3)\times10^{5}$} & {$2.8\pm0.4$} \\
13 & {$251$} & {$80$} & {$4$} & {$1.7\times10^{-4}$} & {$0.042$} & {$(1.2\pm0.0)\times10^{3}$} & {$(2.4\pm0.2)\times10^{5}$} & {$2.4\pm0.4$} \\
14 & {$251$} & {$160$} & {$4$} & {$1.7\times10^{-4}$} & {$0.042$} & {$(1.3\pm0.1)\times10^{3}$} & {$(2.3\pm0.1)\times10^{5}$} & {$2.4\pm0.2$} \\
    \hline
    \hline
  \end{tabular}
  \\  Note.--- Columns denote: (1) simulation identifier; (2) number of zones; (3) number of representative bodies per zone; (4) number of runs; (5) width of a single zone (6) total simulation width; (7) runaway growth timescale; (8) time to grow to 2000 km; (9) minimum filling factor of bodies that make up 50\% of the viscous stirring power of a zone.  The runaway growth timescale $T_\mathrm{rg}$ is defined in \eq{Trg}.  For simulation-id  9--14 where multiple runs have been performed the spread in $T_\mathrm{rg}$ and $T_\mathrm{2k}$ is also given.
\end{table}

\begin{figure}[tp]
  \centering
  \includegraphics[width=0.6\textwidth]{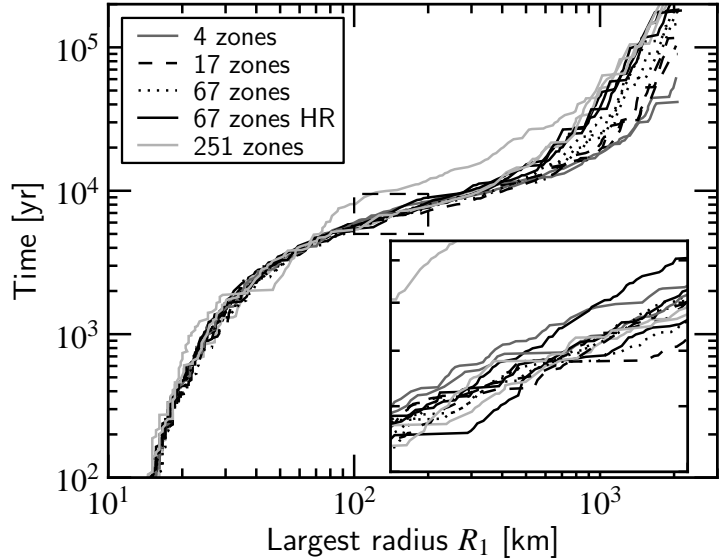}
  \caption{\label{fig:conv}Convergence test.  The evolution of $R_1(t)$, the radius of the largest body at time $t$, is shown for all 14 simulations of \Tb{simruns}. Simulations are identified by their number of zones, $N_\mathrm{zo}$. The inset shows a zoom in for $1\times10^2 < R_1(t) < 2\times10^2$ km.}
\end{figure}
Because of the statistical noise associated with the Monte Carlo method, it is rather difficult to conduct an unambiguous test for convergence.  Rather than focusing on a single measure, perhaps the best approach is to compare the evolution of the curves over an extended period.  This is done in \fg{conv} where the radius of the largest body, $R_1$, is plotted against time for all 14 runs listed in \Tb{simruns}.  The runs are identified by their $N_\mathrm{zo}$ with runs having the same $N_\mathrm{zo}$ sharing the same line style.  For example, the two $N_\mathrm{zo}=4$ runs (\#1 and 2) are both identified by a dark grey line. For the $N_\mathrm{zo}=67$ runs we distinguish between two values of $\Delta a$, $2.5\times10^{-3}$ and $6.3\times10^{-4}$ AU, the lower ones being run at a higher resolution (HR).

From \fg{conv} we conclude the following:
\begin{enumerate}
  \item One of the $N_\mathrm{zo}=251$ simulations (\textit{light} grey curves) shows very erratic behavior.  This concerns the $N_\mathrm{res}=40$ run (\# 12).  The low number of RBs per zone turns out to be too few to accurately resolve the mass distribution. 
  \item Towards the end of the simulation a clear systematic divergence among the curves can be seen with the runs of the smallest $\Delta a$ evolving much faster than those of large $\Delta a$.  In models that do not resolve the disk spatially, the disk appears more homogeneous (in phase space) than it actually is.  In the low-$N_\mathrm{zo}$ models the stirring of planetesimal bodies by the runaway body/oligarch occurs at a slower pace because it has to stir so many of them.  This artificially enhances the accretion.  
  \item However, the $N_\mathrm{zo}=251$ (\# 12-14) and $N_\mathrm{zo}=67$ HR (high resolution, \# 9-10) curves do not separate towards the end of the simulation.  This justifies our criterion for $\Delta a$, \eq{ares}. 
  \item At earlier times these trends are not so obvious.  In \fg{conv} the inset shows a zoom of a region around $R_1\approx10^2$ km.  The final divergence is not apparent here with stochastic behavior seeming to dictate the overall behavior. All curves lie pretty close together, except for simulation \#12.  We also find no clear dependence on $N_\mathrm{res}$.
\end{enumerate}

To make these findings more quantitative, we have in \Tb{simruns} included the timescale to produce a $2\times10^3$ km body, $T_\mathrm{2k}$, and the runaway timescale $T_\mathrm{rg}$.  The latter is a fit over the exponential region of the curve in \fg{conv}, \ie\ for times $t\lesssim10^4$ yr (see below, \fg{stat1Mz}b and \eq{Trg}).  Except for run \#12, we do not see a clear trend of $T_\mathrm{rg}$ over the various simulation runs.  But for $T_\mathrm{2k}$ this trend becomes obvious.

In Col.\ (9) of \Tb{simruns} we have listed the minimum filling factor $\phi_\mathrm{vs,50}^\mathrm{min}$ for the massive bodies that together make up more than 50\% of the stirring power in a certain zone. This quantity is obtained as follows:
\begin{enumerate}
  \item We select the masses above $m_\mathrm{vs,50}$ which together comprise 50\% of the viscous stirring power.  Since the stirring power of the bodies scales as $\Sigma m$ (\eq{rg-oli-trans}), \ie\ as the squares of the masses of the bodies, the criterion for $m_\mathrm{vs,50}$ is
  \begin{equation}
    \sum_{m_i>m_\mathrm{vs,50}}m_i^2 \left/ \sum_i m_i^2 = 0.5\right..
  \end{equation}
  \item We sum up the `spheres of influence' concerning accretion and compare this to the width of the zone $\Delta a$.  Here, we assume that the bodies are dynamically cold and that the net impact radius is on the order of the Hill sphere. Thus,
  \begin{equation}
    \phi_\mathrm{vs,50} = \frac{1}{\Delta a} \sum_{m_i>m_\mathrm{vs,50}} 2R_h(m_i),
    \label{eq:phit}
  \end{equation}
  \item We take the minimum of $\phi_\mathrm{vs,50}$ over the course of the simulation run to arrive at $\phi_\mathrm{vs,50}^\mathrm{min}$
\end{enumerate}

During the runaway growth phase the massive bodies become dynamically very cold, moving on circular orbits.  When scattering dominates we may expect the bodies to become isolated since scattering, together with dynamical friction, leads to orbital repulsion \citep{KokuboIda1995}. If the stirring power resides in the (few) big bodies, their mutual scattering may then cause them to become isolated in the sense that their mutual spacing in terms of semi-major axis becomes too large for collisional interactions.  Then, $\phi_\mathrm{vs,50}$ drops below unity, in which case our statistical assumption -- that the bodies are uniformly distributed over the width of the zone -- breaks down. However, when $\phi_\mathrm{vs,50} \gg 1$ scattering among the bodies cannot result in their isolation; the viscous stirring power is shared among a sufficiently large number of bodies to warrant the validity of the statistical assumption.

We find that, initially, $\phi_\mathrm{vs,50} \gg 1$ since the stirring power is initially determined by the planetesimals ($m_\mathrm{vs,50}\sim m_0$).  However, during the fast runaway growth phase a power-law distribution emerges, in which the stirring power becomes dominated by the bodies at the high-mass end; $\phi_\mathrm{vs,50}$ then quickly decreases and a minimum is obtained. In \Tb{simruns} the minimum of $\phi_\mathrm{vs,50}$ is given.  We see that for the large $\Delta a$ runs it falls below unity, indicating that the simulations do not resolve the spatial structure properly.  Indeed, as can be seen from \fg{conv}, growth in the low $N_\mathrm{zo}$ is the fastest, but this growth -- triggered by merging of big bodies -- is artificial.  However, in the large $N_\mathrm{zo}$ models $\phi_\mathrm{vs,50}$ stays above unity, indicating that the simulation is properly resolved. 

These findings indicate that the criteria outlined in \se{multi-zone}, \ie\ \eqs{ares}{Nz}, do lead to numerical convergence.  For the number of RBs per zone we recommend at least $N_\mathrm{res}=100$.  However, we will typically use a larger $N_\mathrm{res}$ to reduce the MC-noise.

\section{Runaway growth \vs\ oligarchy}
\label{sec:rg-oli}
\subsection{Runaway growth indicators}
\label{sec:rg-indicators}
\begin{figure}[t]
  \centering
  \includegraphics[width=0.45\textwidth]{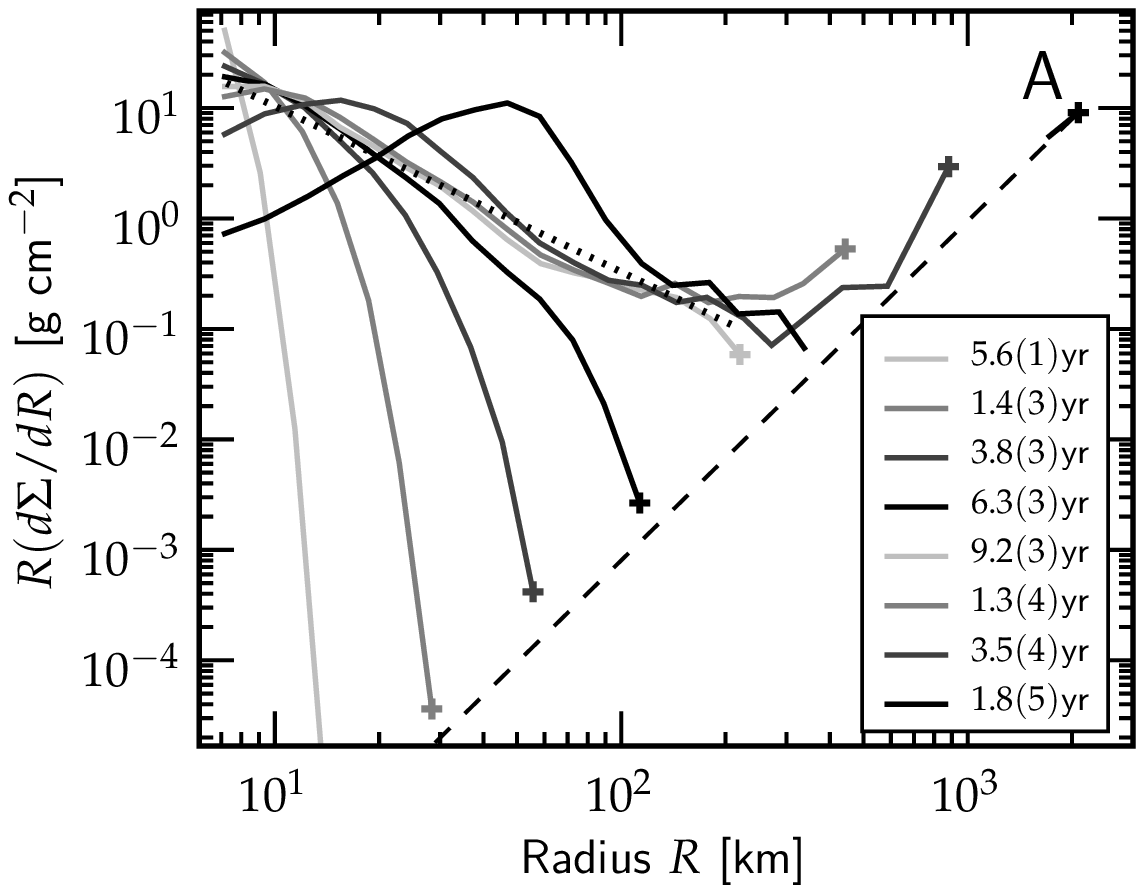}
  \includegraphics[width=0.45\textwidth]{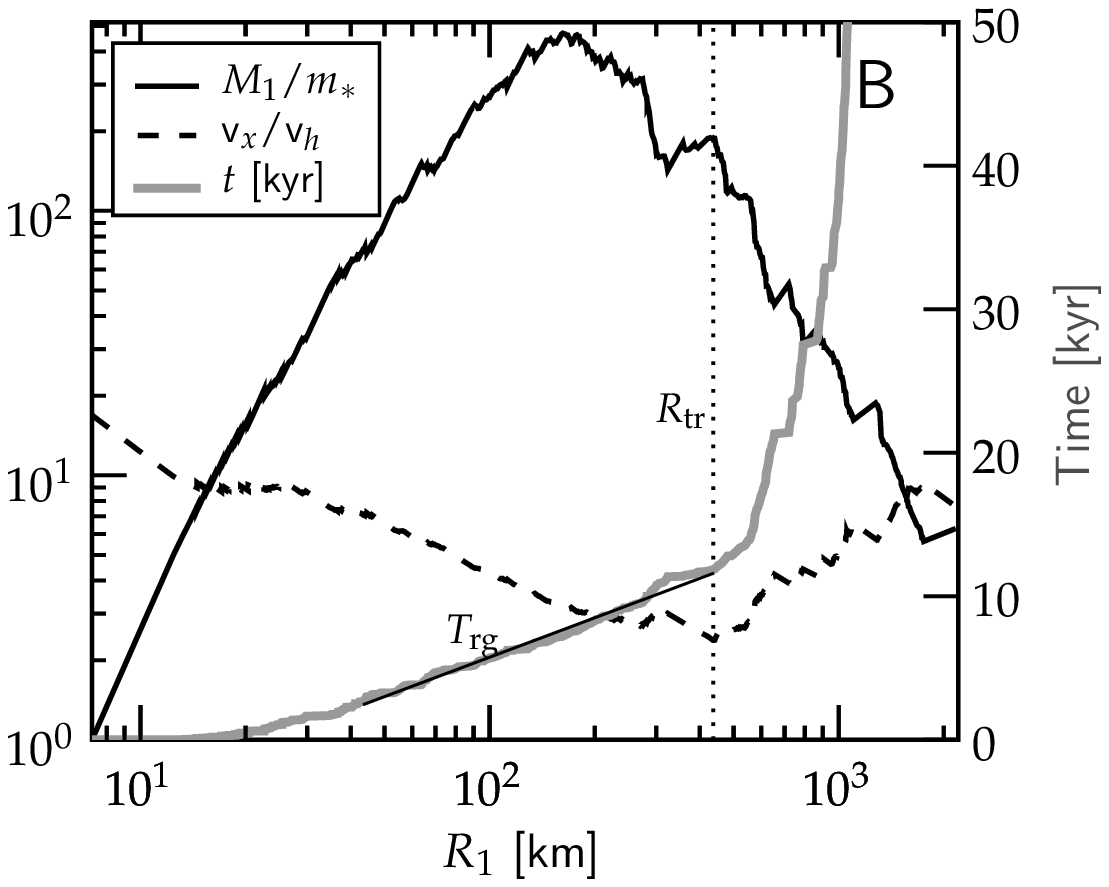}
  \caption{\label{fig:stat1Mz}(\textit{left}) The surface density spectrum, $d\Sigma/dR$ for the 1 AU, gas drag simulations discussed in the previous section.  Curves are plotted at every factor of 2 increase in the radius of the biggest particle, $R_1(t)$. The dotted line in the upper-left of the panels shows the trend of the column density spectra $N_s(m) \propto m^p$ with $p=-2.5$.  The dashed line in the bottom-right corner corresponds to 1 particle per bin.  (\textit{right}) Several statistics shown as function of the \textit{evolutionary parameter} $R_1(t)$: \sumi\ the ratio of $M_1$ to the characteristic mass of the population, $m_\ast$; \sumii\ the maximum velocity within the population to the Hill velocity of the largest body; and \sumiii\ the time (on the second $y$-axis).  The dotted vertical line denotes the minimum of $v_x/v_h$.}
\end{figure}
In \fg{stat1Mz} we again plot results from the previous section -- \ie\ run \#11 of \Tb{simruns}, which results were also presented in \fg{Inaba-1}.  \Fg{stat1Mz}a now shows the mass \textit{spectrum}, instead of the cumulative distribution of \fg{Inaba-1}a.  (Thus, $(d\Sigma/dR)\Delta R$ gives the surface density of bodies within the size interval $[R,R+\Delta R]$).   For our purposes we find it more instructive to show mass spectra like \fg{stat1Mz}a rather than cumulative distributions, since the relevant features turn out more clearly.  However, \fg{Inaba-1}a is in fact just an integrated copy of \fg{stat1Mz}a.

Once the radius of the maximum particle in the distribution has increased by a factor of 2, a curve is plotted and the corresponding time is indicated.  In \fg{stat1Mz}a the dotted auxiliary line in the upper left corner indicates the trend if the column density spectrum, $N_s(m)=(1/m)d\Sigma/dm$, would be a power-law of mass, $N_s(m)\propto m^p$ with exponents $p = -2.5$.  For reference, a flat slope ($p=-2$) would indicate that the distribution contains an equal amounts of mass per logarithmic size bin.  Thus, \fg{stat1Mz}a shows that the high mass tail of the distribution first flattens towards a $p\approx-2.5$ slope, before it breaks, and that most of the mass remains at the initial planetesimal size $R_0$.  This behavior is consistent with $N$-body simulations \citep{KokuboIda1996,BarnesEtal2009}.  Afterwards, the distribution evolves into two components with two identifiable peaks: one bump appears at low-masses that with time evolves to larger sizes and another `spike' is associated with the largest bodies.  Finally, the line in the lower right corner of \fg{stat1Mz}a indicates the size spectrum if there would be only a single body in the mass bin. The curves have to stay above this line.

What would be the best indicator to assess whether a system is in runaway growth (RG)?  Probably the best indicator is $M_1/M_2$, \ie\ the ratio between the mass of the biggest to the second-biggest body in the system.  When this ratio increases, the system is in RG; otherwise it is not.  Unfortunately, the problem is that this quantity behaves very erratically when the bodies are still close to each other in terms of their masses: stochastic processes then interfere to produce a noisy behavior.  For this reason, rather than $M_1/M_2$, we propose to use the ratio of the most massive body over the `characteristic mass', $M_1/m_\ast$, as an indicator for RG, where $m_\ast$ is defined as
\begin{equation}
  m_\ast = \sum_{m_i\neq M_1} N(m_i) m^2_i \left/ \sum_i N(m_i) m_i \right.,
  \label{eq:mpeak}
\end{equation}
which traces the particles that contain most of the mass, excluding the most massive body $M_1$.  For narrow distributions $m_\ast$ approximately corresponds to the peak of the $m^2 N(m)$ mass spectrum.  However, here we will merely use it as a tracer for the `mass flow'. If $m_\ast$ increases more steeply than $M_1$, the mass flow is no longer preferentially directed to $M_1$; the accretion rate of other bodies, \eg\ $m\sim M_2$ or $m\sim m_\ast$, then starts to outweigh that of the most massive one.  In two component systems, previous toy models have shown that $M_1/m_\ast$ increases for RG; otherwise the system is not in RG \citep{WetherillStewart1989,Lee2000,OrmelSpaans2008}.

This ratio is plotted in \fg{stat1Mz}b by the solid black line.  There are still many stochastic fluctuations due to merging of large bodies, particularly at later times, but the general trend is clear.  At $R_1 \approx 100$ km $M_1/m_\ast$ reaches a plateau and starts to decline more visibly after $R_1\approx 400$ km, indicating that RG has terminated.  The mass of the system flows to larger sizes, but not exclusively to one object.

This behavior can be understood from the trend of $v_x/v_h$, plotted by the dashed curve.  Here, $v_x$ is the maximum random velocity in the simulation (which is associated to the low-$m$ planetesimal bodies from which the runaway body is accreting) and $v_h$ is the Hill velocity of the most massive particle.  Thus, the ratio $v_x/v_h$ is a measure of the amount of gravitational focusing (GF) the runaway body experiences when accreting the small bodies: if it decreases, GF increases, whereas if $v_x/v_h$ increases, GF decreases (recall from \se{key-def} that $v_h$ relates to the escape velocity as $v_\mathrm{esc}=v_h\sqrt{6/\alpha}$).  If only accretion would operate (constant $v_x$) $v_x/v_h \propto R_1^{-1}$.  However, by exciting the random motions of the planetesimals, viscous stirring counteracts the decrease of $v_x/v_h$: it increases $v_x$ and decreases the GF.  From \fg{stat1Mz}b it can be seen that accretion dominates during the initial stages but also that the decrease of $v_x/v_h$ is not so steep as in the stirring-free limit.  We find that the scaling is now rather $v_x/v_h \propto R_1^{-0.5}$.  Clearly, there is a positive feedback effect at work: the growth of the largest body increases $v_h$, which in turn increases the GF factor.  This mechanism operates in the initial stages.  However, the relative importance of the viscous stirring becomes more apparent at low $v_x/v_h$, see \fg{rintrad}.  Consequently, a minimum of $v_x/v_h$ is reached.  We denote this point ($R_1 \approx 400$ km) the transition size $R_\mathrm{tr}$. GF factors peak at $R_1 = R_\mathrm{tr}$. 

The transition signifies a different growth phase as can be seen by the gray curve in \fg{stat1Mz}b, which shows the simulation time as function of $R_1$.  At times $R_1(t) < R_\mathrm{tr}$ growth proceeds exponentially, $R_1(t) \propto \exp[t]$, whereas if $R_1(t)\gtrsim R_\mathrm{tr}$ the growth proceeds much slower.  Note the linear spacing of the time-axis in \fg{stat1Mz}b.  We obtain the associated timescale $T_\mathrm{rg}$ empirically by a fit to $R_1(t)$, see \fg{stat1Mz}b, \ie\
\begin{equation}
  M_1(t) \propto \exp \left( \frac{t}{T_\mathrm{rg}} \right).
  \label{eq:Trg}
\end{equation}
(Note that $T_\mathrm{rg}$ is defined in terms of mass, not radius).  We refer to $T_\mathrm{rg}$ as the runaway-growth timescale since it is associated to the RG part of the evolution; \ie\ the phase where $R<R_\mathrm{tr}$. 

The observation that the growth before the transition point proceeds exponentially at a `measured' rate $T_\mathrm{rg}$ is, above all, an empirical finding.  It implies that the accretion rate of the largest particle $dM_1/dt$ proceeds linearly with its mass $M_1$.  From the observed relation $v_x/v_h \propto R_1^{-1/2} \propto M_1^{-1/6}$ one indeed finds that the accretion rate in the dispersion-dominated regime is (approximately) linear, $dM_1/dt \propto R_1^2 (v_h/v_x)^2 \propto M_1$.  In \se{Mcontrib} we will see that during the runaway growth phase all size bins contribute (approximately equally) to the growth of the biggest body.  Mergers among big bodies take place in the s.d.-regime, for which the accretion rate is also linear with the mass of the biggest body.

\subsection{The emergence of oligarchy}
\label{sec:emerge-oli}
Next, we consider the evolution beyond the runaway growth transition $R_1(t)>R_\mathrm{tr}$.  \Fg{scatter1au} shows the spatial distribution of the planetesimal swarms corresponding to the simulation discussed above, \ie\ model \#11 of \Tb{simruns}.  Each symbol indicates a planetesimal swarm (representative body).  In \fg{scatter1au} the total mass in the planetesimal swarms is indicated by the area of the symbols.  Representative bodies that contain a single particle ($N_g=1$) are indicated by diamonds rather than dots.  The Hill spheres of the ten most massive bodies are indicated by red bars.  Finally, the colorbar represents the gravitational focusing factors like in \fg{stat1Mz}: the ratio of the random velocity ($v$) to the Hill velocity of the largest body ($v_h$). 

\begin{figure*}[tbp]
  \includegraphics[width=\textwidth,clip]{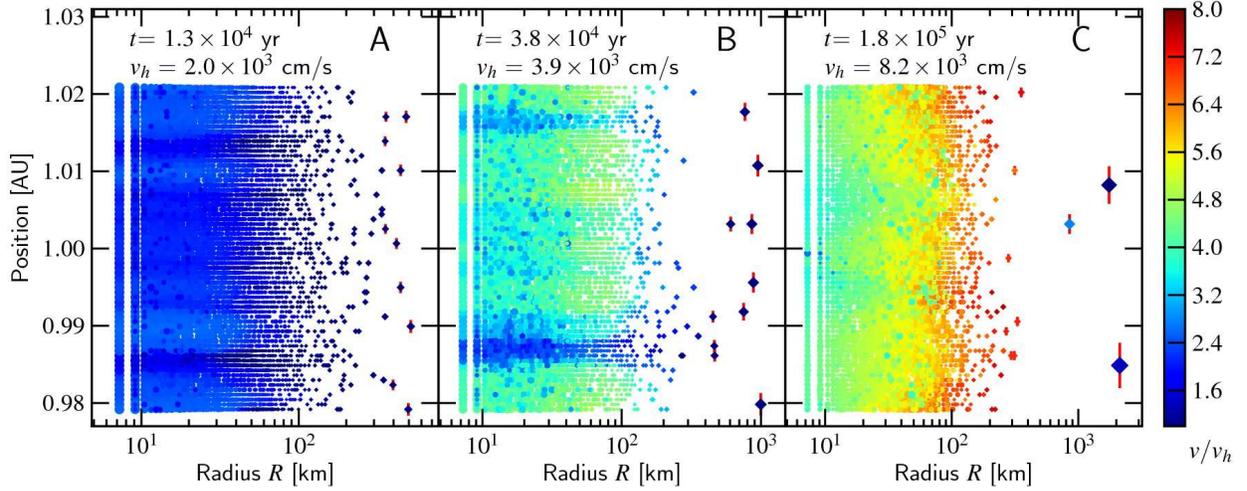}
  \caption{\label{fig:scatter1au}Scatter plot of bodies' mass and position at three times during the oligarchy phase, corresponding to the 1AU gas-drag simulation.  Each dot represents a planetesimal swarm.  The size of the dot is a measure for the total mass of the swarm, such that area(dot) $\propto m_\mathrm{swarm}^{1/3}$.  Individual bodies (\ie\ those that have $N_g=1$) are shown by \textit{diamonds}.  The color is a measure for the random velocity (eccentricity) of the bodies in terms of $v/v_h$ (see the colorbar on the right).  These values are normalized to the Hill velocity ($v_h$) of the largest body, which value is indicated in the panel. The Hill radius of the 10 most massive bodies are indicated by a red bar. Note that the scaling on the $x$-axis differs among the three panels.}
\end{figure*}
\Fg{scatter1au} shows the planetesimal distribution at three different times: at $R_1=500$ km, just after the transition size; at $R_1=10^3$ km; and at $R_1=2\times10^3$ km, the end state of the simulation.  \Fg{scatter1au} very clearly shows that with time: \sumi\ the gravitational focusing factors increase; \sumii\ the number of oligarchs decreases; \sumiii\ the gap in mass between the oligarchs and (leftover) planetesimal increases.  The oligarchy gets more pronounced with time.

For $R_1(t)>R_\mathrm{tr}$, viscous stirring by the high-mass bodies gains the upper hand and accretion times increase due to the fact that gravitational focusing factors decline (increasing $v_x/v_h$).   Other (massive) bodies then catch up.  Indeed, $R_\mathrm{tr}$ also approximately corresponds to the point where $M_1/m_\ast$ starts to decrease noticeably, see \fg{stat1Mz}.  In due time the gas drag should balance the stirring to produce an equilibrium eccentricity that is characterized by a constant $v_x/v_h$ \citep[\eg][]{IdaMakino1993,ThommesEtal2003}.  Our simulation evolves towards this state in a rather erratic way that is caused by the merging among big bodies.  Note that in our case $v_x$ traces the largest random velocity of any body.  Towards the end of the simulation, in \fg{scatter1au}c, a reversal in the random velocity distribution has taken place, in which the low-mass planetesimals have no longer the largest random velocity due to the fact that gas drag is more effective for these bodies (see also \fg{Inaba-1}). This explains why $v_x/v_h$ does not readily approach a constant value.

Since the number of oligarchs declines, a large contribution to the growth during the oligarchy stage should come by the merging of similar-size bodies.  The oligarchs are simply packed too densely to guarantee their mutual existence.  There is some limited diffusion of the oligarchs due to merging of bodies (see \se{multi-zone}).  Scattering among oligarchs is not implemented in our approach.  However, even if implemented, scattering cannot isolate these bodies from each other since there are too many of them.  The only way forward is to merge the oligarchs.  The consequence is that the distance among the oligarchs in our simulations is therefore at least several Hill radii -- a spacing that directly reflects the choice of $R_\mathrm{col} = 2.5R_h$ in the low velocity regime.  Oligarchs that come within this distance have a strong probability to merge.  Although we cannot reproduce the exact spatial structure of the $N$-body simulations, this picture of merging oligarchs is broadly consistent with the $N$-body simulations of \citet{KokuboIda2002}.

Since our model does not treat scattering, gap formation within the planetesimal disk is not observed.  This effect could seriously slow down the growth during the oligarchy stage.  Judging from the calculations of \citet{Rafikov2003ii}, gap formation becomes important for bodies of mass $>10^{25}$ g ($R \approx 10^3$ km).  Yet, again, the oligarchs are likely to be too densely packed to prevent them accreting planetesimals.  With increasing size of the oligarchs scatterings should become more pronounced, however.  In a recent $N$-body simulation involving Earth mass embryos and planetesimals, \citet{LevisonEtal2010} observed that the planetesimals were scattered over AU-distances -- out of the feeding zones of the embryos.  Modeling these effects are beyond the scope of this work.

\subsection{Definitions of runaway growth: local and global}
\label{sec:rg-def}
A system of bodies is in RG when the ratio $M_1/M_2$ increases with time, \ie\ $d(M_1/M_2)/dt > 0$ \citep{WetherillStewart1989}, where $M_1$ and $M_2$ are, respectively, the mass of the most massive and the second-most massive particle in the system. Alternatively, we can compare the accretion timescales; thus, the system is in RG \textit{at time $t$} when the condition 
\begin{equation}
  T_1^\mathrm{ac}(t) < T_2^\mathrm{ac}(t)  
  \label{eq:rg-def}
\end{equation}
is fulfilled, where $T^\mathrm{ac}$ is the accretion timescale
\begin{equation}
  T^\mathrm{ac} = \left( \frac{1}{M} \frac{dM}{dt} \right)^{-1}.
  \label{eq:tac}
\end{equation}

 In a two-component system, the mass accretion rate $dM/dt$ is proportional to the collisional cross section $R_\mathrm{col}^2$, the approach velocity $v_a$, and the number density of particles $N_s/2h_\mathrm{eff}$, \ie\ $dM/dt \propto R^2_\mathrm{col} v_a N_s/h_\mathrm{eff}$.  If we take the dispersion-dominated regime where $R_\mathrm{col} \propto R_s^2 (v_h/v_a)^2$, $v_a/h_\mathrm{eff}=\Omega$ (see \se{interactions}), and take $N_{s1} = N_{s2}$, the condition \eq{rg-def} translates into 
\begin{equation}
  R_1 \left( \frac{v_{x1}}{v_{h1}} \right)^2 < R_2 \left( \frac{v_{x2}}{v_{h2}} \right)^2.
\end{equation}
If the interactions take place in the same zone, \ie\ $v_{x1}=v_{x2}$, the RG-condition is always satisfied since $v_h\propto R$.  In that case the RG-index $\kappa$ as in
\begin{equation}
  \frac{dM}{dt} \propto M^\kappa
  \label{eq:dMdt}
\end{equation}
equals $\kappa = 4/3$.  The usual criterion is that $\kappa>1$ is needed for RG to ensue \citep{WetherillStewart1989}.  However, this is only valid for interaction within the same spatial zone, \ie\ where competitor bodies accrete from the same reservoir of (low-mass) planetesimals.

Similarly, if $v_x$ is constant everywhere the RG-condition holds globally.  However, this is not to be expected since viscous stirring will increase the random velocities of the planetesimal bodies.  We saw above that initially, growth proceeded exponentially, $v_x/v_h \propto R_1^q$ with $q\approx-0.5$, which means that the RG-condition is marginally satisfied.  Nonetheless, we will refer to the initial (exponential) growth state as the \textit{runaway growth phase}.  After the transition size has been reached ($R_1>R_\mathrm{tr}$) $v_x/v_h$ increases with $R_1$ and \eq{rg-def} is no longer \textit{globally} fulfilled (but locally it still is since $v_{x1}=v_{x2}$).  This, is the oligarchy stage.

As the term `runaway growth' is sometimes used rather colloquially in the literature, we summarize a few statements that are in agreement with the formal definition above, \eq{rg-def}:
\begin{enumerate}
  \item \textit{RG is not always synonymous with fast growth}.  Runaway growth is, for understandable reasons, often identified with fast growth rates, and, therefore, sometimes associated with the s.d.-regime, $v_x<v_h$.  However, in the s.d.-regime $\kappa \le 1$ and the growth mode is not runaway\footnote{For a thick planetesimal disk in the shear-dominated regime we find $dM/dt \propto M$ and $\kappa = 1$; for a very thin planetesimal disk, on the other hand, $R_\mathrm{col}$ becomes larger than the scaleheight $h_\mathrm{eff}$. In that essentially 2D setting $dM/dt \propto M^{2/3}$ and $\kappa=2/3$ \citep{KokuboIda1996}.}.  Indeed, \eq{rg-def} does not specify the absolute growth rate of the bodies.   
  \item \textit{RG does not require dynamical friction}.  During RG bodies interact in the d.d.-regime.  When bodies start out at $v \gtrsim v_\mathrm{esc}$ no RG takes place.  In this case dynamical friction, which reduces the random velocities of the most massive bodies, could shift the most massive bodies in the d.d.-regime, such that \eq{rg-def} materializes \citep{WetherillStewart1993}.  However, dynamical friction is not required to sustain RG as long as they take place in the d.d.-regime.  
  \item \textit{RG is not synonymous with immediate mass separation}.  Although \eq{rg-def} implies that masses will separate, this criterion neglects stochastic effects and the actual mass-doubling time of the second-most massive body can still be shorter.  In the long run, however, RG will result in mass-separation but \eq{rg-def} does not specify at which point this occurs.
  \item \textit{In oligarchy, the system is only locally in RG}.  At late times, the runaway body (or oligarch) regulates the velocity dispersion of the planetesimals leading to a positive feedback on the random velocities of the bodies, which slows down the growth.  As explained above, interactions within the same spatial zone, \ie\ the region of the disks where the runaway body have stirred the random velocity of the planetesimals to (the same) $v_x$, always fulfill the RG-condition as long as the d.d.-regime holds.  However, among bodies of different spatial zones \eq{rg-def} is no longer satisfied.  The combined effect of local RG and isolation is better known as oligarchy \citep{KokuboIda1998}.
\end{enumerate}

In the remainder of the paper we will identify systems that experience the initial exponential (global runaway) growth with the \textit{runaway growth phase} and systems that undergo only local-RG with the \textit{oligarchic growth phase}.  That is, we consider an evolutionary sequence in which the oligarchy phase supersedes the runaway growth phase.  We distinguish the runaway growth and the oligarchy phases of the planetesimal accretion process as follows (see \fg{stat1Mz}b):
\begin{itemize}
  \item In the RG phase ($R_1(t)<R_\mathrm{tr}$), $v_x/v_h$ decreases and GF-factors increase.  Growth occurs exponentially at a characteristic timescale $T_\mathrm{rg}$.  In addition, the quantity $M_1/m_\ast$ increases during most of the phase.
  \item In the oligarchy phase ($R_1(t)>R_\mathrm{tr}$), $v_x/v_h$ increases (GF-factors decrease). As a result, accretion timescales increase rapidly and $M_1/m_\ast$ decreases.
\end{itemize}

\section{Parameter study}
\label{sec:rgsims}
\begin{deluxetable}{llllllll}
  \tablecaption{List of simulation runs \label{tab:list}}
  \tabletypesize{\small}
  \tablewidth{0pt}
\tablehead{
\colhead{Radius+Features\tablenotemark{a}} & \colhead{$N_\mathrm{res}$}  & \colhead{$N_\mathrm{zo}$}  &  \colhead{$\Sigma$} & \colhead{$\Delta a$} & \colhead{$R_0$}   & \colhead{$v_0$}   & \colhead{Comments/Figure refs} \\
\colhead{[AU]}              &       &     & \colhead{$\mathrm{[g\ cm^{-2}]}$}   & \colhead{[AU]}      & \colhead{[km]}  & \colhead{[m/s]}   &      \\
\colhead{ (1)}         & \colhead{(2)}               & \colhead{(3)}   &  \colhead{(4)}   & \colhead{(5)} & \colhead{(6)}       & \colhead{(7)}       & \colhead{(8)}
    }
\startdata
\multicolumn{8}{l}{Models at different disk radii} \\ \hline
1Ve & {$500$} & {$67$} & {$16.8$} & {$0.042$} & {$7.3$} & {$4.7$} & Fig.\ \ref{fig:stats1} \\
1VeGd & {$500$} & {$67$} & {$16.8$} & {$0.042$} & {$7.3$} & {$4.7$} & Fig.\ \ref{fig:stats1} \\
1VeGdFr & {$500$} & {$67$} & {$16.8$} & {$0.042$} & {$7.3$} & {$4.7$} & Fig.\ \ref{fig:stats1} \\
6Ve & {$500$} & {$33$} & {$2.0$} & {$0.18$} & {$7.3$} & {$2.7$} & Fig.\ \ref{fig:stats1} \\
6VeGd & {$500$} & {$33$} & {$2.0$} & {$0.18$} & {$7.3$} & {$2.7$} & Fig.\ \ref{fig:stats1} \\
6VeGdFr & {$500$} & {$33$} & {$2.0$} & {$0.18$} & {$7.3$} & {$2.7$} & Fig.\ \ref{fig:stats1} \\
35Ve & {$500$} & {$13$} & {$0.20$} & {$0.97$} & {$7.3$} & {$2.7$} & Fig.\ \ref{fig:stats1} \\
35VeGd & {$500$} & {$13$} & {$0.20$} & {$0.97$} & {$7.3$} & {$2.7$} & Fig.\ \ref{fig:stats1} \\
35VeFr & {$500$} & {$13$} & {$0.20$} & {$0.97$} & {$7.3$} & {$2.7$} & \se{35VeFr}, Fig.\ \ref{fig:stats1} \\
\multicolumn{8}{l}{Models varying fragmentation parameters} \\ \hline
35VeFrHi-afr & {$500$} & {$13$} & {$0.20$} & {$0.97$} & {$7.3$} & {$2.7$} & $a_\mathrm{frag}=10$ cm \\
35VeFrHi-$\epsilon$ & {$500$} & {$13$} & {$0.20$} & {$0.97$} & {$7.3$} & {$2.7$} & $\epsilon=0.1$ \\
35VeFrLo-$\epsilon$ & {$500$} & {$13$} & {$0.20$} & {$0.97$} & {$7.3$} & {$2.7$} & $\epsilon=0.001$ \\
\multicolumn{8}{l}{Models including turbulent stirring} \\ \hline
1VeGdFrTs & {$500$} & {$67$} & {$16.8$} & {$0.042$} & {$7.3$} & {$4.7$} & Fig.\ \ref{fig:stats2} \\
6VeGdFrTs & {$500$} & {$33$} & {$2.0$} & {$0.18$} & {$7.3$} & {$2.7$} & Fig.\ \ref{fig:stats2} \\
35VeGdFrTs & {$500$} & {$13$} & {$0.20$} & {$0.97$} & {$7.3$} & {$2.7$} & Fig.\ \ref{fig:stats2} \\
\multicolumn{8}{l}{35 AU, miscellaneous} \\ \hline
35VeFr1km & {$500$} & {$95$} & {$0.20$} & {$0.99$} & {$1.0$} & {$0.37$} & Fig.\ \ref{fig:stats3} \\
35VeFr50km & {$500$} & {$5$} & {$0.20$} & {$2.6$} & {$50.0$} & {$19.0$} & Fig.\ \ref{fig:stats3} \\
35VeFrLo-v0 & {$500$} & {$13$} & {$0.20$} & {$0.97$} & {$7.3$} & {$1.0$} & Fig.\ \ref{fig:stats3} \\
35VeFrHi-v0 & {$500$} & {$13$} & {$0.20$} & {$0.97$} & {$7.3$} & {$25.0$} & Fig.\ \ref{fig:stats3} \\
35VeFrHi-$\Sigma$ & {$500$} & {$13$} & {$2.0$} & {$0.97$} & {$7.3$} & {$2.7$} & Fig.\ \ref{fig:stats3} \\
35VeFrLo-$\Sigma$ & {$500$} & {$13$} & {$0.020$} & {$0.97$} & {$7.3$} & {$2.7$} & Fig.\ \ref{fig:stats3} \\
\enddata
  \tablenotetext{a}{Featured abbreviations, which make up the simulation name listed in \Col{2}, denote: Ve, velocity evolution (includes viscous stirring, dynamical friction, and collisional cooling); Gd, gas drag; Fr, fragmentation; Ts, turbulent stirring, Lo-v0, low initial random velocity, Hi-v0, high initial velocity, \etc
  }
  \tablecomments{
  Columns denote: (1) disk position (in AU) and features; (2) number of simulation particles per zone; (3) number of zones; (4) surface density in solids; (5) total simulation width; (6) initial radius of bodies; (7) initial random velocity; (8) comments.} 
\end{deluxetable}
\Tb{list} contains the list of runs that have been performed.  The prime goal is to cover a wide range of physical conditions and disk radii to see whether the picture sketched in the previous section holds generally.  In particular, this concerns the behavior of the $v_x/v_h$ indicator and its relation to the growth rate and the $M_1/m_\ast$ indicator.  For this reason we perform simulations at three disk radii: 1, 6, and 35 AU, in which damping by gas drag and fragmentation are varied (first 9 entries in \Tb{list}).  Since we find that fragmentation is an important mechanism, we next test how sensitive the outcome is under variation of the fragment size $a_\mathrm{fr}$ and the coefficient of restitution parameter, $\epsilon$ (see \se{colmod}).  The next class of runs contain turbulent stirring, abbreviated Ts.  Finally, we focus on the gas-free runs at 35 AU and vary additional physical parameters, like the initial planetesimal size $R_0$, the initial random velocity $v_0$, and the surface density $\Sigma$. 

In \Tb{list} \col{1}\ gives the model name, which is a mnemonic abbreviation of the semi-major axis at which the run was performed and the features it includes.  \Col{2} lists the number of superparticles (swarms) used per zone and \col{4} the number of zones that are included, which follows the guidelines outlined in \se{multi-zone}.  \Col{4} provides the surface density of solids, \col{5} lists the total width of the simulation, \col{6} the initial planetesimal radius, and \col{7} the initial random velocity dispersion, which is taken to be half the initial escape velocity of the bodies.  Bodies corresponding to simulations at 1 AU have an internal density ($\rho_s$) of $3\ \mathrm{g\ cm^{-3}}$, whereas the bodies from the 6 and 35 AU simulations have an internal density of $1\ \mathrm{g\ cm^{-3}}$.

The model feature abbreviations (\col{1}) imply that simulations include:
\begin{itemize}
  \item Velocity evolution (Ve). This is shorthand for all processes through which the random velocities are affected by collisions and gravitational stirring:  viscous stirring, dynamical friction, and collisional cooling.  Inclinations are calculated independently from eccentricities.  All runs include these features.
  \item Gas drag (Gd), which damps the random velocities according to \eq{stirring-gd}.
  \item Fragmentation (Fr),  for approach velocities that exceed $v_\mathrm{esc}$ (\ie\ the case without GF), according to the procedure outlined in \se{colmod}.  
  \item Turbulent stirring (Ts), an external source of excitation to the random motions of the bodies, see \se{Tstir}.  We list only runs with $\gamma=10^{-4}$. Runs that were performed at $\gamma=10^{-3}$ did not result in accretion.
\end{itemize}

\begin{deluxetable}{lllllll}
  \tablecaption{Simulation results \label{tab:results}}
  \rotate
\tablewidth{0pt}
  \tabletypesize{\small}
  \tablehead{
  Name  & $R_\ast$        & $R_\mathrm{tr}$             & $T_\mathrm{rg}$   & $T_\mathrm{2k}$ &  $f_\mathrm{frag}^\mathrm{end}$  & $f_\mathrm{tot}^\mathrm{end}$ \\
          & [km]          & [km]                        & [Myr]             & [Myr]           &                                  &      \\
    (1)   & (2)           & (3)                         & (4)               & (5)             &  (6)                             &  (7)
    }
\startdata
1Ve & {$(1.9\pm0.4)\times10^{2}$} & {$(3.8\pm0.7)\times10^{2}$} & {$(1.4\pm0.1)\times10^{-3}$} & {$(4.1\pm0.4)\times10^{-1}$} & {$0$} & {$0$} \\
1VeGd & {$(2.0\pm0.3)\times10^{2}$} & {$(3.6\pm0.5)\times10^{2}$} & {$(1.3\pm0.1)\times10^{-3}$} & {$(2.3\pm0.2)\times10^{-1}$} & {$0$} & {$0$} \\
1VeGdFr & {$(1.9\pm0.2)\times10^{2}$} & {$(3.5\pm0.3)\times10^{2}$} & {$(1.2\pm0.1)\times10^{-3}$} & {$(4.8\pm0.2)\times10^{-2}$} & {$(10.0\pm0.9)\times10^{-2}$} & {$(3.1\pm0.3)\times10^{-1}$} \\
6Ve & {$(3.4\pm0.3)\times10^{2}$} & {$(7.2\pm1.3)\times10^{2}$} & {$(6.9\pm0.3)\times10^{-2}$} & {$(3.0\pm0.7)\times10^{0}$} & {$0$} & {$0$} \\
6VeGd & {$(3.5\pm0.8)\times10^{2}$} & {$(7.7\pm1.3)\times10^{2}$} & {$(6.4\pm0.4)\times10^{-2}$} & {$(2.1\pm0.1)\times10^{0}$} & {$0$} & {$0$} \\
6VeGdFr & {$(3.5\pm0.5)\times10^{2}$} & {$(7.2\pm0.9)\times10^{2}$} & {$(5.9\pm0.3)\times10^{-2}$} & {$(1.2\pm0.1)\times10^{0}$} & {$(4.6\pm0.9)\times10^{-2}$} & {$(6.9\pm1.3)\times10^{-2}$} \\
35Ve & {$(5.0\pm0.5)\times10^{2}$} & {$(1.1\pm0.3)\times10^{3}$} & {$(1.2\pm0.1)\times10^{1}$} & {$(3.2\pm0.7)\times10^{2}$} & {$0$} & {$0$} \\
35VeGd & {$(5.4\pm1.1)\times10^{2}$} & {$(9.5\pm2.1)\times10^{2}$} & {$(5.9\pm0.3)\times10^{0}$} & {$(1.6\pm0.1)\times10^{2}$} & {$0$} & {$0$} \\
35VeFr & {$(4.5\pm0.3)\times10^{2}$} & {$(1.4\pm0.3)\times10^{3}$} & {$(5.6\pm0.2)\times10^{0}$} & {$(1.1\pm0.0)\times10^{2}$} & {$(9.3\pm2.6)\times10^{-3}$} & {$(1.7\pm0.4)\times10^{-2}$} \\
35VeFrHi-afr & {$(4.7\pm0.9)\times10^{2}$} & {$(1.1\pm0.6)\times10^{3}$} & {$(1.1\pm0.1)\times10^{1}$} & {$(1.7\pm0.1)\times10^{2}$} & {$(2.4\pm0.4)\times10^{-2}$} & {$(3.3\pm0.5)\times10^{-2}$} \\
35VeFrHi-$\epsilon$ & {$(2.7\pm0.8)\times10^{2}$} & {$(9.8\pm3.4)\times10^{2}$} & {$(1.3\pm0.1)\times10^{0}$} & {$(3.7\pm0.1)\times10^{1}$} & {$(2.2\pm0.4)\times10^{-2}$} & {$(3.8\pm0.7)\times10^{-2}$} \\
35VeFrLo-$\epsilon$ & {$(4.9\pm0.7)\times10^{2}$} & {$(9.9\pm1.1)\times10^{2}$} & {$(1.2\pm0.1)\times10^{1}$} & {$(2.1\pm0.1)\times10^{2}$} & {$(3.4\pm0.3)\times10^{-3}$} & {$(6.7\pm0.6)\times10^{-3}$} \\
1VeGdFrTs & {$3.5\times10^{2}$} & {$6.2\times10^{2}$} & {$1.0\times10^{-2}$} & {$2.5\times10^{-1}$} & {$3.4\times10^{-2}$} & {$2.0\times10^{-1}$} \\
6VeGdFrTs & {$7.1\times10^{2}$} & {$2.0\times10^{3}$} & {$9.0\times10^{-1}$} & {$2.2\times10^{1}$} & {$2.8\times10^{-2}$} & {$2.5\times10^{-1}$} \\
35VeGdFrTs & {$1.1\times10^{3}$} & {$2.1\times10^{3}$} & {$1.2\times10^{1}$} & {$7.2\times10^{3}$} & {$4.4\times10^{-2}$} & {$3.3\times10^{-1}$} \\
35VeFr1km & {$1.7\times10^{2}$} & {$6.8\times10^{2}$} & {$6.8\times10^{-1}$} & {$1.8\times10^{1}$} & {$3.5\times10^{-2}$} & {$4.8\times10^{-2}$} \\
35VeFr50km & {$1.2\times10^{3}$} & {$1.8\times10^{3}$} & {$5.7\times10^{1}$} & {$5.4\times10^{2}$} & {$2.1\times10^{-3}$} & {$4.6\times10^{-3}$} \\
35VeFrLo-v0 & {$5.0\times10^{2}$} & {$1.7\times10^{3}$} & {$5.9\times10^{0}$} & {$1.0\times10^{2}$} & {$6.7\times10^{-3}$} & {$1.4\times10^{-2}$} \\
35VeFrHi-v0 & {$3.4\times10^{2}$} & {$1.5\times10^{3}$} & {$4.2\times10^{-1}$} & {$3.2\times10^{1}$} & {$1.9\times10^{-2}$} & {$3.0\times10^{-2}$} \\
35VeFrHi-$\Sigma$ & {$7.9\times10^{2}$} & {$1.6\times10^{3}$} & {$4.0\times10^{-1}$} & {$8.9\times10^{0}$} & {$3.5\times10^{-3}$} & {$6.5\times10^{-3}$} \\
35VeFrLo-$\Sigma$ & {$2.4\times10^{2}$} & {$4.8\times10^{2}$} & {$8.8\times10^{1}$} & {$1.5\times10^{3}$} & {$1.6\times10^{-2}$} & {$7.4\times10^{-2}$}
\enddata
  \tablecomments{
  Column entries denote: (1) model name; (2) size at which $M_1/m_\ast$ reaches its maximum; (3) size at which $v_x/v_h$ reaches its minimum; (4) runaway growth timescale; (5) time at the end of the simulation run (at $R_1=2\times10^3$ km); (6) mass fraction in fragments at end of simulation; (6) total mass fraction that has been processed through fragments.  Error bars denote the spread over 5 runs.
  }
\end{deluxetable}
Simulations are continued until a size $R=2\times10^3$ km is reached for the most massive bodies.  \Tb{results} provides statistical quantities that characterizes the outcome of the runs.  These include, the radii at which $M_1/m_\ast$ and $v_x/v_h$ have their extrema, denoted $R_\ast$ and $R_\mathrm{tr}$, respectively, and the runaway growth timescale, $T_\mathrm{rg}$ (see \se{rg-indicators}).  When error bars are given these indicate the spread in the quantities over 5 independent simulation runs.  \Tb{results} further gives the mass fraction fragments constitute at the end of the simulation, $f_\mathrm{frag}^\mathrm{end}$, and the total fraction of the mass that has once been in fragments, $f_\mathrm{tot}^\mathrm{end}$.\footnote{Formally, this number can exceed unity since we do not adjust for multiple fragmentation-accretion cycles.}  By comparing these fractions one gets an estimate of the importance of fragmentation \textit{and} of the importance of the re-accretion of these fragments.

\subsection{Runs including gas drag and fragmentation}
\label{sec:outer}
\begin{figure}[tbp]
  \centering
  \includegraphics[width=\figw]{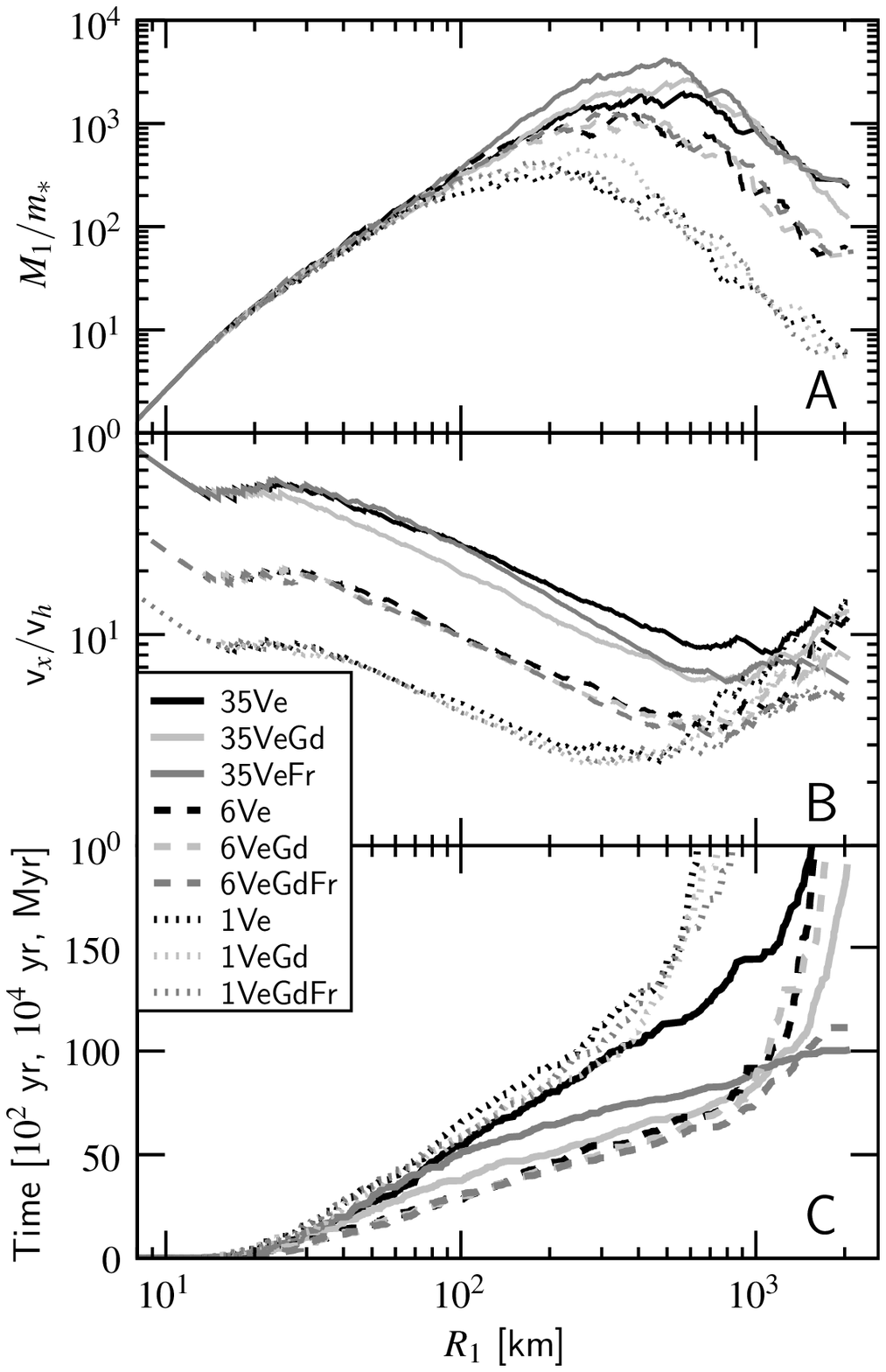}
  \caption{\label{fig:stats1}The statistical quantities for runs at several disk radii.  As function of evolutionary parameter $R_1(t)$ -- the radius of the most massive particle -- are shown: (A) $M_1/m_\ast$, (B) $v_x/v_h$, and (C) time.  Black curves correspond to runs that include neither gas drag nor fragmentation, light gray curves indicate runs that include gas drag, and dark grey curves indicate fragmentation.  In (C) times are normalized to $10^2$ yr (for the 1 AU runs), $10^4$ yr (6 AU), and $10^6$ yr (35 AU), respectively.}
\end{figure}
\Fg{stats1} presents the quantities $M_1/m_\ast$, $v_x/v_h$, and time for a variety of simulations including gas drag and/or fragmentation.  Runs at three different disk radii are shown: 1 AU, 6 AU, and 35 AU.  Note that in \fg{stats1}c time is plotted in units that depend on the radial distance: Myr for the 35 AU runs, $10^4$ yr for the 6 AU runs, and $10^2$ yr for the 1 AU runs.

The general trend of the curves is the same for all the disk radii and reflects the discussion of \se{rg-indicators}.  First, in the runaway growth phase, $M_1/m_\ast$ rises and $v_x/v_h$ decreases.  The outer disk models start out at a larger value of $v_x/v_h$ since Hill velocities for the same mass decrease with increasing disk radius.  Very generally, one can say that interactions in the inner disk are more prone to take place in the shear-dominated regime.  At the point where $v_x/v_h$ has reached a minimum, accretion timescales increase.  A major difference is that for the 1 AU run the transition occurs much sooner (in terms of the evolutionary parameter $R_1(t)$) than for the outer disk models. 

From \fg{stats1} it is seen that gas drag does not influence the growth during the initial runaway growth phase ($R_1<R_\mathrm{tr}$), indicating that it acts on a longer timescale.  The exception is the 35 AU model, where the divergence occurs at a relatively low $R_1$.  The likely reason for this phenomenon is the different gas drag law ($C_D>1$) small bodies experience at 35 AU, which, somewhat paradoxically, increases the (relative) effectiveness of gas drag.  However, in our simulations we do not reduce the gas density at large timescales;  for $t \simeq 10^7\ \mathrm{yr}$ the gas would surely have dissipated from the nebula.  On the other hand, for the longer timescales that characterize the oligarchic stage, gas drag increases in importance.  Balancing stirring with damping, it is found that GF-factors approach an equilibrium \citep[\eg][]{KokuboIda2002}.  Thus, the black curves (no gas drag) should separate from the other curves in the oligarchy stage in \fg{stats1}b.  Although we can see the start of this process, its overall signature is not very clear yet due to the strong fluctuations in the curves caused by merging among similar-size big bodies (\cf\ also our discussion in \se{emerge-oli}.  

Fragmentation, although not playing a major role in the initial runaway growth phase, has the tendency to smooth the features associated with the transition size.  There is still a clear peak in $M_1/m_\ast$ but the signatures of the transition are not so evident (but still present) in $v_x/v_h$ and the timescales plot.  Compare, for example, the behavior of the dashed curves in \fg{stats1}.  The 1 AU and 6 AU curves visibly steepen after the RG/oligarchy transition (at $R_1\approx 10^3$ km) but for the \simu{6VeGdFr} curve this effect is much less obvious.  Fragments do not conform to the self-regulated aspect of oligarchy, since their random velocity is not strongly affected by (the growth of) the biggest bodies.

Our simple fragmentation model contains two free (uncertain) parameters that affect the behavior of fragments: the size of the fragments $a_\mathrm{fr}=1$ mm and the coefficient of restitution, $\epsilon=0.01$.  We have tested the influence of these canonical values by running models with $a_\mathrm{fr}=10$ cm and varying $\epsilon$ with respect to the (gas-free) \simu{35VeFr} run, see \Tb{list}.  A larger fragment size results in less efficient cooling among the fragments due to their reduced total cross section for interaction.  As a result, the fragments are not so efficiently re-accreted during the runaway growth phase and the runaway growth timescale, $T_\mathrm{rg}$, is rather like the non-fragmentation \simu{35Ve} run.   The influence of varying $\epsilon$ is quite significant.  This can be understood by considering the extreme limits: $\epsilon=0$ implies that each collision is fully inelastic and no fragmentation takes place, whereas $\epsilon=1$ implies that every $v>v_\mathrm{esc}$ collision completely shatters both bodies.  Thus, our $\epsilon=0.001$ run lies closer to the non-fragmentation run, whereas in the $\epsilon=0.1$ run more fragments are produced, resulting in shorter overall accretion timescales. 

\subsection{The effects of turbulent stirring}
\label{sec:Tstir2}
\begin{figure}[tbp]
  \centering
  \includegraphics[width=\figw,clip]{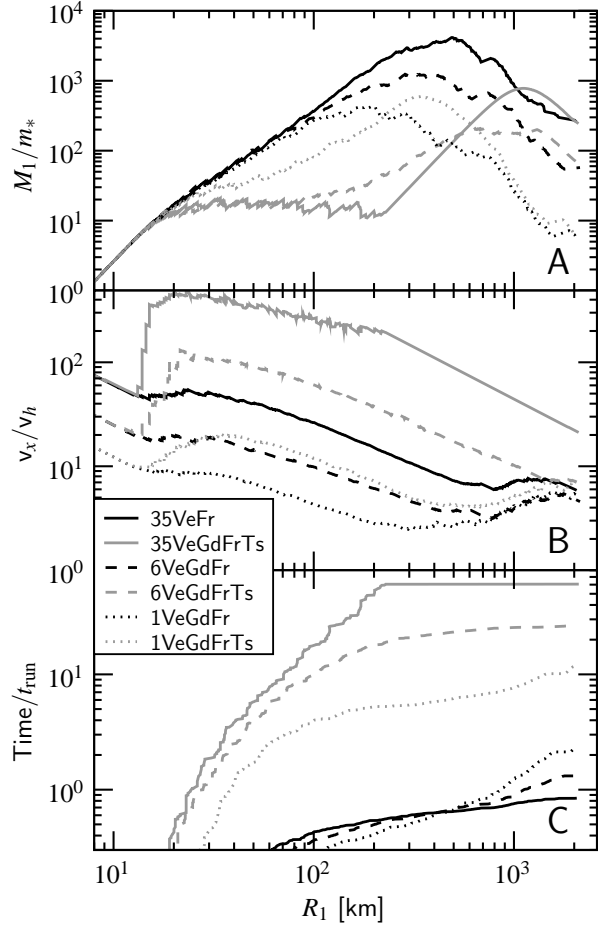}
  \caption{\label{fig:stats2}Statistics of runs including turbulent stirring (Ts). The non-Ts runs are plotted for comparison (\textit{black} lines).}
\end{figure}
 
\Fg{stats2} presents the key indicators for models including turbulent stirring (Ts) and fragmentation (Fr).  Models without stirring are plotted for comparison (black lines).  Here, only runs with a value of $\gamma=10^{-4}$ for the turbulent stirring parameter are shown since it was found that when the stirring parameter $\gamma$ was set to $\gamma=10^{-3}$ no accretion takes place.  In \fg{stats2}c times are normalized by the fiducial
\begin{equation}
  t_\mathrm{run} = \frac{R_0 \rho_s}{\Sigma \Omega} = 1.6\times10^4\ \mathrm{yr} \left( \frac{R_0}{10\ \mathrm{km}} \right) \left( \frac{\Sigma}{10\ \mathrm{g\ cm^{-2}}} \right)^{-1} \left( \frac{\rho_s}{\mathrm{g\ cm^{-3}}} \right) \left( \frac{a}{\mathrm{AU}} \right)^{-3/2}
  \label{eq:trun}
\end{equation}
and plotted on a logarithmic $y$-axis.  In \eq{trun}, $t_\mathrm{run}$ is, upon neglect of a numerical constant, equal to the initial collision timescale between the bodies of size $R_0$, internal density $\rho_s$, and surface density $\Sigma$. That is, $t_\mathrm{run} = (n_0 \sigma_0 v_0)^{-1}$ with $n_0 = \Sigma_0/m_0 h_\mathrm{eff,0}$, $m_0 \sim \rho_s R_0^3$, $h_0 \sim v_0/\Omega$, \etc\  For exponential growth or runaway growth one can expect the initial timescale to be a characteristic timescale of the system.

We find that turbulent stirring at fixed $\gamma$ affects the outer disk more than the inner disk.  This can be understood since gas damping is more effective in the inner nebula.  As can be seen from \fg{stats2}, at 1 AU the effects of turbulent stirring are relatively minor.  However, adding even a relatively small amount of stirring does increase accretion timescales by a factor of 10 or more due to the lower focusing factors (\fg{stats2}c).  In addition, the accretion of the biggest bodies is dominated by fragments.  This has the tendency of erasing the imprints of the runaway/oligarchy transition as present in the conventional models (without Ts).  This effect was already seen in the non-turbulent models, but becomes now more pronounced.

\begin{figure}[tbp]
  \centering
  \includegraphics[width=\figw,clip]{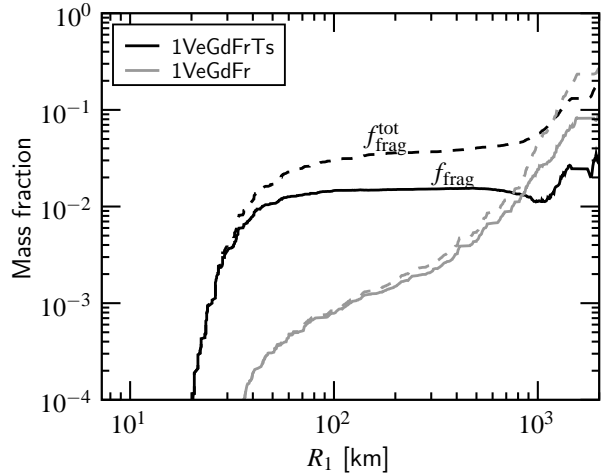}
  \caption{\label{fig:fragstat}The evolution of the quantities $f_\mathrm{frag}$ (the fraction of the total mass that resides in fragments at that time; \textit{solid} curve) and $f_\mathrm{tot}$ (the total mass fraction that has once been in fragments; \textit{dashed} curves).  The difference between $f_\mathrm{tot}$ and $f_\mathrm{frag}$ is due to accretion of fragments.}
\end{figure}
To further illustrate this behavior, \fg{fragstat} plots the mass fraction in solids as function of time, or rather, the evolutionary parameter, $R_1(t)$.  Here, $f_\mathrm{frag}$ is the mass fraction that resides in fragments at $R_1(t)$ and $f_\mathrm{tot}$ the cumulative mass fraction, \ie\ the total amount of mass that has once been in fragments.  The difference between these curves then provides a measure for the amount of fragments that are accreted.  From \fg{fragstat} it can be seen that in the turbulent stirring run (\simu{1VeGdFrTs}) collisions quickly put $\sim$1\% of the mass into fragments \textit{and} that these fragments are accreted efficiently since $f_\mathrm{tot}$ keeps increasing while $f_\mathrm{frag}$ levels out.  On the other hand, in the \simu{1VeGdFr} run the importance of fragmentation (and subsequent accretion) only gradually increases.  It is rather unimportant during the runaway-growth phase, but gathers pace towards the end of the simulation, even overtaking the turbulent stirring curves.  Therefore, the tabulated fractions towards the end of the simulation run (\ie\ Cols.\ 4, 5 of \Tb{results}) do not reflect the importance of the fragmentation during the earlier runaway growth phase. 

\subsubsection{Delayed onset of runaway growth}
\label{sec:D-rg}
The Ts-runs conducted at 6 and 35 AU (\fg{stats2}) show very interesting behavior.  Relative velocities are initially so large that GF is unimportant.  However, these collisions are just not yet violent enough to be fully destructive: there is net accretion and the mass spectrum is characterized by an exponential tail at the high mass end.  Accretion timescales become very long, though, when GF is absent.  A self-similar size distribution emerges in which $M_1/m_\ast \approx 10$.  However, at the point where $v_M \sim v_{\mathrm{esc},M}$ collisions enter the d.d.-regime.  This occurs first for interactions among the biggest bodies but later also between big and small bodies. The transition to RG is initiated and growth accelerates. 

For the 6 AU run the behavior can be regarded as a `delayed' onset of runaway growth, \ie\ RG takes off at $R_1\sim100$ km but its general characteristics are not much different than the non-Ts run.  However, the behavior of the \simu{35VeGdFrTs} model is more interesting.  Here, RG also takes off at $R_1\sim100$ km but then displays very extreme properties.  Growth very rapidly produces a $\sim$$2\times10^3$ km oligarch (see \fg{stats2}c).  The clue to the understanding of this behavior lies in the sizable number of fragments that are produced and in the different way these are treated in the 6 and 35 AU runs.  In the 6 (and 1) AU runs the fragments are assumed to move with the gas at a fixed approach velocity $v_a = \eta v_k$.  However, at 35 AU such a restriction was not applied, \ie\ fragments decouple from the gas and cool themselves efficiently through inelastic collisions.  Thus, we have the peculiar situation that the random velocity of the big bodies ($v_M$) is larger than that of the smaller bodies (here: fragments), $v_f$.

At the point where the massive particle fulfills the condition $v_M<v_{\mathrm{esc},M}$ dynamical friction with the low-mass bodies (and fragments) starts to further decrease its random motions $v_M$.  Due to the enhanced GF, these bodies quickly accrete the fragments.  The situation is exacerbated because the relative velocity between the fragments and the runaway body is set by $v_M$ (\ie\ the random velocity of the runaway body), and not $v_f$ (the random velocity of the fragments), since $v_M>v_f$.  Since $v_M$ is decreasing, the growth displays more extreme characteristics with RG-index $\kappa>4/3$ (see \se{rg-def}).  

Note again that for the 35 AU Ts run we have treated an academic case since turbulent stirring by density fluctuations in the gas disk cannot operate on timescales longer than the disk dissipation timescale.  Another word of caution concerns the (artificial) sharpness of the transition between the superescape and dispersion-dominated regime; in our model dynamical friction is suddenly `switched on' at $v=v_\mathrm{esc}$, whereas in reality the transition occurs smoothly.  Notwithstanding these concerns, this simulation can be regarded as representative for runs that are characterized by initially superescape velocities, but which are nonetheless accretionary.  Accretion timescales are generally long; however, at the point where $v_M \lesssim v_{\mathrm{esc},M}$ RG sets in and a very rapid evolution follows, due to sweep-up of a population of dynamically cold fragments that has been produced during the preceding superescape phase.

\subsection{35 AU gas-free models}
\label{sec:35VeFr}
\begin{figure*}[tbp]
  \includegraphics[width=0.48\textwidth,clip]{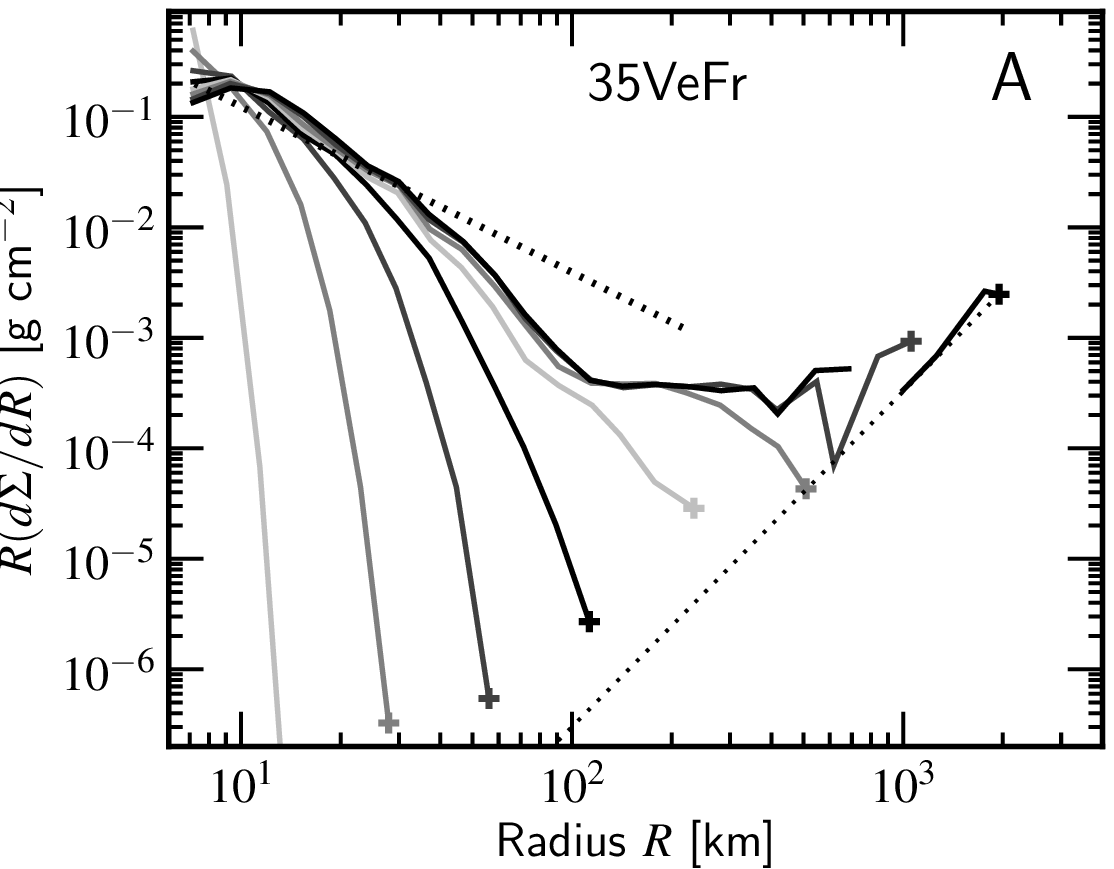}
  \includegraphics[width=0.48\textwidth,clip]{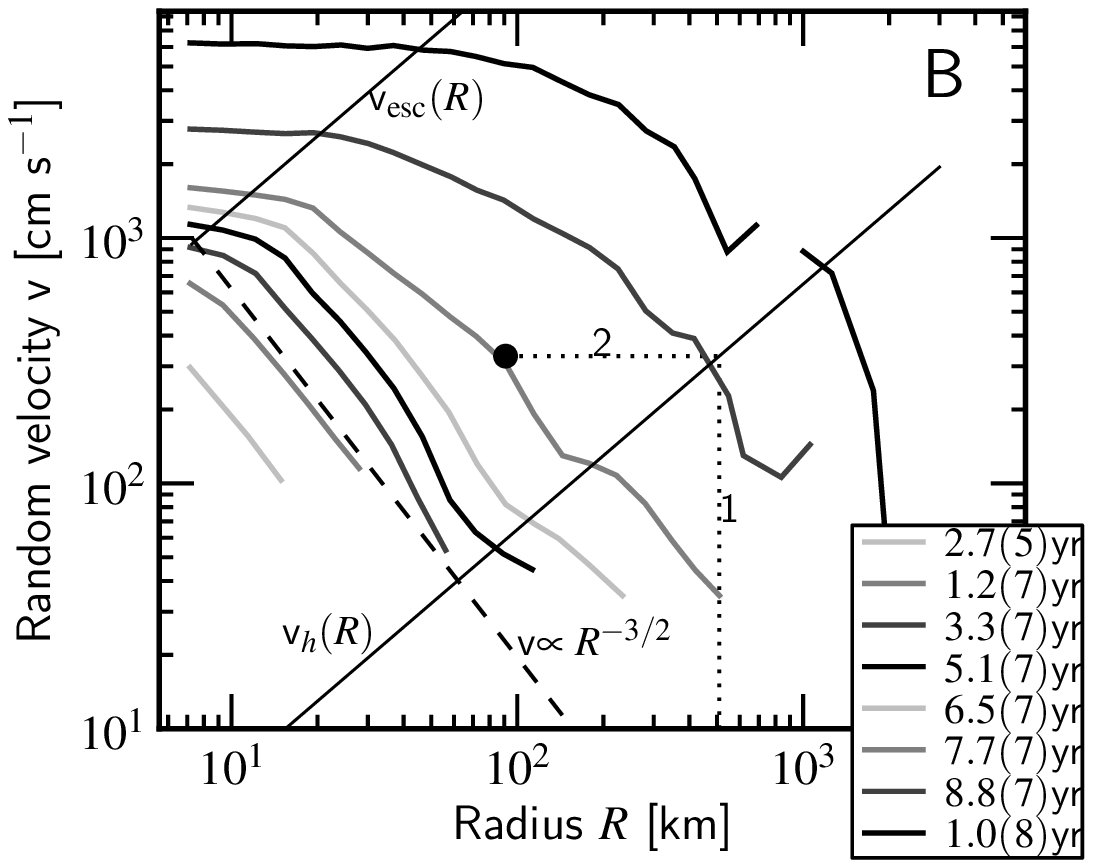}
  \caption{\label{fig:standard1} (A) The surface density distribution at eight times during the runaway growth and oligarchy stages of the \simu{35VeFr} run.  The fragment distribution is not shown here. (B) The velocity distribution for the same times as in (A). The dashed line indicates the slope for dynamical equilibrium, while the thin solid lines gives $v_h(R) = R\Omega/\alpha$ and $v_\mathrm{esc}(R)$. The times are indicated in the legend.  For the $t=7.7\times10^7\ \mathrm{yr}$ curve two auxiliary lines are plotted, illustrating (1) the position of $R_1(t)$ and (2) $v_h(R_1)$ at this point.  The most massive particle accretes bodies to the right of the dot in the shear-dominated regime.}
\end{figure*}
We analyze the \simu{35VeFr} run, which includes fragmentation but no turbulent stirring, in some more detail.  \Fg{standard1}a provides the mass spectrum at several times.   During the runaway phase the high mass tail of the column density distribution is characterized by a power-law slope with index $p<-2$.  This slope flattens during the RG phase and, like in \fg{stat1Mz}, approaches a value $p\approx-2.5$.  However, the evolution to $p=-2.5$ does not fully complete at the higher masses; compared to \fg{stat1Mz} it seems to break at an earlier stage.  We attribute this difference to the presence of fragments that start to dominate the accretion behavior (see below).

In \fg{standard1}b the velocity distribution is plotted. It can be seen that initially the distribution is in dynamical equilibrium, for which $v \propto m^{-1/2} \propto R^{-3/2}$ (dashed line).  However, towards the end of the simulation the low-$m$ bodies no longer obey this relation; the velocity distribution flattens out at low $R$.  The reason for this behavior is that these bodies are stirred faster by the bigger bodies than they can equilibrate by dynamical friction with other small bodies.  The random velocities among the small bodies lie close to the escape speed, which implies that GF and subsequently the interaction rates are weak.  Dynamical friction among these bodies is therefore suppressed.  On the other hand, the stirring by the runaway bodies/oligarchs does not discriminate between the mass of the small bodies: the big bodies regard all bodies at lower $m$ as (massless) test particles.

The lower, thin solid line gives the Hill velocity, $v_h(R)$. Bodies of size $R$ that lie below this curve can accrete other, less massive bodies in the s.d.-regime. These are mostly particles of similar mass, but for the most massive body the s.d.-regime applies for almost one order of magnitude in size.  Similarly, fragments (not shown in \fg{standard1}b) are mostly accreted in the s.d.-regime, because collisions among fragments keep their random velocity low. 

\begin{figure*}[tbp]
  \includegraphics[width=\textwidth,clip]{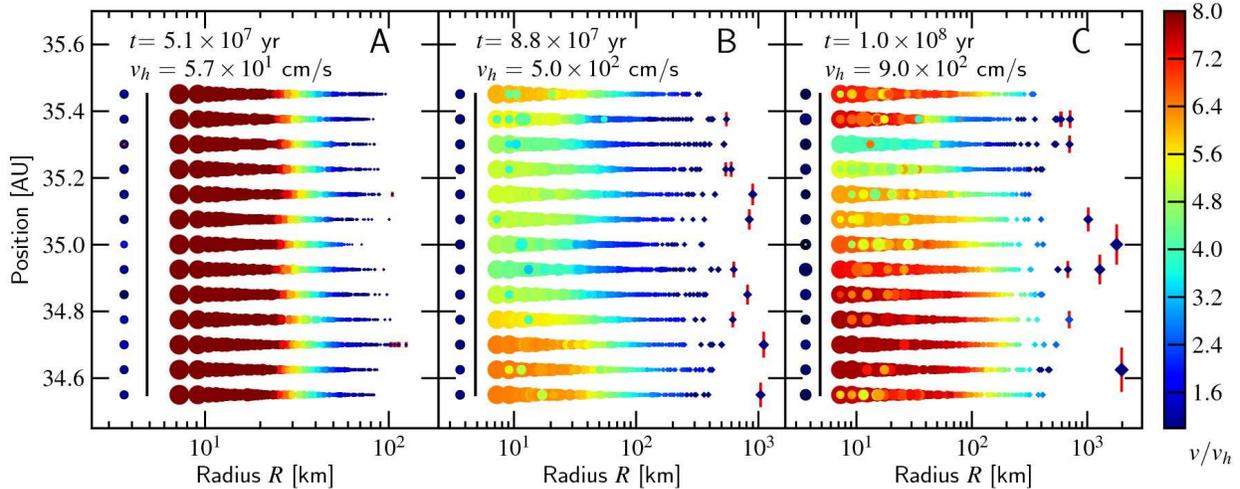}
  \caption{\label{fig:standard}Scatter plot of bodies' mass and position at three times during the 35VeFr simulation. See the caption of \fg{scatter1au} for the description of the symbol- and color-coding.} 
\end{figure*}
\Fg{standard} presents the spatial distribution of all particle groups the simulation contains at three different times.  See the caption of \fg{scatter1au} for the coding of the symbols and colors.  The decreasing $v/v_h$ during the runaway growth stage can clearly be seen from the bluer colors in \fg{standard}b compared to \fg{standard}a.  However, the random velocity at a given $R$ always increases with time (\fg{standard1}b) and collisions among low-mass bodies result in fragmentation since $v$ exceeds the escape velocity of these bodies.  Runaway growth is clearly fast ($v_h$ grows faster than $v$) but it is also clear that strong particle separation -- a key signature of RG -- does not take place due to the fact that there are so many competing bodies.  In \fg{standard}b, and especially in \fg{standard}c, this separation is more obvious, however. But by now the system is in oligarchy: a few bodies have separated from the main distribution and these are dynamically heating the remainder. The colors turn red again.  

\subsubsection{Which bodies contribute to the growth?}
\label{sec:Mcontrib}
\begin{figure}[tbp]
  \centering
  \includegraphics[width=\figw,clip]{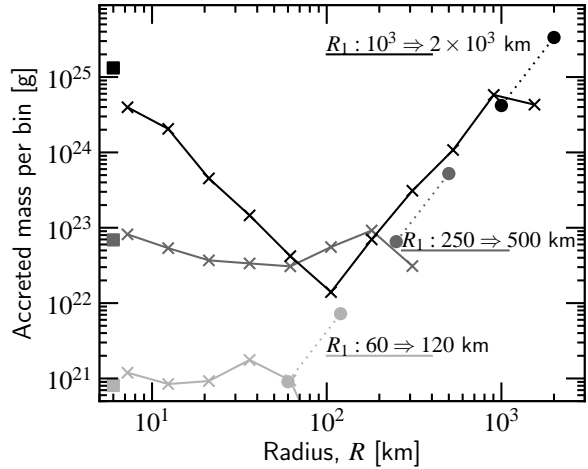}
  \caption{\label{fig:Mcontrib} The contribution by mass of bodies accreted by the maximum particle.  The \textit{dotted lines} indicate the radius-mass relation for a factor of 10 increase in the mass of the most massive particle.  The corresponding starting and end points of $R_1(t)$ are indicated by \textit{circles}. For each case, the distribution of collision partners during the growth of $R_1(t)$ are shown (\textit{crosses}).  The contribution of fragments is shown by \textit{squares}.  The data are averaged over 5 simulations.}
\end{figure}
It is instructive to see what kind of collisions dominate the growth of the most massive particles during the RG and oligarchy stages. This information is presented in \fg{Mcontrib} at three intervals, during which the size of the most massive particle, $R_1$, doubles.  For example, the lower (light grey) symbols give the contribution by mass of the particles that accrete with the largest body over its growth from $R_1=60$ km to $R_1=120$ km.  The detached square gives the contribution from the mm-size fragments.   What can be seen from \fg{Mcontrib} is that during the RG stages (\textit{grey curves}) bodies at each $R<R_1$ contribute approximately equally to the growth of the largest body.  The assumption that only the bodies that dominate the mass (\ie\ those at $R\sim R_0$) contribute, would be wrong.  Instead, collisions that take place in the shear-dominated regime, \ie\ those in the tail of the distribution, contribute a sizable fraction to the growth of $R_1$, despite their low abundance.  

In the final stages (\textit{black curves}) this trend reverses.  Here, bodies at $\sim$$R_0$ and $\sim$$R_1$ contribute most with the contribution from intermediate-size bodies being suppressed.  This signifies that the transition to the (classical) two component oligarchy state has taken place.  In fact, the contribution from fragments (indicated by \textit{squares} in \fg{Mcontrib}) starts to dominate.  But the key insight is that during the RG-phase the two-component approximation is invalid: all mass ranges contribute to the growth. As explained in \se{conv} and \se{emerge-oli} scattering among the high-mass bodies should not seriously affect the validity of these conclusions as long as the density of oligarchs is high enough.

\citet{MakinoEtal1998} recognized that the large focusing factors among particles in the high mass tail of the size distribution compensates for their low numbers.  Under the assumptions that the relevant quantities, \ie\ the velocity spectrum and the collision radii, can be given as power-laws of their masses, \citet{MakinoEtal1998} solved for the steady-state value of the mass-distribution and showed that it was consistent with a power-law index of $p=-8/3$.  This is consistent with what we find, although we would like to emphasize the dynamic nature of the process.  During RG, a power-law at the high mass tail of the distribution emerges until a point is reached at which it breaks \citep[\cf][]{Wetherill1990}.

In the \citet{MakinoEtal1998} model collisions among the largest bodies contribute, most to the growth of the most massive body. This is in contrast to our results, where we find that all bodies contribute roughly equally.  The reason for this discrepancy can be found in the power-law assumptions that \citet{MakinoEtal1998} employ, \ie\ that thermal equilibrium holds at all sizes and, more critically, the neglect of the shear-dominated regime.  Under these (idealized) conditions, GF-factors become infinite, which in reality is not possible.  

\subsection{Miscellaneous models}
\begin{figure}[tbp]
 \centering
 \includegraphics[width=\figw,clip]{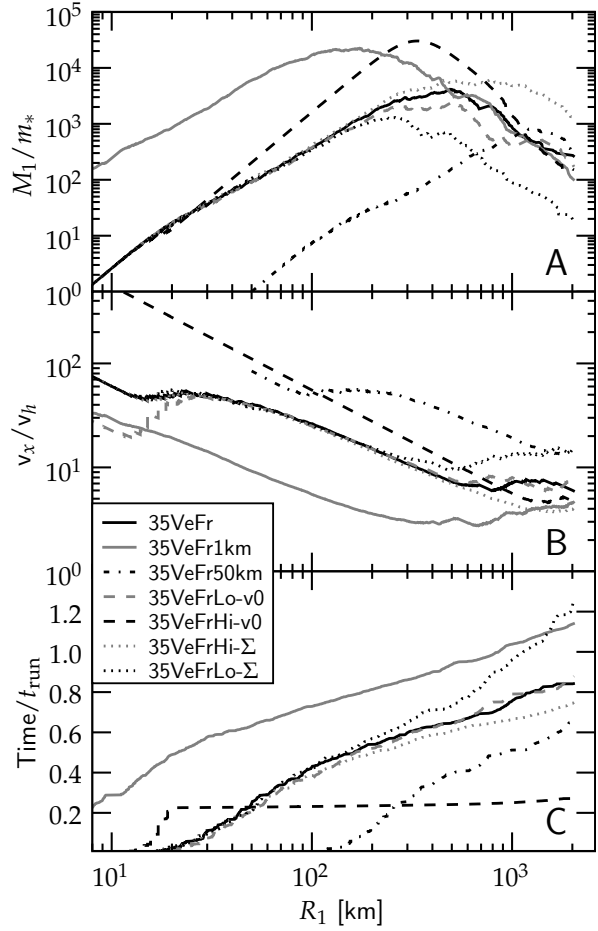}
 \caption{\label{fig:stats3}Outputs of various simulations without gas drag but with fragmentation, performed at 35 AU.}
\end{figure}
To test the robustness of our simulation results against the initial conditions (random velocity, planetesimal size, surface density) we performed additional runs at 35 AU in which these quantities ($\Sigma$, $R_0$ and $v_0$) are varied.  The results are presented in \fg{stats3}.  The $y$-axis in \fg{stats3} is again normalized to $t_\mathrm{run}$ (\eq{trun}) which is a fixed but different number for each run.

Perhaps the most striking feature is that the normalized timescales are all of order unity for all models, indicating that $t_\mathrm{run}$ (or $T_\mathrm{rg}$) is the appropriate timescale to characterize the runaway growth phase.  Furthermore, the shapes of the curves are generally similar: all runs show a peak in $M_1/m_\ast$ (\fg{stats3}a) and a minimum of $v_x/v_h$ (\fg{stats3}b).  The associated steepening of the timescale curves is not always so clear in \fg{stats3}c: growth becomes fragment-dominated in many simulations and remains fast.

The radius where the minimum of $v_x/v_h$ takes place varies.  Compared to the \simu{35VeFr} curve it seems to occur at a lower size ($\sim$500 km) for the low $R_0$ and low $\Sigma$ runs but at a larger size (really towards the end of the simulation) for the high $R_0$ and high $\Sigma$ runs.  For example, comparing the low and high $\Sigma$ runs in \fg{stats3}c (dotted curves) one sees that they diverge at around $R_1 \sim300$ km.  The high-$\Sigma$ run stays longer in the RG phase.

Regarding the variation in initial velocity $v_0$, note that the low-$v_0$ run (dashed grey line) quickly catches up with the standard run (the black solid line in \fg{stats3}b).  The reason is that planetesimal-planetesimal stirring is very effective for $v<v_\mathrm{esc}$.  This justifies our choice to determine the width of the zones from the escape velocity of the bodies (\eq{ares}, \se{multi-zone}).  

The high velocity run (black dashed curve), on the other hand, shows very different behavior.  Initially it is (expectedly) much slower than any other model but then at $R_1\approx 20$ km there is a sudden transition to a much faster growth mode.  The behavior of this run falls in the same category as the 35 AU turbulent-stirring one, explained in \se{D-rg}. The superescape velocities produce copious amounts of small fragments and once the big body is able to use its GF to sweep them up, (extreme) runaway growth follows.

\section{Discussion}
\label{sec:discussion}
\subsection{Transition from the runaway growth to the oligarchy phase}
\label{sec:Rtr}

In \se{rg-oli} we have identified the minimum of $v_x/v_h$, the point where gravitational focusing (GF) peaks, as the transition between the runaway growth (RG) and oligarchy accretion phases.  The corresponding transition size, $R_\mathrm{tr}$, is also found to be close to the point where $M_1/m_\ast$ peaks.  We have tabulated $R_\mathrm{tr}$ for a variety of simulations, corresponding to varying physical conditions at several disk radii (\Tb{results}).  

We now compare $R_\mathrm{tr}$ with the prediction by \citet{IdaMakino1993}, \eq{rg-oli-trans}.  Using $\Sigma_M = M/(2\pi a \Delta a_\mathrm{st})$, $\Sigma_m \approx \Sigma$ the density in solids, with $\Delta a_\mathrm{st} =A R_h$ the width of the heating region where $A\approx5$ reflects the spacing among oligarchs, one obtains for the RG/oligarchy transition of \eq{rg-oli-trans} (\cf\ \citealt{ThommesEtal2003})
\begin{align}
  R_\mathrm{rg/oli} = \left[ \frac{3Aa\Sigma R_0^3}{4\rho_s \alpha} \right]^\frac{1}{5} = 94\ \mathrm{km} \left( \frac{A}{5} \right)^\frac{1}{5} \left( \frac{\rho}{\mathrm{g\ cm^{-3}}} \right)^{-\frac{2}{15}}
  \left( \frac{\Sigma}{10\ \mathrm{g\ cm^{-2}}} \right)^\frac{1}{5} \left( \frac{a}{\mathrm{AU}} \right)^\frac{2}{5} \left( \frac{R_0}{10\ \mathrm{km}} \right)^\frac{3}{5},
  \label{eq:Roli}
\end{align}

Comparing the theoretical prediction \eq{Roli} with the `measured' transition points from our `indicators' ($R_\ast$ and $R_\mathrm{tr}$, see \Tb{results}) we see a clear discrepancy; typically $R_\mathrm{tr}$ is larger than \eq{Roli} by several factors in radius but the discrepancy increases for the outer disk models.  For example, whereas according to \eq{Roli} RG will stall at the $\sim$100--200 km size, we see from \Tb{results} that $R_\mathrm{tr}$ is rather $\sim$300 km for the 1 AU models, $\sim$600 km for the 6 AU models, and $\sim$$10^3$ km for 35 AU models.  Clearly, the domain of pure RG extends a little further than \eq{Roli} predicts, especially in the outer disk.

The underlying reason for this discrepancy is the fact that RG is irreconcilable with the two component picture \eq{Roli} relies on.  For example, we have seen in \se{Mcontrib} that the biggest body accretes from all mass bins, not just from the ones that dominate the mass of the distribution.  The same holds for the stirring of small bodies.  For a $p\approx-2.5$ mass spectrum the biggest bodies dominate the stirring, but it is not one big body that dominates.  In other words, \eq{Roli}, which presumes that the two component approximation is valid, cannot be applied to the RG stage.

However, for oligarchy a two component approximation becomes valid.  Indeed, it may be defined as such. Therefore, the start of oligarchy can be defined at the point where the relevant timescales in the two component approximation match the RG timescale $T_\mathrm{rg}$.  Initially, $T_\mathrm{rg}$ is the dominant (shortest) timescale and the two component picture is invalid.  However, due to the increasing GF-factors there will be a `tipping point' after which a two-component picture does become applicable.  We have recently addressed this issue quantitatively \citep{OrmelEtal2010} and found an analytic estimate for the transition size, $R_\mathrm{tr}$:
\begin{equation}
  R_\mathrm{tr} \approx 320\ \mathrm{km} \left( \frac{K_\mathrm{rg}}{0.1} \right)^{3/7} \left( \frac{\rho_s}{1\ \mathrm{g\ cm^{-3}}} \right)^{-1/7} \left( \frac{R_0}{10\ \mathrm{km}} \right)^{3/7}
  \left( \frac{a}{\mathrm{AU}} \right)^{5/7} \left( \frac{\Sigma}{10\ \mathrm{g\ cm^{-2}}} \right)^{2/7}.
  \label{eq:R-rg}
\end{equation}
Using the values for $\Sigma$ and $R_0$ listed in \Tb{list}, \eq{R-rg} gives transition radii of $R_\mathrm{tr}$ $\sim$300 km, 600 km, and $\sim$10$^3$ km for the runs at 1, 6, and 35 AU, respectively.  This corresponds reasonably well with the obtained transition radii in \Tb{results}.  Note the rather weak dependence of $R_\mathrm{tr}$ on $K_\mathrm{rg}$.

\begin{figure}[t]
  \centering
  \includegraphics[width=\figw,clip]{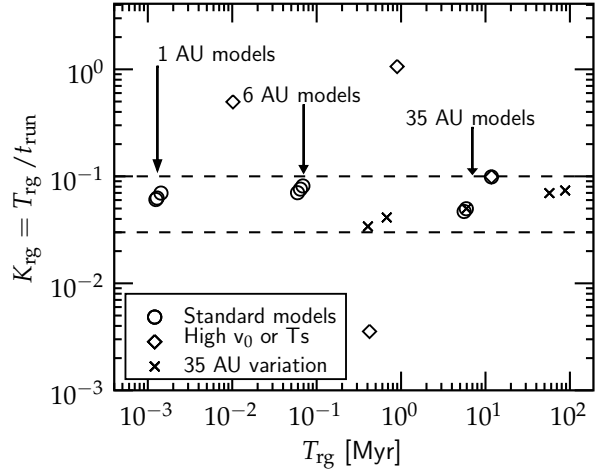}
  \caption{\label{fig:tscale}The runaway growth timescale $T_\mathrm{rg}$ normalized to $t_\mathrm{run} = R_0 \rho_s/\Sigma \Omega$ for various runs listed in \Tb{list}.  The standard runs including gas drag and fragmentation (first 9 rows of \Tb{list}) are indicated by \textit{circles}.  Runs characterized by large random velocities are indicated by \textit{diamonds} and other 35 AU runs are indicated by \textit{crosses}.  The majority of the runs are well represented by $K_\mathrm{rg} \approx 0.03$--$0.1$.  But the high velocity models deviate significantly from this trend. }
\end{figure}
In \eq{R-rg} $K_\mathrm{rg}$ is related to $T_\mathrm{rg}$ as in $K_\mathrm{rg} = T_\mathrm{rg}/t_\mathrm{run}$ where $t_\mathrm{run}$ has been defined in \eq{trun}.  In a dimensionless form, $K_\mathrm{rg}$ characterizes the outcome of our RG simulations.  In \fg{tscale} we plot its value for many of the simulation runs listed in \Tb{results}.  Overall, we find that $K_\mathrm{rg}$ lies in the range of 0.03--0.1 for simulations that include the relevant physics.  Since $K_\mathrm{rg}$ is reasonably well constrained, \eq{R-rg} is a robust prediction.  However, simulations that are characterized with very large (initial) random velocities or include turbulent stirring do not obey this trend.  But these simulations are (initially) not in RG.

\subsection{The importance of scattering and migration of bodies (neglected effects)}
Scattering and migration have been ignored in this study.  In \ses{conv}{emerge-oli} we have already discussed the effect of gravitational scattering of bodies.  In our simulations we treat only a small patch of the disk and we do not expect scattering to significantly change the outcome.  However, the assumption that a local representation of the disk is applicable itself may be questionable.  Scattering of bodies or long-range interactions from the inner disk (where evolution timescales are shorter) may affect the evolution of bodies in the outer disk.  In future studies, we may include the change in semi-major axis due to the scattering of bodies in our model, following, for example, the prescriptions outlined in \citet{TanakaIda1996,TanakaIda1997}.

 Similarly, we have neglected migration of big bodies and subsequent shepherding of small bodies.  We can, however, assess the timescale due to type-I migration \citep{TanakaEtal2002}:
\begin{equation}
  T_\mathrm{migr} \approx 60\ \mathrm{Myr}\ \left( \frac{\Sigma_g}{10^3\ \mathrm{g\ cm^{-2}}} \right) \left( \frac{R_\mathrm{pp}}{10^3\ \mathrm{km}} \right)^{-3} \left( \frac{c_g}{10^5\ \mathrm{cm\ s^{-1}}} \right)^2 \left( \frac{a}{\mathrm{AU}} \right)^{3/2},
  \label{eq:Tmigr}
\end{equation}
for a solar mass star, and a protoplanet size $R_\mathrm{pp}$ of internal density $\rho_s = 3\ \mathrm{g\ cm^{-3}}$.  Since $R_\mathrm{pp}=2\times10^3$ km is the maximum size we reach in our runs, the effects of type-I migration are rather minor as $T_\mathrm{migr}$ is much larger than any relevant timescale we consider.  However, for the subsequent core-accretion phase, where $R_\mathrm{pp}$ grows to Earth-size proportions or larger, type-I migration becomes important \citep{TanakaEtal2002}.

Gas drag-induced migration is also not included in our manuscript.  The timescale for migration of km-size, planetesimal bodies, may still be reassuringly long not to affect any of our key conclusions, but for small fragments it is another story. These could be removed from the region where they are produced in as little as a few hundred years \citep{Weidenschilling1977}.  We will now take a more closer look to the behavior of the fragments.

\subsection{The importance of fragmentation}
\label{sec:rel-frag}
In simulations where a lot of fragments are produced, like in the runs that include turbulent stirring or a large initial velocity dispersion, the shear-dominated (s.d.) regime can become important.  In \app{Hill} we obtained $dM/dt$ and, using \eq{tac}, we find a timescale \citep[\cf][]{GoldreichEtal2004,Rafikov2004,Weidenschilling2005,Chambers2006} 
\begin{equation}
  T_\mathrm{ac}^\mathrm{2D-s.d.} \approx \frac{\alpha^{3/2} \rho_s R_\mathrm{pp}}{3 \Sigma} \Omega^{-1} \approx 3\times10^4\ \mathrm{yr}\ \left( \frac{\rho}{1\ \mathrm{g\ cm^{-4}}} \right)^{1/2} \left( \frac{\Sigma}{1\ \mathrm{g\ cm^{-2}}} \right)^{-1} \left( \frac{R_\mathrm{pp}}{10^3\ \mathrm{km}} \right),
  \label{eq:Tac-sd}
\end{equation}
\Eq{Tac-sd} assumes that the fragments constitute a cold, very thin disk ($v \lesssim v_h$ and $v_z  \lesssim  \alpha^{1/2}$).  The striking feature of \eq{Tac-sd} is that, except for $\Sigma$, it does not depend on disk radius $a$.   In the outer disk, s.d.-growth by sweepup of fragments is the only fast growth mode available.  Despite the scaling with $R_\mathrm{pp}$, which shows that growth is not in the runaway mode, it is here much faster than its runaway counterpart, \ie\ $T_\mathrm{ac}^\mathrm{2D-s.d.} \ll T_\mathrm{rg}$. 

However, there are many caveats regarding our treatment of fragments.  First, there is the uncertainty in the collision model, reflected in the parameters $a_\mathrm{fr}$ and $\epsilon$ (see \se{outer}): changing these parameters may increase or decrease the significance of fragmentation (and its re-accretion).  Another concern is the presence of gas drag.  \Eq{Tac-sd} assumes that the fragments, like the protoplanet, move at the Keplerian orbital velocity.  We have already mentioned the fast radial orbital decay these particles experience.  Furthermore, if the particles are small enough they move with the gas at a relative motion of $\eta v_k \sim 30\ \mathrm{m\ s^{-1}}$ with respect to the planetesimal bodies.  Interactions between massive bodies and fragments then probably take place in the high-velocity regime (`dispersion-dominated' is perhaps unsuited here since the velocity difference is systematic) and \eq{Tac-sd} is inapplicable.  In our simulations, we have tried to pre-empt this effect by fixing the random velocities of the 1 and 6 AU runs at $v = \eta v_k$.  We are currently investigating in more detail the effects of gas drag on the gravitational cross sections for small particles (Ormel \& Klahr, submitted).

It has previously been reported that the settling/damping process of gas drag allows for a very efficient growth of fragments in the s.d.-regime via \eq{Tac-sd} \citep{KenyonBromley2009}.  However, these fragments should occupy a very special niche: they cannot be too small as they would otherwise couple too strongly to the gas to allow efficient accretion, but cannot be too big either since then the assumption of a thin disk and efficient cooling becomes problematic.  Thus, it is unclear whether \eq{Tac-sd} can materialize in gas-rich environments.  On the other hand, it is likely that the collisional cross section $R_\mathrm{col}$ is enhanced above the 2-body limit due to the dissipative nature of the encounter \citep{InabaIkoma2003,MutoInutsuka2009,TanigawaOhtsuki2010}.  This is an active area of research. 

\section{Summary and Conclusions}
\label{sec:summary}
We have developed a new statistical code to study the evolution of the planetesimal size distribution.  The key novel element is that the code treats interactions between particles, rather than mass bins.  We have tested the code against existing literature $N$-body and statistical studies.  Starting from an initially monodisperse distribution of $\sim$km-size planetesimals, we have performed a large parameter study with the aim to identify the transition between the initial runaway growth phase and the later oligarchy phase and to assess the sensitivity of key physical processes like gas drag, fragmentation, and turbulent stirring on the planetesimal accretion process.  Our main conclusions are the following:

\begin{enumerate}
  \item Interactions that take place in the dispersion dominated regime, where random velocities $v$ lie between the Hill velocity $v_h$ and the escape velocity $v_\mathrm{esc}$ of the largest body, fulfill the conditions for runaway growth, \eq{rg-def}.  However, for spatially resolved systems, one must distinguish between the local and the global variant of the runaway growth definition.  In oligarchy, the system is only locally in runaway growth, \ie\ \eq{rg-def} no longer holds between bodies that are spatially isolated.
  \item In simulations that start out at the very typical condition where the initial random velocity $v_0$ of the bodies is less then their escape velocity, $v_\mathrm{esc,0}$, runaway growth ensues.  We find that during the runaway growth phase all masses contribute -- approximately equally -- to the growth of the largest body, which occurs exponentially at a characteristic timescale $T_\mathrm{rg}$.  
  \item The runaway growth timescale $T_\mathrm{rg}$ is empirically determined and expressed in terms of the initial parameters as $T_\mathrm{rg} = K_\mathrm{rg} \rho_s R_0/\Sigma \Omega$.  We find a value of $K_\mathrm{rg}$ of 0.03--0.1 is typical.  Simulations that are characterized by high initial velocities ($v_0 \gg v_\mathrm{esc,0}$) do not obey this trend.
  \item We find that at a transition size $R_\mathrm{tr}$ the runaway (exponential) growth is over and becomes much slower (a considerable distribution of fragments may mitigate this effect, though).  At this transition point the gravitational focusing factors peak.
  \item During the runaway growth phase the mass distribution at the high mass end gradually changes into a power-law $N_s(m)\propto m^p$ with $p$ approaching $\approx$$-2.5$ near the end of the RG phase.  During the oligarchic phase the distribution breaks and becomes characterized by two components: the oligarchs (at high $m$) and leftover planetesimals.
  \item Due to the gravitational stirring, interactions among low-mass bodies reach the fragmentation regime $v>v_\mathrm{esc}$.  In this study we have treated such collisions as erosive, producing copious amounts of $\sim$mm-size fragments.  The ability of fragmentation to influence the growth of the biggest bodies (through re-accretion of fragments) depends somewhat on the adopted collision parameters.  Fragments do not dominate the evolution during the runaway growth phase, except in models that include (significant amounts of) external stirring.  However, during the oligarchic phase collisions become violent enough (and timescales long enough) for fragments to become important. 
  \item Simulations that start out at $v>v_\mathrm{esc}$ but where velocities are still sufficiently low to be accretionary are characterized by long accretion timescales and copious amounts of fragment production.  At the point where the escape velocity of the largest bodies starts to exceed $v$ the growth mode turns to runaway.  It becomes especially fast if the fragments are kept dynamically cold due to mutual (elastic) collisions.
  \item Sweepup of such a dynamically cold population of fragments takes place in the shear-dominated regime, which can lead to very rapid growth rates (see \se{rel-frag}) .  However, we have questioned the viability of this mechanism in gas-rich systems since particles are tied to the gas and suffer radial orbital decay.
\end{enumerate}

\begin{table}[t]
  \caption{\label{tab:scenarios} Summary of runaway growth scenarios}
\begin{tabular}{llp{12cm}}
\hline \hline
    & Case & Description \\
    \hline
    \multicolumn{3}{l}{(A) Classical regime} \\
    \cline{1-2}
    & Prerequisites:        & Velocity dispersion below escape velocity of biggest body, ($v<v_\mathrm{esc}$)\\
    & Key characteristics:  & Column density evolves into a power-law distribution, $N_s(m)\propto m^{-p}$ with $p\approx-2.5$\\
    &                       & Collisions between all masses contribute to the growth of the biggest body\\
    &                       & Runaway growth fast with timescale given by $T_\mathrm{rg} = K_\mathrm{rg} t_\mathrm{run}$ (see \eq{trun}) with $K_\mathrm{rg}\sim0.03$--$0.1$ \\
    & Outcome:              & Transition to oligarchy at radius given by \eq{R-rg} \\
    \hline
    \multicolumn{3}{l}{(B) Fragmentation-dominated } \\
    \cline{1-2}
    & Prerequisites:        & Production of a sizable amount of \textit{dynamically cold} fragments with $v_f<v_h$ \\
    & Key characteristics:  & Very fast growth possible (shear-dominated regime) at timescale given by \eq{Tac-sd}\\
    & Outcome:              & (most likely) 2 component oligarchy of protoplanets and fragments \\
    \hline
    \multicolumn{3}{l}{(C) Superescape regime } \\
    \cline{1-2}
    & Prerequisites:        & Velocity dispersion above escape velocity of the biggest body ($v>v_\mathrm{esc}$), but net accretion \\
    & Key characteristics:  & Continuous size distribution, declining exponentially at high-$m$; no gravitational focusing, slow growth, fragmentation\\
    & Outcome:              & Transition to runaway growth (scenario A) at point where $v<v_\mathrm{esc}$; possibly strong fragment-dominated growth (scenario B) \\
    \hline
\end{tabular}
\end{table}
From these general findings, we construct three scenarios through which planetesimal growth could have proceeded, see \Tb{scenarios}.  In the first scenario, the classical regime, the system starts out in the dispersion-dominated regime, $v_h < v <v_\mathrm{esc}$, and runaway growth ensues according to points 2--5 listed above.  At $R_1(t)=R_\mathrm{tr}$ (\eq{R-rg}) the runaway growth phase is over and is superseded by oligarchic growth.  This is the point where a full 2-component approximation (oligarchs and smaller bodies) becomes first applicable.  Most studies that deal with planetesimal accretion have focused on this scenario.

Planetary accretion may have deviated from the above-sketched contours of the `classical regime'. For example, if a populous reservoir of dynamically cold particles (fragments) is present, the accretion takes place in the shear-dominated regime ($v_x<v_h$, scenario B).  In the shear-dominated regime gravitational focusing factors are not determined by the random velocity of the particles, \ie\ the self-regulated effect which is the hallmark of oligarchic growth is absent.  We obtain such a situation if the initial conditions features an already mature body embedded in a `sea' of small particles \citep{Weidenschilling2005} or when collisions among the (rubble-pile) planetesimals lead to fragmentation (\ie\ $v>v_\mathrm{esc}$).    Provided the fragments are able to cool themselves (through mutual highly dissipative inelastic collisions) they could quickly dominate the contribution to the mass gain of the biggest body.  The shear-dominated regime marks a very fast growth mode, especially for the outer solar system and for gas-free environments.  However, fragments should be kept dynamically cold and we have raised a note of caution on the viability of this mechanism in gas-rich environments.

The final scenario (C) is characterized by random velocities of the planetesimal bodies that (initially) exceed the escape velocity ($v_x>v_\mathrm{esc}$) but are nonetheless accretionary.  External stirring, \eg\ as a result of density inhomogeneities in the gas disk, may provide these conditions (provided it is not too strong to shut-off accretion altogether).  Since gravitational focusing is negligible when $v>v_\mathrm{esc}$, accretion timescales are very long.  As the system is not in runaway growth, the size distribution remains continuous with an exponentially-declining tail at the large masses, rather than a power-law.  However, at the point where the largest body fulfills $v_{\mathrm{esc},M}>v_M$, growth will enter the runaway regime (dispersion-dominated regime).  If a lot of fragments has been produced and mutual elastic collisions are able to keep the random velocities of these particles low, the runaway effect is especially pronounced since the gravitational focusing factors are determined by the random velocity of the biggest body ($v_M$).  Eventually, growth could enter the shear-dominated regime (scenario B). 

\vspace{1cm}
\noindent
\textbf{Acknowledgments}:
The comments of the referees, Eiichiro Kokubo and Stuart Weidenschilling, resulted in major improvements to the manuscript and led us to fine-tune the numerical model.  The authors highly appreciate the inquisitive nature of their comments.  C.W.O.\ further appreciates stimulating discussions with Jeff Cuzzi, Shigeru Ida, Christoph Mordasini, and Rainer Spurzem  on various parts of this manuscript and acknowledges the Alexander von Humboldt foundation for critical financial support.

\footnotesize
\bibliographystyle{icarus}
\bibliography{ads,arXiv}
\normalsize

\appendix

\section{Determination of $N_g^\ast$}
\label{app:app-rg}
In \se{MC} we introduced $N_g$, the number of physical particles that are associated to a representative body (RB).  The group number, $N_g$, is determined by a function that depends on the distribution (the distribution method; \citealt{OrmelSpaans2008}).  As the distribution changes with time the amount of grouping likewise adjust.  We denote this function by $N_g^\ast$.  Here, we sketch how it is obtained.

\begin{figure*}[t]
  \centering
  \includegraphics[width=13cm,clip]{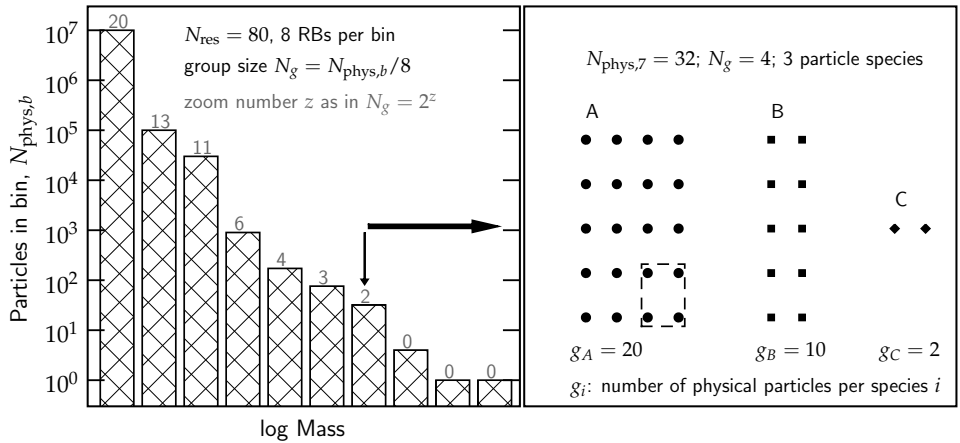}
\caption{Sketch illustrating how the zoom factors are determined.  (\textit{left}) The particle distribution of a certain zone at time $t$ is binned in terms of log mass. The group size $N_g$ and the \textit{zoom factor} $z$ of the swarms are determined from the number of particles in the bin $N_{\mathrm{phys},b}$ and the particle resolution $N_\mathrm{res}$. In this example, $N_\mathrm{res}=80$ and 8 RBs are assigned per mass bin.  For the $z=0$ bins the high-mass bodies are individually resolved. \textit{(right)} Illustration of the state in the indicated bin, for which the group size is 4 ($z=2$, \textit{dashed} rectangle). One has that $g_A+g_B+g_C=N_{\mathrm{phys},7}$ with $g_i$ the true number of particles for a species group $i$.  Particle swarms A and B are resolved, C is not. }
  \label{fig:group-sketch}
\end{figure*}
We illustrate the distribution method using \fg{group-sketch} as an example (the numbers given here are entirely arbitrary).   The distribution method assigns the RBs equally over logarithmic mass.  Therefore, we first bin the particles by mass.  This binning is just an auxiliary feature and not fundamental to the program, \ie\ it is only used here for the determination of $N_g^\ast$.  The zone that \fg{group-sketch} corresponds to is covered by in total 80 RBs ($N_\mathrm{res}=N_\mathrm{rb}/N_\mathrm{zo}=80$).  These are equally assigned over the 10 logarithmic mass bin, \ie\ 8 RBs per bin. Therefore, the size of the groups equals $N_g = N_{\mathrm{phys},b}/8$ where $N_{\mathrm{phys},b}$ is the number of physical particles in mass bin $b$.  Numbers are rounded where needed; in fact, $N_g$ is taken to be a multiple of 2, \ie\ $N_g = 2^z$ with $z$ integer (see \citealt{OrmelSpaans2008} for rationale).  Thus, the result is that the numerous low-mass bodies are highly grouped (large $N_g$), whereas the high-mass bodies are individually resolved ($N_g=1$ or $z=0$). This property allows RG to be accurately modeled.

In the right panel of \fg{group-sketch} we further illustrate the relation between the RBs and physical particles. We focus on bin \#7, which has $N_\mathrm{phys,7}=32$ physical particles that are represented by 8 RBs of $N_g=4$.  In principle, each physical body is different from another, \ie\ it is characterized by unique physical properties (mass, velocities, \etc).  However, the program only deals with encounters between RBs -- \ie\ particle groups that share the same properties -- which preserves an inherent level of `graininess' to the distribution.  We will refer to these ensembles of identical particles as `species'; \eg\ there are $g_A$ physical particles of species A that have identical properties which are different from those of species B.

Thus, we see that since the group size equals $N_g=4$, 5 RBs are assigned to the particles of species A, 2.5 to those of species B, and 0.5 to those of species C. The total amount of RBs add up to 8, while the total amount of physical particles add up to $N_{\mathrm{phys},7}=32$.  The fact that the total number of RBs assigned to the B-particles is not an integer does not pose a problem, as long as $N_g$ particles can take part in the group collision.  However, for the `C-particles' the number of physical particles falls below the group size.  These situations are strictly forbidden; under resolved groups are merged with other swarms which are closest in mass and velocity space (and also in radial position: merging occurs only for groups within the same zone) according to the criteria discussed in \se{MC}.  For example, the C-group may be merged with the B-group, provided their properties (masses, positions, eccentricities, \etc) do not greatly differ.

\section{Calibration of the model}
\label{app:calib}
The interaction timescale of a single body with a group of bodies $j$ is $t_\mathrm{int} = (n_j \sigma_\mathrm{int} v_a)^{-1}$, where $n_j$ is the number density of $j$-bodies, $\sigma_\mathrm{int}$ the interaction cross section, and $v_a$ the approach velocity.  Inverting this expression, the \textit{interaction rate} is defined as
\begin{equation}
  \lambda_j^\mathrm{(1)} = \frac{ \pi R_x R_z v_a}{2 h_\mathrm{eff}} N_{sj}\qquad  [\mathrm{s^{-1}}],
  \label{eq:1rate}
\end{equation}
where we have substituted $\sigma_\mathrm{int} \equiv \pi R_x R_z$ and $n_j = N_{sj}/2h_\mathrm{eff}$. In \eq{1rate} $N_{sj}$ is the column density of $j$-bodies, $h_\mathrm{eff}$ the effective scaleheight, and $R_x$, $R_z$, respectively, the planar and vertical interaction radii.  These latter quantities follow from $R_\mathrm{int}$ \textit{and} from the geometry of the encounter. For, example, $R_z$ can never exceed the scaleheight $h_\mathrm{eff}$, \ie\ $R_z = \textrm{min}(h_\mathrm{eff}, R_\mathrm{int})$. For each type of interaction, furthermore, the interaction radii depend on the velocity regime.  The superscript `1' in $\lambda_j^\mathrm{(1)}$ emphasizes that \eq{1rate} is the collision rate for a single particle.

Using \eq{1rate} accretion rates, $dM/dt$, and stirring rates, $dv^2/dt$, can be constructed.  For accretion we simply multiply \eq{1rate} by $m_j$, while for stirring we multiply by the change in $v^2$ that the body suffers during the encounter, see \app{stir-delv2}, \ie\
\begin{equation}
  \frac{dv^2}{dt} = \frac{ \pi R_x R_z v_a}{2 h_\mathrm{eff}} N_{sj} (\Delta v^2)_\mathrm{int},
  \label{eq:stir}
\end{equation}
where `int' refers to either dynamical friction or viscous stirring.  Using these expressions and the procedure sketched in \se{interactions}, we will verify that the geometrical model is, within factors of unity, consistent with more refined literature treatments regarding accretion and stirring.  However, concerning the numerical application of the model we do regard these offsets as important.  Retrieving these order unity factors by comparing with existing literature expressions is what we understand under `calibration' and is the purpose of this section.

A swarm of bodies of individual mass $m$ is often characterized by a distribution in random velocities.  In particular, the Rayleigh distribution is frequently adopted; \ie\ 
\begin{equation}
  P(v',v_z') \mathrm{d}v' \mathrm{d}v_z'= \frac{4v' v'_z}{v^2 v_z^2} \exp \left[ -\left( \frac{v'}{v} \right)^2 -\left( \frac{v_z'}{v_z} \right)^2 \right] \mathrm{d}v' \mathrm{d}v_z'.
  \label{eq:rayleigh}
\end{equation}
\citep{GreenzweigLissauer1992} is the probability that a body has random velocities $v', v_z'$ given that $v$ and $v_z$ are the rms-values of the distribution.   One would naturally expect that the mutual interactions within a population produces a distribution, and $N$-body experiments indeed indicate that this is the case \citep{Wetherill1980,IdaMakino1992}.  For this reason, accretion and stirring rates are often given as distribution-averaged quantities.  When calibrating our model we will follow this convention -- \ie\ calibrate against the distribution average. 

\subsection{Collision rates}
\label{sec:colrates}
\subsubsection{The superescape regime, $w>v_\mathrm{esc}$}
In this regime we assume $\beta=i/e=v_z/v=0.5$ \citep{IdaEtal1993,TanakaIda1996} and that the scaleheight of the disk $h_\mathrm{eff}$ exceeds $R_\mathrm{col}$ such that $R_z=R_x=R_\mathrm{col}=R_s$.  Furthermore, $v_a=v$ and $v_a/h_\mathrm{eff} = v/(v_z/\Omega) = \Omega/\beta = 2\Omega$ and the accretion rate \eq{1rate} becomes
\begin{equation}
  \lambda_j^\mathrm{(1)} = A_1 \pi R_s^2 \Omega N_\mathrm{sj},
  \label{eq:CR1}
\end{equation}
where we have augmented \eq{1rate} by $A_1$, the calibration factor.  We can compare \eq{CR1} with the one-body accretion rate, averaged over the distribution, obtained by \citet{GreenzweigLissauer1990,GreenzweigLissauer1992}
\begin{equation}
  \langle \lambda_j^\mathrm{(1)} \rangle = R_s^2 \frac{\cal F(\beta)}{2\pi} N_{sj},
\end{equation}
where $\cal F(\beta)$ is an integral expression of $\beta$.  A numerical evaluation gives ${\cal F}(0.5) = 16.1$, from which we find $A_1 = 0.90$.

\subsubsection{The dispersion-dominated regime, $2.5v_h<w<v_\mathrm{esc}$}
In the dispersion-dominated regime, $R_\mathrm{col} = R_s v_\mathrm{esc}/w$.  For our calibration we again assume $R_\mathrm{col}=R_x=R_z=R_s < h_\mathrm{eff}$ and $v_a/h_\mathrm{eff} = 2\Omega$.  Then,
\begin{equation}
  \frac{dN_\mathrm{acc}}{dt} = A_2 \pi R_s^2 \left( \frac{v_\mathrm{esc}}{w} \right)^2 \Omega N_\mathrm{sj},
  \label{eq:CR2}
\end{equation}
with $A_2$ the calibration factor for this regime.  The equivalent expression by \citet{GreenzweigLissauer1992}, their Eq.\ 17, reads
\begin{equation}
  \langle \frac{dN_\mathrm{acc}}{dt} \rangle = \frac{R_s^2 \Omega}{2\pi} \left[ {\cal F}(\beta) +  {\cal G}(\beta)\left( \frac{v_\mathrm{esc}}{w} \right)^2 \right].
\end{equation}
A further numerical evaluation gives ${\cal G}(0.5)= 43.0$.  Ignoring the $\cal F$ term, then gives $A_2 = 1.5$.

\subsubsection{The Hill regime, $w<v_h$}
\label{app:Hill}
\begin{figure}[t]
  \centering
  \includegraphics[width=\figw,clip]{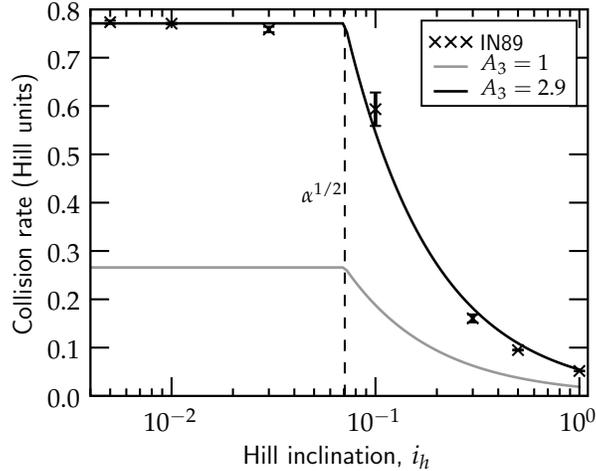}
  \caption{\label{fig:crates}The collision rate $\lambda_j^\mathrm{(1)}/N_{sj}$ expressed in Hill units as function of Hill inclination, $i_h=v_z/v_h$.  Symbols show the numerical results of \citet{IdaNakazawa1989}, where we have only included values $i_h,e_h <1$, \ie\ the low-velocity regime. For each $i_h$, the eccentricity-values were averaged with the error bars indicating the scatter (which is small).  The gray line shows \eq{myrate} with $A_3=1$ in $f_\mathrm{hit}$ (\eq{fhit}).  Setting the calibration factor to $A_3=2.9$ produces an excellent fit. }
\end{figure}
In the Hill regime, the random relative velocity $w$ of the particles falls below $v_h$; the approach velocity is now determined by the Keplerian shear, $\sim$$v_h$.   Also, for the calibration we assume that the vertical accretion radius is limited by the disk scaleheight, $R_z = h_\mathrm{eff}$.  Not all particles approaching at impact parameters $|b|<R_x=2.5 R_h$ manage to enter the Hill sphere, however.  Particles approaching at small impact parameters move on horseshoe orbits and their trajectories will move away from the Hill sphere.  Numerical studies have shown that only particles approaching at impact parameters $1.7R_h<b<2.5R_h$, \ie\ only a fraction $f_\mathrm{in}\approx 0.3$ of $R_\mathrm{col}$, will make it into the Hill sphere \citep[\eg][]{Nishida1983,PetitHenon1986,IdaNakazawa1989,GreenbergEtal1991}.  The average approach velocity of these particles is $\overline{ \frac{3}{2}\Omega b} = 3.2 v_h$ and we take this value as the approach velocity in the Hill regime (see \fg{rintrad}).  Finally, only a fraction $f_\mathrm{hit}$ of the particles that enters the Hill sphere will be accreted. Combining these expressions, we obtain
\begin{equation}
  \lambda_j^\mathrm{(1)} = f_\mathrm{in} f_\mathrm{hit} A_3 \frac{\pi R_x R_z v_a N_{sj}}{2 h_\mathrm{ef}} = 3.77 R_h v_h N_\mathrm{sj}  f_\mathrm{hit}.
  \label{eq:myrate}
\end{equation}
We estimate $f_\mathrm{hit}$ rather crudely by assuming that within the Hill sphere the particle motion is random at an average velocity of $w\sim 2.5v_h$.  This translates to a 2-body impact parameter of $b_\mathrm{col} = R_s v_\mathrm{esc}/2.5 v_h \approx \alpha^{1/2}R_h$.  The hit probability is then simply the ratio of this factor to the two relevant length scales,
\begin{equation}
  f_\mathrm{hit} = A_3 \times \frac{b_\mathrm{col}}{R_h} \times \textrm{min}\left( 1, \frac{b_\mathrm{col}}{h_\mathrm{eff}} \right), 
  \label{eq:fhit}
\end{equation}
where $A_3$ is the three-body calibration factor. In the last term we have taken care that the hit rate is limited to the scaleheight of the particle layer (in which case the interaction becomes truly 2D).

We compare \eq{fhit} with the findings of \citet{IdaNakazawa1989}.  \citet{IdaNakazawa1989} have numerically integrated trajectories of particles in the restricted three-body problem as function of inclination and eccentricity and obtained the collision rate for $\alpha = R/R_h = 10^{-3}$.  Their results are presented by the symbols in \fg{crates} as function of the reduced (Hill) inclination, $i_h = v_z/v_h$.  In Hill units lengths are normalized to $R_h$ and times to $\Omega^{-1}$ such that the accretion rate \eq{myrate} becomes $\lambda_{1j,h} = 3.77 f_\mathrm{hit} N_{sj}$ and \eq{fhit} $f_\mathrm{hit} = A_3 \alpha^{1/2}\textrm{min}(\alpha^{1/2}/i_h, 1)$.  Since in the Hill regime the velocities are determined by the Keplerian shear, the outcome is insensitive to $e_h$; therefore, we have in \fg{crates} for each $i_h$ averaged the \citet{IdaNakazawa1989} results over the eccentricity-values that have $e_h<1$.  \Fg{crates} shows that when $A_3=2.9$ we obtain excellent agreement with the numerical results.  Besides, our simple model correctly predicts the transition to the 2D regime when $h_\mathrm{eff}<b_\mathrm{col}$ ($i_h<\alpha^{1/2}$).  

In the case of a very thin disk, $v_z/v_h < \alpha^{1/2}$, the accretion rate then becomes $dM/dt \approx 11\alpha^{1/2} R_h v_h \Sigma_j$, in agreement with previous estimates \citep[\eg][]{GoldreichEtal2004,Rafikov2004,Chambers2006,Weidenschilling2005}.  This 2D result is the fastest possible accretion rate in our model, but it does not qualify for runaway growth since the mass exponent, $\kappa$ (\se{rg-def}), is less than unity, $\kappa=2/3$ \citep[\cf][]{KokuboIda1996}.

\subsection{Stirring rates: velocity change at interaction radii}
\label{app:stir-delv2}
Before a comparison of \eq{stir} with the literature can be made, we first have to specify the changes in velocity, $\Delta v^2_m$ and $\Delta v^2_M$, for dynamical friction and viscous stirring.  This is the topic of this section.  The results are summarized in \Tb{delv-table}.  In \app{stir-cal} we then perform the calibration.
\begin{deluxetable}{llll}
  \tablecaption{\label{tab:delv-table}Summary of the velocity changes that particles experience upon an interaction}
  \tablewidth{0pt}
  \tablehead{
  Case\tablenotemark{a} & \multicolumn{3}{c}{Velocity change $\Delta v_m^2$ or $\Delta v_M^2$} \\
  \cline{2-4}
                              & Collisions (Bouncing)    & Dynamical friction                         & Viscous stirring\tablenotemark{b}
  }
  \startdata
  
    General                  & \Eq{dvel2-ic}              & \Eq{dvel2-ec}                               &  \Eq{delv2-vs}             \\ 
    $v_m: (m \ll M)$         & $(v_M^2-v_m^2)$            & $4v_M^2 -4(m/M) v_m^2$                      & $v_m^2$ \\
    $v_m: (m = M)$           & $(-3v_m^2+v_M^2)/4$        & $v_M^2 - v_m^2$                             & $v_m^2/4$ \\
    $v_M: (m \ll M)$         & $-2(m/M)v_M^2 +(m/M)^2v_m^2$ & $-4(m/M)v_M^2 +4(m/M)^2 v_m^2$            & $(m/M)^2 v_m^2$ \\
  \enddata
  \tablenotetext{a}{For example, the special case `$v_m: (m \ll M)$' means that the $\Delta v_m^2$ changes are given under the condition that $m\ll M$.}
  \tablenotetext{b}{For viscous stirring, if the interaction takes place in the shear-dominated regime, $v_m^2$ has to be replaced by $(2.5v_h)^2$.}
\end{deluxetable}

\subsubsection{Dynamical friction (elastic collisions)}
We model the velocity change that results from dynamical friction with a 1D fully elastic collision.  The velocity after such a collision is given by
\begin{subequations}
  \label{eq:elas}
  \begin{equation}
    v_M' = \frac{v_M(M-m) + 2 m \overline{v_m}}{(M+m)},
    \label{eq:elas-a}
  \end{equation}
  \begin{equation}
    v_m' = \frac{v_m(m-M) + 2 M \overline{v_M}}{(M+m)}.
    \label{eq:elas-b}
  \end{equation}
\end{subequations}
In \eq{elas} a bar reflects the orientation of the collision; \eg\ $\overline{v_m} = \pm v_m$, dependent on whether the collision is \textit{head-on} (negative values) or \textit{tail-on} (positive).  \Eq{elas} leads to a change in the kinetic energy of
\begin{subequations}
  \label{eq:elas-En}
  \begin{equation}
    \Delta(M v_M^2) = \frac{4mM}{(M+m)^2} (\overline{v_m}-v_M) (M v_M + m \overline{v_m});
  \end{equation}
  \begin{equation}
    \Delta(m v_m^2) = \frac{4mM}{(M+m)^2} (\overline{v_M}-v_m) (M \overline{v_M} + m v_m).
  \end{equation}
\end{subequations}
These expressions add up to 0, reflecting conservation of energy, regardless of whether the bars indicate positive values (tail-on collisions) or negative values (head-on collisions). In order to get a mean change we simply average these expressions over the head-on/tail-on collisions;
\begin{subequations}
  \label{eq:dvel2-ec}
  \begin{equation}
    \Delta(v^2_M)_\mathrm{df} = -\frac{4m}{(M+m)^2} (M v_M^2 - m v_m^2),
  \end{equation}
  \begin{equation}
    \Delta(v^2_m)_\mathrm{df} = \frac{4M}{(M+m)^2} (M v_M^2 - m v_m^2).
  \end{equation}
\end{subequations}
These expressions are used for the velocity changes that result from dynamical friction.

\subsubsection{Inelastic collisions (bouncing or accretion)}
We optionally make the assumption that physical collisions fully dissipate their collision energy.   When implemented, accretion and bouncing are modeled as a perfectly inelastic collision. For a fully inelastic collision we have that the velocity after the collision is the same for both particles
\begin{equation}
	v_M' = v_m' = \frac{M v_M + m v_m}{M+m},
  \label{eq:vu-el}
\end{equation}
although in the case of bouncing they will not stick. Repeating a similar approach as above we end up with the velocity changes as
\begin{subequations}
  \label{eq:dvel2-ic}
  \begin{equation}
    \Delta (v_M^2)_\mathrm{ic} = \frac{m}{(M+m)^2} (-2M v_M^2 + m(v_m^2-v_M^2)),
  \end{equation}
  \begin{equation}
    \Delta (v_m^2)_\mathrm{ic} = \frac{M}{(M+m)^2} (-2m v_m^2  - M(v^2 - v_M^2)),
  \end{equation}
\end{subequations}
which shows that the total energy change $M\Delta(v_M^2) + m\Delta(v_m^2)$ is always negative.

\subsubsection{Viscous stirring: distinction between 2D and 3D interactions}
\label{app:vs}
In free space, the absolute relative velocity before and after a two-body encounter is equal in magnitude.  In that case, only dynamical friction operates.  However, in the protoplanetary disks, due to the influence of the sun, the 2-body energy is not conserved.  This holds even for encounters in the dispersion-dominated (d.d.-) regime, where the Keplerian potential acts as a reservoir with which energy can be exchanged.  From this perspective encounters are in fact always three-body interactions and solar gravitational energy can be converted into random motion, and vice-versa on thermodynamic grounds.  The energy exchange with the solar body preserves the total energy of the system, but not the energy of the two orbiting bodies as a subsystem. The process of extracting energy from (or adding to) the potential is better known as viscous stirring.

To understand viscous stirring physically, we follow a geometrical argument that originated from \citet{Safronov1969}, see also \citet{GoldreichEtal2004}.  We consider the case of a small body with random velocity $v_m$ experiencing a close encounter with a more massive body moving on a circular orbit.  Due to the encounter the phase angle of the smaller body will shift by, on average, 90 degrees but the magnitude of the \textit{local} relative velocity $w$ will stay the same.  The point is that this latter quantity is not equal to $v_m$; at quadrature it is, but if the encounter takes place at perihelion or aphelion the relative velocity of the elliptical orbit with the local Keplerian velocity is $w=v_m/2$.  Then, if the encounter takes place at one of the latter locations and changes the orbit towards quadrature, this will have circularized the orbit of the smaller body.  For example, if $w_\mathrm{a/p}$ denotes the relative velocity at aphelion or perihelion and $w_q$ that at quadrature, the above reasoning reads $w_{a/p}=v_m/2=w'_q=v_m'$ where primes denote the velocity after the encounter. Therefore, $v_m'=v_m/2$.  However, if the encounter takes place at quadrature, $v_m$ could increase by a factor two if the orbit is re-oriented towards perihelion or aphelion ($v_m=w_q=w'_{a/p}=v_m'/2$ and $v_m'=2v$).  On average, these reorientations result in a gain in random energy: the (absolute) change is larger at quadrature than at aphelion or perihelion.

From this discussion it is clear that viscous stirring results from the (random) re-orientation of the phase angle of the bodies motion.  We have defined the viscous stirring radius $R_\mathrm{vs}$ such that $\Delta v_m \sim v_m$ for the small particle.  For the heavy particle, the encounter will lead to a response that is approximately a factor $\sim$$m/M$ less due to its larger inertia.  More formally, we will weigh the change in velocity $\Delta v$ by the masses of the collision partners, \ie\ an amount $M/(M+m) v_m$ goes to the small particle and an amount $m/(m+M) v_m$ to the big particle.  Next, we recognize that these impulses lead to \textit{random} changes in the phase angle and that only the squares add up, \ie\ we define
\begin{subequations}
  \label{eq:delv2-vs}
  \begin{equation}
    \left( \Delta v_m^2\right)_\mathrm{vs} = \left(\frac{M}{m+M} v_m\right)^2
  \end{equation}
  \begin{equation}
    \left(\Delta v_M^2\right)_\mathrm{vs} = \left(\frac{m}{m+M} v_m\right)^2.
  \end{equation}
\end{subequations}
as the change in random energy resulting from a viscous stirring encounter.

In the shear-dominated (s.d.-) regime, scatterings that take place within the Hill sphere can be strong;  velocities of the small particles are boosted to $\sim$$v_\mathrm{h}$.  Consequently, instead of the previous expression which holds for the d.d.-regime, we have $(\Delta v_m^2)_\mathrm{vs} = ([M/(M+m)]2.5 v_h)^2$ and $(\Delta v_M^2)_\mathrm{vs} = ([m/(m+M)] 2.5v_\mathrm{h})^2$.  Here, we normalize to $2.5v_h$ since this is the characteristic velocity for interaction in the s.d.-regime.

In the next section, we will calibrate the resulting stirring rates against the stirring rates obtained by \citet{OhtsukiEtal2002} and invoke an order of unity correction, $f_\mathrm{vs}$.  However, rather than merely a constant, we will see that $f_\mathrm{vs}$ is different for horizontal ($v$) and vertical ($v_z$) velocities and also depends on $\beta=v_z/v$.  Viscous stirring is highly dependent on the geometry of the collision. In the s.d.-regime ($w/v_h \ll 1$) interactions are typically 2D since the radius at which the interaction takes place ($R_\mathrm{vs}$) is typically larger than $h_\mathrm{eff}$.  This means that eccentricities are more strongly excited than inclinations. The latter's amplitudes are correspondingly reduced by a factor $(h_\mathrm{eff}/R_\mathrm{vs})$. Of course in a fully 2D setting one would not stir the inclinations.

However, in the d.d.-regime, interactions are modeled as 3D.  Viscous stirring then equalizes the random velocity components, meaning that $\beta$ is being driven to an equilibrium value of $\beta=0.5$ for a Keplerian disk \citep{IdaEtal1993,TanakaIda1996}. This can be thought of as a kind of equipartition, although as noted before there is no two-body energy conservation for viscous stirring.  It is even possible for viscous stirring to produce negative rates when the ratio $\beta = v_z/v \ll 1$, for which $f_\mathrm{vs}$ will also become negative.

We will therefore distinguish between the 2D and 3D regimes when next discussing the magnitude (and sign) of the viscous stirring factors $f_\mathrm{vs}$ and $f_\mathrm{vs-z}$.

\subsection{Stirring rates: literature calibration}
\label{app:stir-cal}
We compare our viscous stirring expression that follows from \eq{stir} and the discussion above, against previous literature studies \citep{IdaEtal1993,TanakaIda1996,StewartIda2000,OhtsukiEtal2002}.  For dynamical friction we will not introduce a calibration factor as we will find that the present formulation models the dynamical friction stirring rates reasonably well (see \app{stirring-plots}).

According to \citet{OhtsukiEtal2002},  the stirring of the $m$-particle in the limit of $m \gg M$ is given by,\footnote{
Here, we have rewritten Ohtsuki's expression in our notation, see Eq.\ 6 of \citet{OhtsukiEtal2002}. For the eccentricity stirring, under the assumptions outlined above, we have (in Ohtsuki's notation)
\begin{equation}
  \frac{d\langle e_1^2\rangle_\mathrm{vs} }{dt}= a_0^2\Omega N_{sj} h_{12}^4 \langle P \rangle_\mathrm{vs}, 
  \label{eq:Ohtsuki6}
\end{equation}
where we have used that $(m=) m_1 \ll m_j (=M)$ with $e_1 = v_m/v_k$ the eccentricity, $a_0$ the semi-major axis, $N_{sj}$ the column density, and $h_{12} = (R_h/a)^4$.  Multiplying both sides by $(a\Omega)^2$ then gives \eq{du2-osi}.}
\begin{equation}
  \frac{dv^2_{m,\mathrm{vs}}}{dt} = \langle P_\mathrm{vs} \rangle \Omega^3 N_{sj} R_h^4,
  \label{eq:du2-osi}
\end{equation}
where $\langle P \rangle_\mathrm{vs}$ represents the dimensionless viscous stirring factor for eccentricity, averaged over the distribution.  Similarly, $\langle Q \rangle_\mathrm{vs}$ encapsulates the stirring of inclinations due to viscous stirring.  These functions are in turn functions of the (reduced) inclination and eccentricity, $i_h$ and $v_h$.  In addition, $\langle P \rangle_\mathrm{vs}$ and $\langle Q \rangle_\mathrm{vs}$ depend on the velocity regime, \ie\ the shear-dominated ($v_m\ll v_h$) and dispersion-dominated ($v_m\gg v_h$) regimes.  The s.d.-regime is modeled as 2D; interaction take place at a small angle, $\theta = h_\mathrm{eff}/R_\mathrm{vs} \ll 1$. The d.d.-regime is assumed 3D ($\theta\sim1$).

We introduce the symbol $f_\mathrm{vs}(\beta)$ as the calibration constant for viscous stirring.  Since $\langle P \rangle_\mathrm{vs}$ and $\langle Q \rangle_\mathrm{vs}$ are different in the 2D and 3D regime we discuss the regimes separately.

\label{app:app-vel}
\subsubsection{2D regime (shear-dominated), $h_\mathrm{eff} < R_\mathrm{vs}$}
We assume the disk is flat compared to $R_\mathrm{vs}$, such that $R_z = h_\mathrm{eff}$.  Inserting $R_\mathrm{int} = 2.5 R_h$, $R_z=h_\mathrm{eff}$, $v_a=3.2 v_h$, and $\Delta v_m^2 = f_\mathrm{vs} (2.5v_h)^2$ into \eq{stir}, we obtain a stirring rate of
\begin{equation}
  \frac{dv^2_{m,\mathrm{vs}}}{dt} = f_\mathrm{in} \frac{\pi (2.5R_h) (3.2v_h) }{2} N_{sM} f_\mathrm{vs} (2.5v_h)^2 = 23 f_\mathrm{vs} N_{sM} R_h v_h^3.
  \label{eq:stirring-vs}
\end{equation}
where $f_\mathrm{in}=0.3$ is the fraction of particles that enter the Hill sphere, see \app{Hill}.  \citet{OhtsukiEtal2002} argue on basis of numerical integrations \citep{StewartIda2000} that $\langle P_\mathrm{vs} \rangle = 73$ in the s.d.-regime. Taking this value and equating with \eq{du2-osi}, we arrive at a value of $f_\mathrm{vs} \approx 3.0$.  

For the vertical excitations, we have argued that $f_\mathrm{vs-z}$ should contain a factor $\theta^2 = (h_\mathrm{eff}/R_\mathrm{vs})^2 = (v_z/2.5v_h)^2 = 0.16 \beta^2 (v/v_h)^2$, reflecting the geometry of the encounter. Similarly, \citet{OhtsukiEtal2002} gives $\langle Q_\mathrm{vs} \rangle \approx 4\beta^2 (v/v_h)^2$. Comparing these expressions we find
\begin{equation}
  f_\mathrm{vs-z} = 1.1 \left( \frac{h_\mathrm{eff}}{R_\mathrm{vs}} \right)^2,
  \label{eq:stirring-vsz}
\end{equation}
since then $23f_\mathrm{vs-z}$ equals $\langle Q_\mathrm{vs} \rangle$.

\subsubsection{3D regime (dispersion-dominated), $v_m > 2.5v_h$}
When $v_m\gg v_h$ it can be assumed that $R_z = R_x = R_\mathrm{vs} \ll h_\mathrm{eff}$ and $\beta=0.5$. Inserting $R_x^2 = 36R_h (v_h/v_m)^4$, $v_a/2h_\mathrm{eff} = \Omega^{-1}$, and $\Delta v_m^2 = f_\mathrm{vs}(\beta) v_m^2$, \eq{stir} becomes
\begin{equation}
  \frac{dv_m^2}{dt} = 36\pi f_\mathrm{vs}(\beta) v_m^2 R_h^2 \left( \frac{v_h}{v_m} \right)^4.
  \label{eq:dv2dt-me}
\end{equation}
Now \citet{OhtsukiEtal2002} gives for the high velocity regime
\begin{equation}
  \frac{d\langle P \rangle_\mathrm{vs}}{dt} = \frac{72 I_\mathrm{vs}^P}{\pi e_h i_h} \left[ \log (1+\Lambda^2) - \frac{\Lambda^2}{1+\Lambda^2} \right],
  \label{eq:Pvs}
\end{equation}
where $I_\mathrm{vs}^P(\beta)$ is an integral expression that depends only on $\beta$ and $\Lambda$ is the Coulomb factor.  A similar relation exists for $\langle Q \rangle_\mathrm{vs}$ but then with $I_\mathrm{vs}^Q(\beta)$.  The functions $I_\mathrm{vs}^P$ and $I_\mathrm{vs}^Q$ are plotted in \fg{IvsPQ}.  In evaluating the $I_\mathrm{vs}^{P/Q}(\beta)$ terms we have approximated the formal definition (which involves an integral expression that cannot be analytically solved) by an exponential fit, $I_\mathrm{vs}^{P/Q} \approx a_0 + a_1 \exp[-a_2 \beta]$, see \fg{IvsPQ}.\footnote{\citet{Chambers2006} applies a similar fit to the $P_\mathrm{vs}$ and $Q_\mathrm{vs}$ expressions.}

\begin{figure}[tbp]
  \centering
  \includegraphics[width=8cm,clip]{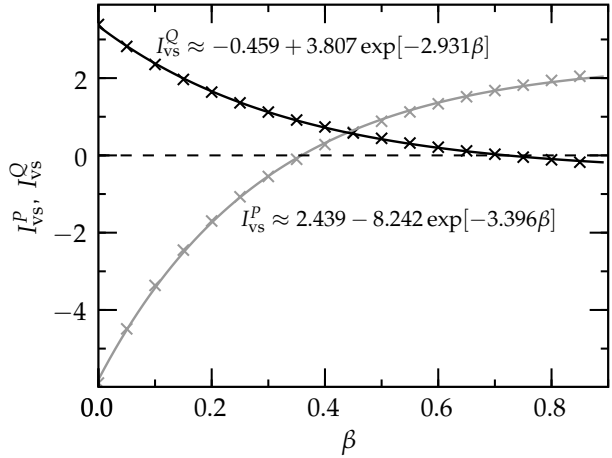}
  \caption{\label{fig:IvsPQ}The functions $I_\mathrm{vs}^P(\beta)$ and $I_\mathrm{vs}^Q(\beta)$ as defined by \citet{OhtsukiEtal2002}. Crosses denote numerical evaluations at discrete values of $\beta$.  The curves present a fit to $I_\mathrm{vs}^{P/Q}$ of the form $f(\beta) = a_0+a_1\exp[-a_2\beta]$.  The fit parameters are indicated.}
\end{figure}
We approximate the term in the square brackets in \eq{Pvs} as $\approx$$2\log \Lambda$, which is appropriate if $\Lambda \gg 1$.  Inserting \eq{Pvs} into \eq{du2-osi} with $e_h = v_m/v_h$ and $i_h=\beta e_h/v_h=v_m/2v_h$ then gives
\begin{equation}
  \frac{dv_m^2}{dt} = \frac{288 I_\mathrm{vs}^P v_h^4 R_h^2 N_{sM}}{\pi v_m^2} (\log \Lambda) \Omega^{-1}
  \label{eq:dv2dt-osi}
\end{equation}
and a similar term (but then with $I_\mathrm{vs}^Q$) for the vertical velocities.  Comparing \eq{dv2dt-osi} with \eq{dv2dt-me} gives
\begin{equation}
  \arrayii{f_\mathrm{vs}}{f_\mathrm{vs-z}} = \frac{8}{\pi^2} \arrayii{I_\mathrm{vs}^P(\beta)}{I_\mathrm{vs}^Q(\beta)},
  \label{eq:fvs-dd}
\end{equation}
where we have ignored the Coulomb term (see \app{stirring-plots}).

Again, we see that our calibration factors are of order of unity.  However, for low $\beta$-values $I_\mathrm{vs}^P$ becomes negative: eccentricities are strongly damped and inclinations strongly excited.  The net effect of the negative eccentricity stirring is then to (rapidly) increase $\beta$, until an equilibrium value $\beta^\ast$ is reached, for which ${\beta^\ast} = \sqrt{I_\mathrm{vs}^Q/I_\mathrm{vs}^P} \approx 0.55$. This behavior is quite the reverse from the s.d.-regime, where any inclination-stirring is strongly suppressed with respect to the eccentricity.  The key physical reason is that in the s.d.-regime the interactions take place at a radius much larger than the disk height, $R_\mathrm{vs} > h_\mathrm{eff}$, which enforces the 2D geometry.  In the d.d.-regime, on the other hand, $R_\mathrm{vs} < h_\mathrm{eff}$ and interactions are 3D.  Of course, our simplified treatment -- to identify the s.d.-regime with the 2D case and the d.d.-regime with the 3D -- does not fully do justice to the full complexity at the transition $v_m\sim v_h$.  When a system moves from the s.d.-regime into the d.d.-regime, interactions are initially still 2D.  However, the outcome will qualitatively be the same: $R_\mathrm{vs}$ decreases, interactions become increasingly 3D, and any stirring only enhances this trend.  Indeed, \citet{RafikovSlepian2010} have recently studied this specific setting and found that the 2D d.d.-transition regime is extremely short-lived.

\subsection{Final expressions concerning stirring and comparison with \citet{OhtsukiEtal2002} stirring curves}
\label{app:stirring-plots}
We summarize the expressions for stirring rates that result from the geometrical model.  These can be concisely written as
\begin{equation}
  \label{eq:stir-gen}
  \frac{d}{dt} \arrayii{v^2}{v_z^2} = \pi R_x R_z N_{sj} \arrayii{f_\mathrm{vs}(\beta)}{f_\mathrm{vs-z}(\beta)} (\Delta v^2)_\mathrm{int} \log \Lambda.
\end{equation}
In the s.d.-regime $f_\mathrm{vs}=3.0$ and $f_\mathrm{vs-z}$ is given by \eq{stirring-vsz}. In the d.d.-regime $f_\mathrm{vs}(\beta)$ is given by \eq{fvs-dd}.  Furthermore, the d.d.-regime is characterized by a Coulomb term, $\log \Lambda$.  The Coulomb term takes account of the encounters that occur at $R_\mathrm{int}<b<h_\mathrm{eff}$, which collectively can give a contribution -- which is of order unity, but important for the numerical implementation of our work.  We estimate the magnitude of the contribution to be determined by the ratio of $R_\mathrm{int}$ to the scaleheight,
\begin{equation}
  \log \Lambda = \log \left(\exp[1] + \frac{h_\mathrm{eff}}{R_\mathrm{int}} \right),
  \label{eq:coulomb}
\end{equation}
(the inclusion of the $\exp[1]$ term ensures that $\log \Lambda\ge 1$ for all values of $h_\mathrm{eff}/R_\mathrm{int}$).

\begin{figure}[tbp]
  \centering
  \includegraphics[width=0.7\textwidth, clip]{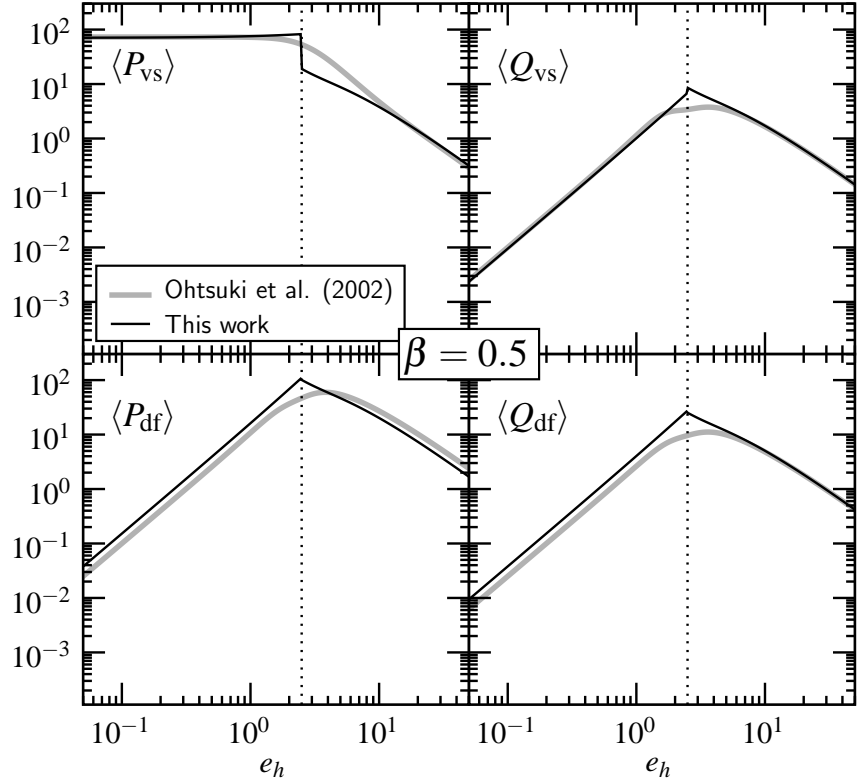}
  \caption{\label{fig:osi-b0.5}. The dimensionless stirring rates for eccentricities ($\langle P \rangle_\mathrm{vs}$, $\langle P \rangle_\mathrm{df}$) and inclinations ($\langle Q \rangle_\mathrm{vs}$, $\langle Q \rangle_\mathrm{df}$) as function of Hill eccentricity, $e_h = v/v_h$.  The thick gray lines represents the stirring rates according to \citet{OhtsukiEtal2002}, while the thin black lines are obtained using our geometrical model.  The dotted line at $e_h=2.5$ identifies the transition between the s.d.\ and d.d.-regimes.  }
\end{figure}
These expressions can be compared with other works, \eg\ \citet{OhtsukiEtal2002}.  \Fg{osi-b0.5} presents the stirring rates that follow form \eq{stir-gen} for $\beta=0.5$.  Rates are given in Hill units, where lengths are normalized to $R_h$ and times to $\Omega^{-1}$ and $N_\mathrm{sj}=1$, see the discussion in \app{Hill}.  One then arrives at the dimensionless expressions denoted by $P_\mathrm{vs}, P_\mathrm{df}$, \etc, for which \citet{OhtsukiEtal2002} gives analytical fits.  These are plotted in \fg{osi-b0.5}. 

Comparing the curves, one observes a satisfactory correspondence.  For viscous stirring the agreement is not so surprising, since we have calibrated the expressions against \citet{OhtsukiEtal2002}.  For dynamical friction, however, we have not done so, and the close match vindicates our geometrical model. Note, finally, that the accuracy of our geometrical model breaks down near $v\sim 2.5v_h$ ($e_h\sim2.5$) which indicates the transition between the s.d.- and d.d.-regimes, which is in our case sharp by construction.

\section{Calculation of interaction rates in multi-zone setting}
\label{app:calc-ints}
\noindent
The interaction rates between two particle swarms over a region of space $V$ is given by
\begin{equation}
   \lambda_{12} = \int_V d^3\mathbf{x}\ n_1(\mathbf{x}) n_2(\mathbf{x}) \sigma_\mathrm{int}(\mathbf{x}) v_a,\qquad \mathrm{[s}^{-1}] 
  \label{eq:2rate}
\end{equation}
where $n_{1}(\mathbf{x})$ and $n_{2}(\mathbf{x})$ are the particle number densities, $\sigma_\mathrm{int} = \pi R_x R_z$ the interaction cross section, $v_a$ the approach (relative) velocity, and $V$ the volume that is under consideration. We approximate the $n_1(\mathbf{x})$ as
\begin{equation}
  n_1(\mathbf{x}) = n_1(x,y,z) = N_1 \frac{P_1(x)}{(2\pi a) 2h_1},\qquad (|z|\le h_1)
  \label{eq:n1}
\end{equation}
where $N_1$ is the total number of particle in the swarm, $h_1$ the scaleheight, $P_1(x)$ the probability density of the first particle over the radial coordinate $x$. The number density is zero for heights $|z|>h_1$. Using \eq{n1} we rewrite \eq{2rate} as
\begin{equation}
  \lambda_{12} = \frac{\pi v_a N_1 N_2}{4(2\pi a)} \phi_x \phi_z,
  \label{eq:lam-phi}
\end{equation}
where the integration of the azimuthal coordinate ($y$-direction) has canceled one factor of $2\pi a$ and $\phi_x$, $\phi_z$, the filling factors, are defined as
\begin{equation}
    \phi_x =  \int dx\ 2R_x(x) P_1(x) P_2(x);
    \label{eq:phix}
\end{equation}
\begin{equation}
   \phi_z = \int dz\ \frac{2R_z}{2h_1 2h_2}.
  \label{eq:phiz}
\end{equation}
 The filling factors have the following geometrical interpretation: it gives the fraction of the space with respect to radial or vertical dimension of $V$ over which the interaction can take place.  We therefore have that $0 \le \phi_{x,z} \le 1$.  If the filling factor is unity, the interaction between the two particle swarms can take place at any point.  For example, if $R_z \gg h_1, h_2$ the particles can interact irrespective of their $z$-coordinates and $\phi_z=1$.  However, in the more general case of fractional filling factors the interactions are restricted to take place in a fraction of the volume that is under consideration -- \ie\ when the particles are within a distance $R_z$.  Let us first consider \eq{phiz}.  The integration over $z$ proceeds over a length given by the minimum of the scaleheights, since interactions can only take place if both particles are present.  Therefore, \eq{phiz} integrates to $R_z/h_\mathrm{eff}$, where the effective scaleheight, as defined before, is the maximum of the two scaleheights.  However, there is one caveat: $\phi_z$ is not allowed to exceed unity (indeed, we have restricted the interaction range before).  Thus we have
\begin{equation}
  \phi_z = \min(h_\mathrm{eff}/R_z, 1).
\end{equation}

\begin{figure}[t]
  \centering
  \includegraphics[width=10cm]{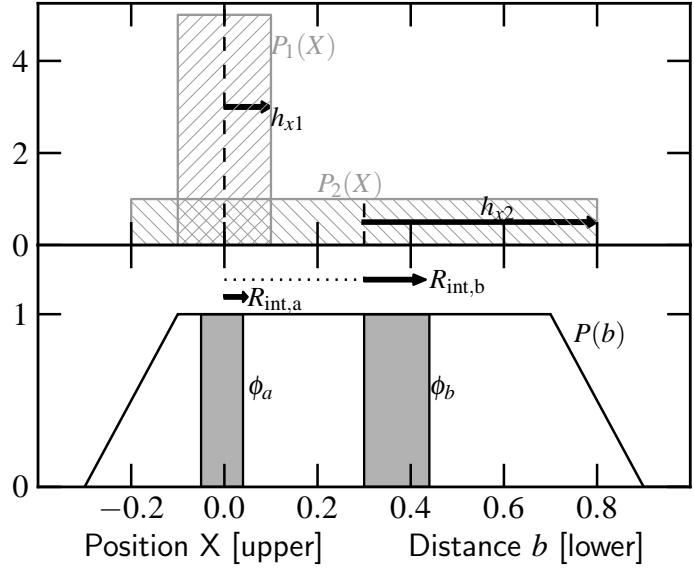}
  \caption{\label{fig:Pd}. (\textit{top} panel) Position density distribution $P(X)$ of two particle groups.  The distributions are assumed to be uniform with scalelengths $h_{x1}$ and $h_{x2}$, respectively. (\textit{bottom} panel) The \textit{distance} probability density function obtained from the cross-correlation of $P_1(X)$ and $P_2(X)$ (black curve).  Distributions integrate to unity.  The filling factors for two interaction radii $R_{\mathrm{int},a}$ and $R_{\mathrm{int},b}$ are indicated.}
\end{figure}
Next, we consider the radial filling factor $\phi_x$.  The situation is somewhat more complicated here, since the particle distributions are not centered at the same position $X$, see \fg{Pd}.  In \fg{Pd}a (top panel) the distribution density function of two groups is given.  The unit of length is arbitrary.  It is assumed that the distribution is uniform with horizontal widths $h_{x1}$ and $h_{x2}$, respectively.  In the figure the first particle group is, without loss of generality, centered at $X=0$ and characterized by a scalelength $h_{x1}=0.1$.  The second particle swarm is extended over a total length of unity and centered at $X=0.3$.

We intuitively recognize from \fg{Pd} that the amount of overlap between the populations is given by the distance distribution $P(d)$ among the bodies.  That is, the filling factor $\phi_x$ represents the fraction of particles that comes within reach of the interaction radius $R_\mathrm{int}$.  This interpretation refines and supersedes \eq{phix}.  The procedure to compute $\phi_x$ is outlined in  \fg{Pd}b. the distribution of relative distances can be obtained from
\begin{equation}
  P(b) = \int P_1(x) P_2(x+b) \mathrm{d}x.
  \label{eq:Pd}
\end{equation}
\Eq{Pd} is mathematically equivalent to the cross correlation of the distributions.  For the parameters given above, $P(b)$ is given in the lower panel of \fg{Pd} by the black solid line. The range in distances, or impact parameters, spans from $b=-0.3$ to $d=0.9$, corresponding to the cases where the particles are furthest apart (negative values mean here that the second particle is to the left of the first one). $P(b)$ also integrates to unity.  The arrow in \fg{Pd} indicates the interaction radius, $R_\mathrm{int}$.  Two interaction radii are drawn.  The first, $R_{\mathrm{int},a}$ operates at distances $|b|<0.1$, while $R_{\mathrm{int},b}$ does not operate at distances $|b|<0.3$.  If particles come within the interaction range, the particles are allowed to interact.  In this particular example we have chosen $R_{\mathrm{int},a}=0.05$ and $\phi_{x,a}$ integrates to 0.1: the area of the shaded box in \fg{Pd}.

In the special case where the distributions are centered at $X=0$ we have, like the $z$-direction, $P(x)=1/2h_{x}$ and arrive at $\phi_x = R_x/h_{x,\mathrm{eff}}$ (where $h_{x,\mathrm{eff}}$ is the largest scalelength).  Inserting these expressions into \eq{lam-phi} we obtain $\lambda_{12} = \pi v_a N_1 N_2 R_x R_z /8\pi a h_\mathrm{eff} h_{x,\mathrm{eff}}$.   If we consider the interaction of a single protoplanet ($N_1=1$) and a swarm planetesimals, the \textit{surface density} of the latter is $N_{s2} = N_2/(2\pi a) (2h_{x,\mathrm{eff}})$ and \eq{1rate} is retrieved.

The advantages of the filling factor approach is that the formalism can be extended to more general situations where it concerns the particle distributions or interaction radius $R_\mathrm{int}$.  The filling factor formalism becomes especially advantageous for interactions in the Hill regime, where only interactions at $1.7 < b/R_h <2.5$ are allowed (particles on smaller impact parameters move on horseshoe orbits and do not enter the Hill sphere).  The algorithm then neatly takes account of this inner `gap', as is sketched in \fg{Pd}.  In fact the function $P(b)$ can be expressed as a series of step functions, that can be integrated analytically, which gives a cumulative density function, from which the filling factors are readily obtained.

\section{List of frequently used symbols and abbreviations}
\label{app:symbols}
\small
\begin{deluxetable}{lp{10cm}l}
  \tablecaption{List of symbols and abbreviations \label{tab:caption}}
  \tablewidth{0pt}
  \tablehead{ \colhead{Symbol} & \colhead{Description} & \colhead{Reference} }
  \startdata
  $\alpha$                & ratio $R/R_\mathrm{h}$                            & \eq{alpha} \\
  $\beta$                 & ratio inclination:eccentricity ($=i/e=v_z/v$) \\
  $\gamma$                & turbulent stirring parameter                      & \se{Tstir}\\
  $\Delta a$              & grid resolution                                   & \fg{geopic2} \\
  $\Delta a_\mathrm{sim}$ & total simulation width                            & \fg{geopic2} \\
  $\Delta v_\mathrm{int}$ & velocity change upon interaction                  & \app{stir-delv2} \\
  $\eta$                  & nebula pressure parameter                         & \Tb{disk} \\
  $\theta$                & deflection angle                                  & \se{interactions} \\
  $\kappa$                & accretion rate index as in $dM/dt \propto M^\kappa$ & \se{rg-def} \\
  $\lambda_j^\mathrm{(1)}$          & single particle interaction rate                  & \eq{1rate} \\
  $\lambda_{12}$          & group interaction rate                            & \eq{2rate} \\
  $\rho_s$                & internal density of bodies (including pores)      & \se{diskprops} \\
  $\rho_g$                & gas density                                       & \se{diskprops} \\
  $\sigma_\mathrm{int}$   & interaction cross section ($\equiv \pi R_x R_z$)  & \\
  $\Sigma$                & surface density in solids                         & \se{diskprops} \\
  $\Sigma_g$              & surface density in gas                            & \se{diskprops} \\
  $\phi_\mathrm{vs,50}$   & filling factor for the bodies that comprise 50\% of the stirring power & \se{conv} \\
  $\phi_x, \phi_z$        & filling factor for interactions                   & \app{calc-ints}\\
  $\Omega$                & local orbital frequency \\
  $A_i$                   & calibration constant                              & \se{colrates}\\
  RB                      & representative body                               & \se{MC} \\
  $a$                     & semi-major axis \\
  $a_\mathrm{fr}$         & fragment size                                     & \se{colmod}\\
  $b$                     & impact parameter \\
  $C_D$                   & drag coefficient                                  & \se{diskprops} \\
  $c_g$                   & sound speed                                       & \se{diskprops} \\
  d.d.                    & dispersion-dominated \\
  $e$                     & eccentricity ($=v/v_k$)                           & \se{key-def} \\
  $e_h$                   & Hill eccentricity ($=v/v_h$)                      & \app{Hill} \\
  $f_\mathrm{vs}$         & order of unity factor that enters into the viscous stirring rate expression for the planar component (eccentricities) & \app{vs} \\
  $f_\mathrm{vs-z}$       & order of unity factor that enters into the viscous stirring rate expression for the vertical component (inclinations) & \app{vs} \\
  $G$                     & Newton's constant                                 & \\
  GF                      & gravitational focusing                                 & \\
  $g_i$                   & number of physical bodies belonging to species $i$ & \app{app-rg} \\
  $h_\mathrm{eff}$        & effective scaleheight of interaction ($=w_z/\Omega$)      & \se{interactions} \\
  $h_x$                   & scalelength over which particles are distributed   & \fg{geopic2}\\
  $h_z$                   & scaleheight                                               & \fg{geopic2}\\
  $i$                     & inclination                                               & \se{key-def} \\
  $m$                     & particle mass  \\
  $K_\mathrm{rg}$         & dimensionless runaway growth timescale ($=T_\mathrm{rg}/t_\mathrm{run}$)                & \se{Rtr} \\
  $M$                     & particle mass (of big bodies) \\
  $M_1$                   & mass of most massive body in population                   & \se{rg-indicators}\\
  $M_2$                   & mass of second-most massive body in population                  & \se{rg-indicators} \\
  $M_\mathrm{tot}$        & total (solid) mass in system                              & \se{MC} \\
  $m_\ast$                & characteristic mass of distribution                       & \eq{mpeak}\\
  $N_{sj}$                & surface density of $j$-particles \\
  $N_s(m)$                & surface density spectrum                                   & \se{rg-indicators}\\
  $N_g$                   & group size (property of RB)                               & \se{MC} \\
  $N_g^\ast$              & desired group size (as given by algorithm)                & \se{MC}/\app{app-rg} \\
  $N_\mathrm{rb}$         & total number of RBs (computational particles) in simulation & \se{MC} \\
  $N_\mathrm{res}$        & total number of RBs per zone ($=N_\mathrm{rb}/N_\mathrm{zo}$) & \se{Inaba} \\
  $N_\mathrm{zo}$         & number of zones                                           & \fg{geopic2} \\
  $f_\mathrm{in}$         & fraction of particles within $|b|<2.5R_h$ that enters the Hill sphere  & \app{Hill} \\
  $f_\mathrm{hit}$        & collision probability of particles in Hill sphere         & \app{Hill} \\
  $f_\mathrm{frag}$       & fraction of mass in fragments                             & \se{rgsims} \\
  $f_\mathrm{tot}$        & cumulative fraction of mass in fragments                  & \se{rgsims} \\
  $p$                     & index of power-law mass distribution                      & \se{rg-indicators}\\
  $R_\ast$                & radius corresponding to peak of $M_1/m_\ast$              & \se{rgsims} \\
  $R_0$                   & initial planetesimal radius\\
  $R_1(t)$                & radius of most massive body at time $t$                   & \se{rg-indicators} \\
  $R_\mathrm{df}$         & interaction radius for dynamical friction                 & \se{interactions}\\
  $R_\mathrm{int}$        & interaction radius                                        & \se{interactions}\\
  $R_f$                   & final radius of most massive body \\
  $R_h$                   & Hill radius                                               & \eq{Rh} \\
  $R_\mathrm{pp}$         & Radius of protoplanet   & \\
  $R_\mathrm{rg/oli}$     & transition radius between runaway growth and oligarchy in 2-component approximation   & \eq{Roli} \\
  $R_s$                   & combined radius ($=R_1+R_2$)\\
  $R_\mathrm{tr}$         & radius between runaway growth and oligarchy phase         & \eq{R-rg} \\
  $R_\mathrm{vs}$         & interaction radius for viscous stirring                   & \se{interactions}\\
  $R_\mathrm{vs-d}$       & interaction radius for viscous stirring (distant interactions) & \se{interactions}\\
  $R_x$                   & geometrically constrained interaction radius in planar direction\\
  $R_z$                   & geometrically constrained interaction radius in vertical direction\\
  RG                      & runaway growth \\
  s.d.                    & shear-dominated \\
  $T_\mathrm{ac}$         & accretion (growth) timescale                              & \eq{tac} \\
  $T_\mathrm{ac}^\mathrm{2D-sd}$  & accretion timescale in 2D, shear-dominated regime & \eq{Tac-sd} \\
  $T_\mathrm{rg}$         & runaway growth timescale                                  & \eq{Trg} \\
  $t$                     & time \\
  $t_\mathrm{drag}$       & friction (stopping) time of bodies                        & \se{colmod} \\
  $t_\mathrm{run}$        & fiducial timescale that depends on initial conditions only & \eq{trun} \\
  $v$                     & planar rms-velocity dispersion                            & \se{key-def} \\
  $v_M$                   & planar rms-velocity dispersion of heaviest particle in interaction      & \se{key-def} \\
  $v_a$                   & approach velocity                                                       & \se{interactions} \\
  $v_\mathrm{esc}$        & escape velocity                                                         & \se{key-def}\\
  $v_h$                   & Hill velocity                                                           & \eq{Rh} \\
  $v_k$                   & local orbital Keplerian velocity \\
  $v_m$                   & planar rms-velocity dispersion of lightest particle in interaction      & \se{key-def} \\
  $v_x$                   & maximum random velocity of the particle distribution                    & \se{rg-indicators}\\
  $v_z$                   & vertical rms-velocity dispersion (usually of the large body(ies))       & \se{key-def}\\
  $w$                     & relative rms-velocity of two particles (excluding Keplerian shear)      & \se{interactions} \\
  $w_z$                   & relative rms-velocity of two particles (vertical components)            & \se{interactions} \\
  \enddata
\end{deluxetable}
\end{document}